\documentclass[reprint,showpacs,superscriptaddress, english]{revtex4-1}
\usepackage[utf8]{inputenc}
\usepackage{color}
\usepackage{array}
\usepackage{verbatim}
\usepackage{multirow}
\usepackage{amsmath}
\usepackage{amssymb}
\usepackage{dsfont}
\usepackage{graphicx}
\usepackage{esint}
\usepackage[unicode=true,
 bookmarks=true,bookmarksnumbered=true,bookmarksopen=false,
 breaklinks=true,pdfborder={0 0 0},backref=false,colorlinks=true]
 {hyperref}
\usepackage{cleveref}
\usepackage{longtable}

\crefname{section}{Section}{Sections}
\crefname{appendix}{Appendix}{Appendices}
\crefname{equation}{Eq.}{Eq.}
\crefname{table}{Table}{Tables}
\crefname{figure}{Figure}{Figures}
\newcommand{\figref}[2][]{Figure \hyperref[#2]{\ref*{#2}}#1}
\renewcommand{\onlinecite}[1]{Ref. \cite{#1}}

\begin{document}
\title{Quantitative mappings between symmetry and topology in solids}
\author{Zhida Song}
\affiliation{Beijing National Research Center for Condensed Matter Physics,
and Institute of Physics, Chinese Academy of Sciences, Beijing 100190, China}
\affiliation{University of Chinese Academy of Sciences, Beijing 100049, China}
\author{Tiantian Zhang}
\affiliation{Beijing National Research Center for Condensed Matter Physics,
and Institute of Physics, Chinese Academy of Sciences, Beijing 100190, China}
\affiliation{University of Chinese Academy of Sciences, Beijing 100049, China}
\author{Zhong Fang}
\affiliation{Beijing National Research Center for Condensed Matter Physics,
and Institute of Physics, Chinese Academy of Sciences, Beijing 100190, China}
\author{Chen Fang}
\email{cfang@iphy.ac.cn}
\affiliation{Beijing National Research Center for Condensed Matter Physics,
and Institute of Physics, Chinese Academy of Sciences, Beijing 100190, China}
\affiliation{CAS Center for Excellence in Topological Quantum Computation, Beijing, China}
\begin{abstract}
The study of spatial symmetries was accomplished during the last century, and had greatly improved our understanding of the properties of solids. Nowadays, the symmetry data of any crystal can be readily extracted from standard first-principles calculation. On the other hand, the topological data (topological invariants), the defining quantities of nontrivial topological states, are in general considerably difficult to obtain, and this difficulty has critically slowed down the search for topological materials. Here, we provide explicit and exhaustive mappings from symmetry data to topological data for arbitrary gapped band structure in the presence of time-reversal symmetry and any one of the 230 space groups. The mappings are completed using the theoretical tools of layer construction and symmetry-based indicators. With these results, finding topological invariants in any given gapped band structure reduces to a simple search in the mapping tables provided.
\end{abstract}

\maketitle

\section{Introduction}\label{sec:introduction}

Distinct phases do not always differ from each other in their symmetries as expected in the Ginzburg-Landau paradigm.
Two gapped phases having the same symmetry may be distinguished by a set of global quantum numbers called topological invariants \cite{Thouless1982,Haldane1983,Affleck1987,Wen1990,Kitaev2001,Kane2005,Moore2007,Fu2007a}.
These invariants are quantized numbers, whose types (integer, Boolean and others) only depend on the symmetry group and the dimension of the system \cite{Schnyder2008,Kitaev2009}.
The invariants fully characterize topological properties that are unchanged under arbitrary adiabatic tuning of the Hamiltonian that preserves the relevant symmetry group.
Materials having nonzero topological invariants are loosely called topological materials (whereas the technical term is symmetry protected topological states \cite{Gu2009,Chen2012,Lu2012}), a new kind of quantum matter that hosts intriguing physical observables such as quantum anomaly on their boundaries \cite{Bardeen1969,Haldane1988,Ryu2012}, and are considered candidate materials for new quantum devices \cite{Bernevig2006,Chang2013,Liu2013,Qian2014,Fang2017}.
Success in finding these materials largely depends on the numerical evaluation (prediction) of the topological invariants in a given candidate material.
However, even for electronic materials having weak electron correlation, where the topological invariants are best understood and expressed in terms of the wave functions of the valence bands, these calculations still prove quite challenging.
In fact, numerically finding a new topological material has proved so hard that a single success \cite{Bernevig2006,Zhang2009,Hsieh2012,Wang2016,Wieder2017} would have triggered enormous interest \cite{Konig2007,Chen2009,Hsieh2009,Tanaka2012,Dziawa2012,Xu2012,Ma2017}.

On the other hand, mathematicians and physicists have since long developed, via the representation theory of space groups, a complete toolkit for the study of the symmetry properties of bands in solids \cite{Hahn2002,Bradley2010}.
Given any point in momentum space, each energy level corresponds to an irreducible representation (irreps) of the little group at that momentum, depending on the Bl{\"o}ch wave functions at the level.
Modern implementations of the density functional theory output both the energy levels and their Bl{\"o}ch wave functions for any given crystal momenta, such that finding the irreps of all valence bands in a band structure is now considered a solved problem that can be automated.

It has been eagerly hoped that quantitative relations exist between the topological invariants and the irreps in the valence bands at high-symmetry points in the Brillouin zone, i.e., the symmetry data of valence bands.
These relations, if exist, would reduce the difficult task of finding the former to a routine calculation of the latter.
But the examples are rare \cite{Fu2007,Turner2010,Hughes2011,Fang2012}.
Fu-Kane formula \cite{Fu2007} for topological insulators protected by time-reversal symmetry (TI for short from now) is an exemplary one, mapping the four topological $\mathbb{Z}_{2}$-invariants to inversion eigenvalues at eight high-symmetry points.
This simple golden rule considerably expedited the search for TI in all centrosymmetric materials via first principles numerics \cite{Zhang2009}.
Nevertheless, for general topological states in three dimensions protected by any one of the 230 space groups with and without time-reversal, or topological crystalline insulators \cite{Fu2011} (TCI), explicit formulae relating their topological invariants to symmetry data have so far been missing.

Recently, a solid step along this direction is made in \onlinecite{Po2017,Bradlyn2017,Bradlyn2017a,Kruthoff2016}, where the authors systematically study the connectivity of bands in a general gapped band structure, and identify the constraints on the symmetry data in the form of linear equations called the compatibility equations. \onlinecite{Bradlyn2017} explicitly provides these relations for each space group, and observes that if a symmetry data satisfying all compatibility relations cannot decompose into elementary band representations (sets of symmetry data of atomic insulators, given in the same paper),  the material must be topologically nontrivial.
\onlinecite{Po2017} shows that the symmetry data of any gapped band structure can be compressed into a set of up to four $\mathbb{Z}_{n=2,3,4,6,8,12}$ numbers called symmetry-based indicators (SI).
(See \cref{sec:methods} for a brief review of SI.)
The set of SI is a lossless compression of symmetry data as far as topological invariants are concerned: 
all topological invariants that may be extracted from symmetry data can be inferred from the corresponding SI.
The theory presented in \onlinecite{Po2017} does not, however, relate SI to the topological invariants, the defining quantities of topological states: a band structure having nonzero SI is necessarily topological, but the type of the topology in terms of invariants is unknown.
The explicit expressions of the SI in terms of symmetry data are also missing in \onlinecite{Po2017}.

This paper aims to complete the mapping between symmetry data and topological invariants in systems with time-reversal symmetry and significant spin-orbital-coupling (the symplectic Wigner-Dyson class or class AII in the Altland-Zirnbauer system \cite{Altland1997}).
To achieve this, we first derive the explicit expression of each SI in all space groups (\cref{tab:z8,tab:formula,tab:Indicator}), and then given any nonzero set of SI in every space group, we enumerate all possible combinations of topological invariants that are compatible with the SI (\cref{tab:lattice,tab:eLC,tab:eLC2,tab:SI2TOP,tab:TOP2}).
These invariants include: three weak topological invariants $\delta_{\mathrm{w},i=1,2,3}$ \cite{Fu2007a}, mirror Chern number $C_\mathrm{m}$ \cite{Hsieh2012,Teo2008}, glide plane (hourglass) invariant $\delta_\mathrm{h}$ \cite{Wang2016}, rotation invariant $\delta_\mathrm{r}$ \cite{Shiozaki2014,Song2017,Fang2017}, the inversion invariant $\delta_\mathrm{i}$ \cite{Turner2010,Hughes2011,Fang2017},  a new $\mathbb{Z}_2$ topological invariant protected by screw rotations $\delta_\mathrm{s}$, and finally a new $\mathbb{Z}_2$ topological invariant protected by $S_4$-symmetry $\delta_{S_4}$. The last two invariants are theoretically established in \cref{sec:Layer}.
In the main results, the strong time-reversal invariant $\delta_\mathrm{t}$ \cite{Fu2007a} is assumed to vanish, so that the results are restricted to TCI only, or states that can be adiabatically brought to atomic insulators in the absence of crystalline symmetries; the $\delta_\mathrm{t}=1$ cases are briefly discussed in the end of the Results section.
The exhaustive enumeration maps 478 sets of SI to 3133 linearly independent combinations of topological invariants, as tabulated in \cref{tab:SI2TOP}.
A guide for reading this table is offered in \cref{sec:table}.

\section{Results}\label{sec:results}

\begin{figure}
\begin{centering}
\includegraphics[width=1\linewidth]{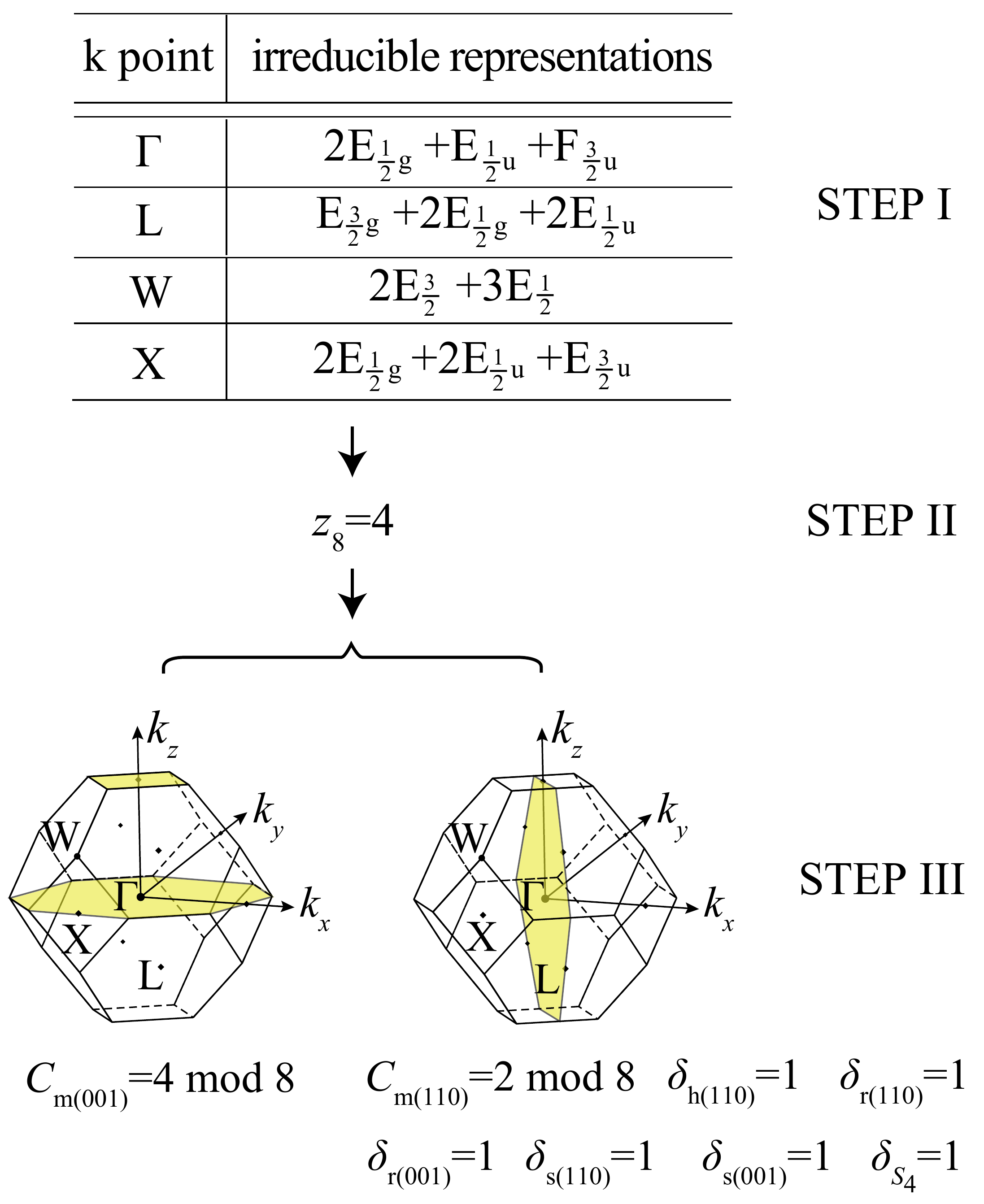}
\par\end{centering}
\protect\caption{\label{fig:1}
A demonstration of the diagnosis for tin telluride of space group $Fm\bar{3}m$ (\#225) using our results.
The table on the top shows the symmetry data obtained in the first principles calculation (details given in text), where the numbers of appearance of each irrep in the valence bands are listed for each high-symmetry point in the face-centered-cubic Brillouin zone.
From the data one finds the SI $z_8=4$ using \cref{tab:z8,tab:formula}, and then by searching for this indicator in \cref{tab:SI2TOP}, two possible sets of topological invariants are found, listed at the bottom left and bottom right, respectively. 
The yellow planes in the Brillouin zone are where the mirror Chern numbers $C_{\mathrm{m}(001)}$ and $C_{\mathrm{m}(110)}$ are defined.
Indices in the parentheses in subscript represent the directions of the corresponding symmetry elements. 
The real material has been shown in \onlinecite{Hsieh2012} to have the topological invariants listed on the bottom right.}
\end{figure}

\textit{An example showing the usage of our results.}
Before entering into the derivation of the results, we use tin telluride (SnTe) crystal having space group $Fm\bar{3}m$ (\#225) to illustrate how the results should be used in \cref{fig:1}.
One should first compute the symmetry data of the material, finding the numbers of appearances for each irrep in the valence bands at the high-symmetry momenta, namely, $\Gamma$, X, L and W.
This can be done in any modern implementation of first principles numerics, and here we use Vienna ab-initio simulation package (VASP \cite{VASP-1,VASP-2}).
From the symmetry data obtained in the top of \cref{fig:1}, we apply the formulae given in \cref{tab:z8,tab:formula} to find the SI, which in this case is a single $\mathbb{Z}_8$ number, and we find $z_8=4$.
After this, we can use \cref{tab:SI2TOP} and find that $z_8=4$ corresponds to two and only two possible sets of topological invariants shown on the bottom of \cref{fig:1}: it either has nonzero mirror Chern number $C_{\mathrm{m}(001)}=4$ (mod 8) for the $k_z=0$-plane (and symmetry partners), or has mirror Chern number $C_{\mathrm{m}(110)}=2$ (mod 8) for the $k_x+k_y=0$-plane (and symmetry partners).
It is impossible, however, to distinguish these two cases using symmetry data, but advanced tools such as Wilson loops must be invoked.
Further analysis shows that the latter state appears in the real material \cite{Hsieh2012}.

\textit{Layer construction as an general approach.}
A remarkable feature of all known TCIs is that any TCI can be adiabatically (without gap closing) and symmetrically tuned into a simple product state of decoupled, identical layers in real space, each of which decorated with some 2D topological state\cite{Isobe2015,Fulga2016,Ezawa2016,Song2017a,Huang2017,Fang2017}. 
This form of fixed-point wave function for a TCI is called its layer construction (LC). 
An analogy to atomic insulators can be drawn to help understand the physical nature of LC in the following aspects:
While an atomic insulator is built from decoupled point-like atoms, the building blocks of an LC are decoupled layers.
Each atom in an atomic insulator is decorated with electrons occupying certain atomic orbitals, while each layer is decorated with electrons forming a 2D topological state.
The atomic orbitals of an atom in lattice correspond to the irreducible representations of the little group at that atomic position, while the possible topological states on a layer also depends on the little group leaving the layer invariant.
In an LC, there are only two possible decorations: if the layer coincides with some mirror plane of the space group, the state for decoration is a 2D mirror TCI with mirror Chern number $C_\mathrm{m}$ and if not coincide, a 2D TI.
We define elementary layer construction (eLC) as an LC generated by a single layer in real space 
\begin{equation}
(mnl;d)\equiv\{\mathbf{r}|(m\mathbf{b}_1+n\mathbf{b}_2+l\mathbf{b}_3)\cdot\mathbf{r}=2\pi{d} \mod 2\pi \}    
\end{equation}
Here $(mnl)$ are the Miller indices, and $\mathbf{b}_i$'s the reciprocal lattice vectors; generation here means we take all elements $g\in{G}$ to obtain the set of layers $E(mnl;d)\equiv\{g(mnl;d)|g\in{G}\}$ by acting $g$ on $(mnl;d)$.
Every LC is a superposition of a finite number of eLCs, and thanks to the additive nature of all known topological invariants, the topological invariants of any LC is the sum of the invariants of all constituent eLCs.

For any space group, we exhaustively find all eLCs using the method detailed in \cref{sec:find-eLC}.
While the calculation of topological invariants is difficult for an arbitrary band structure, it is easy for an eLC, thanks to its simple structure.
In fact, the invariants only depend on how many times each symmetry element is occupied. (See \cref{sec:Layer}, wherein the occupation for glide plane or screw axis is subtle.)
A symmetry element is a manifold in real space, where each point is invariant under some symmetry operation.
It could be a discrete point such as an inversion center or a center of $S_4:(x,y,z)\rightarrow(-y,x,-z)$, a line such as a rotation axis, or a plane such as a mirror plane.
In this way, topological invariants for each eLC are calculated and tabulated in \cref{tab:eLC,tab:eLC2}.
On the other hand, the SI of an eLC are also easily calculated, detailed in \cref{sec:LCind} again due to the decoupled nature of the layers.
Matching the SI with invariants for each eLC, we hence find the full mapping between SI and topological invariants for TCI.
%For intuitive understanding, we also plot a set of figures (Supplementary Figures 1-8) showing the invariants, SI, and phase transitions of eLCs.

%Below we illustrate the described method by explicitly deriving the mapping from SI to topological invariants for the simplest space group $P\bar1$ (\#2), where a step-by-step example of layer construction is given, and for space group $P2/m$ (\#10), where the one-to-many nature of the mapping is demonstrated.

\begin{figure}
\begin{centering}
\includegraphics[width=1\linewidth]{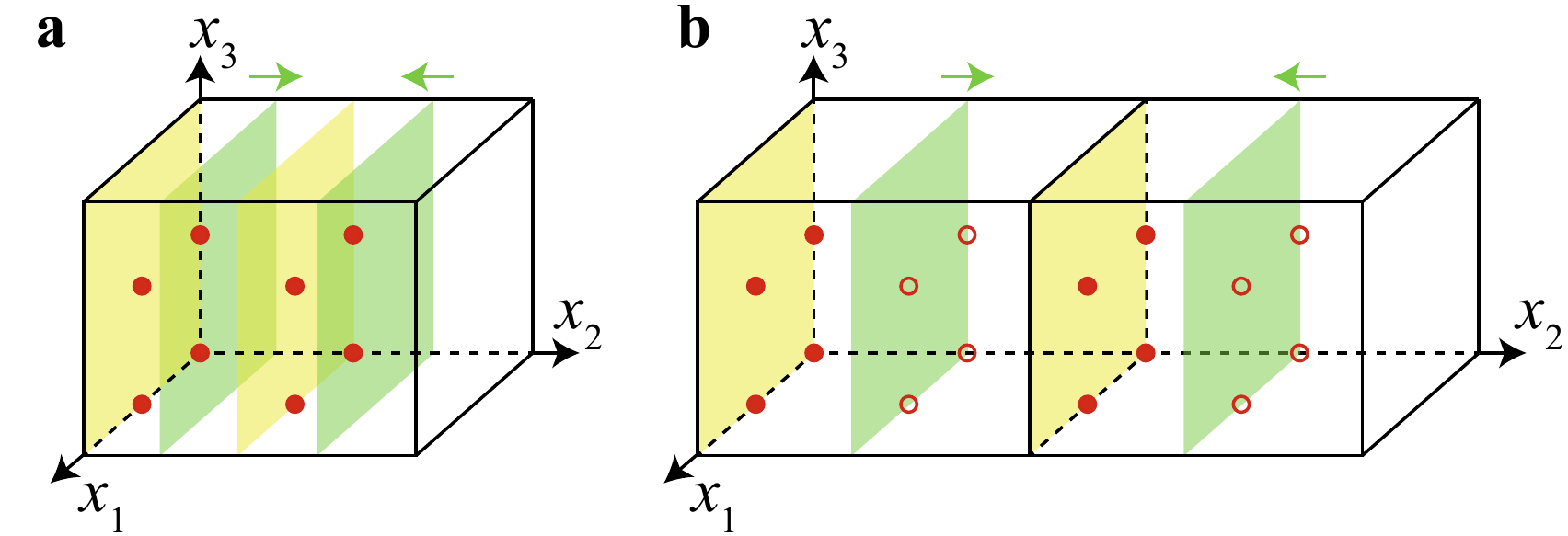}
\par\end{centering}
\protect\caption{\label{fig:2}Layer constructions for space group $P\bar1$ (\#2). 
{\bf a} The yellow planes are $(010;0)$ and $(010;\frac12)$ respectively, and the two green planes are $(010;d)$ and $(010;1-d)$ with $d\neq0,\frac12$. 
The arrows mean that the two green planes can move toward each other without breaking inversion. 
{\bf b} After doubling the unit cell along $x_2$-direction, the open dots are no longer inversion centers as they were, while the solid dots remain. Again the arrows mean that two green planes can move toward each other without breaking the inversion symmetry, after unit cell doubling.}
\end{figure}

\textit{From indicators to invariants.}
Here we take space group $P\bar1$ as an example to show the mapping between indicators and invariants and leave the general discussion in the \cref{sec:Layer,sec:LCind}.
The space group $P\bar1$ has non-orthogonal lattice vectors $\mathbf{a}_{i=1,2,3}$ and inversion symmetry. 
Within a unit cell, there are eight inversion centers at $(x_1,x_2,x_3)/2$ in the basis of lattice vectors (the red solid circles in \cref{fig:2}), where $x_i=0,1$.
These inversion centers are denoted by $V_{x_1x_2x_3}\equiv{}(x_1,x_2,x_3)/2\;\mathrm{mod}\;1$.
A generic layer $(mnl;d)$ is given by $L=\{\mathbf{r}|(m\mathbf{b}_1+n\mathbf{b}_2+l\mathbf{b}_3)\cdot\mathbf{r}=2\pi{d} \mod 2\pi \}$, where $d\in[0,1)$, and at least one of $m,n,l$ is odd (or they would have a common factor).
If $d\neq0,\frac12$, we have $d\neq-d\;\mathrm{mod}\;1$, then under inversion a generated plane $L'=(mnl;1-d)\neq{L}$ is a different plane symmetric to $L$ about the origin.
In that case, the two planes $L$ and $L'$ can adiabatically move towards each other without breaking any symmetry until they coincide, a process illustrated in \cref{fig:2}a.
The state decorated on $L$ and $L'$ are 2D TIs, and due to the $\mathbb{Z}_2$-nature, when $L$ and $L'$ coincide, the resultant double-layer becomes topologically trivial.
The eLC generated by $(mnl;d\neq0,\frac12)$ is hence a trivial insulator.
For $d=0,\frac12$, $L$ is invariant under inversion, and always passes four of the eight inversion centers, that is, $V_i$'s that satisfy the equation $mx_1+nx_2+lx_3=2d\;\textrm{mod}\;2$.
For examples, if $(mnl;d)=(010;0)$, then $V_{000,001,100,101}$ are on $L$ (the left yellow plane in \cref{fig:2}a). 
Since each layer is decorated with 2D TI, $\mathrm{eLC}(mnl;d)$ ($d=0,\frac12$) is the familiar weak TI having weak invariants 
\begin{equation}
\delta_{\mathrm{w},1}=m\;\mathrm{mod}\;2\quad 
\delta_{\mathrm{w},2}=n\;\mathrm{mod}\;2\quad
\delta_{\mathrm{w},3}=l\;\mathrm{mod}\;2 \label{eq:weak_indices}
\end{equation}
Now we turn to the inversion invariant $\delta_{i}$, which is a strong invariant robust against all inversion preserving perturbations.
Let us consider a perturbation that doubles the periodicity in the $(mnl)$-direction, while preserving the inversion center at origin.
After the doubling, four of the eight inversion centers satisfying $mx_1+nx_2+lx_3=1\;\textrm{mod}\;2$ are no longer inversion centers, so that the plane $(mnl,\frac12)$ after the doubling no longer passes through any inversion center, and the generated eLC by $(mnl,\frac12)$ can be trivialized after the doubling by pairwise annihilating with its inversion partner.
Therefore, eLC$(mnl,\frac12)$ can be trivialized while keeping the inversion symmetry about the origin, thus having $\delta_{i}=0$.
In \cref{fig:2}b, we take $(mnl)=(010)$ as example and doubled the unit cell.
We see that the four inversion centers marked by empty circles are not inversion centers in the new cell, and the blue plane at $(010;\frac12)$ in the original cell becomes $(010;\frac14)$ in the new cell.
The two blue planes can move and meet each other at the yellow plane, denoted by $(010;\frac12)$ in the new cell.
The eLC generated by $(mnl;0)$-plane, however, passes all eight inversion centers in the enlarged unit cell and cannot be trivialized without breaking inversion (yellow planes in \cref{fig:2}b), so that the inversion invariant $\delta_\mathrm{i}=1$. 

After finding the invariants for all possible eLCs, we turn to the SI for each eLC.
The SI group of $\mathrm{P\bar1}$ takes the form $\mathbb{Z}_2\times\mathbb{Z}_2\times\mathbb{Z}_2\times\mathbb{Z}_4$, wherein the first three are the weak TI indicators $z_{2\mathrm{w},i=1,2,3}$ and the last one is the $z_4$ indicator.
The calculation method is briefly described in \cref{sec:LCind}, and here we only give the results.
For eLC$(mnl;0)$ and eLC$(mnl;\frac12)$, their values are found to be 
 $(m\;\mathrm{mod}\;2,n\;\mathrm{mod}\;2,l\;\mathrm{mod}\;2,2)$ 
 and $(m\;\mathrm{mod}\;2,n\;\mathrm{mod}\;2,l\;\mathrm{mod}\;2,0)$, respectively. 
For this space group, the mapping from SI set to topological invariants is therefore one-to-one: 
 $z_{2\mathrm{w},i}=\delta_{\mathrm{w},i}$ and $z_{4}=2\delta_\mathrm{i}$.

\textit{Convention dependence of topological invariants.}
A subtle but important remark is due at this point. 
There are always eight inversion centers in a unit cell in the presence of inversion symmetry, and when translation symmetry is broken, only one, two, or four of them remain.
In the definition of inversion invariant $\delta_\mathrm{i}$, one of the eight is chosen as the inversion center that remains upon translation breaking.
In the above example, when the unit cell is doubled, the origin was chosen as the center that remains, but if we chose $V_{010}=(\frac12,0,0)$, which is a completely valid choice, the four open circles in \cref{fig:2}b are still inversion centers but the solid circles are not after the doubling.
In that case, we would find that eLC$(mnl,\frac12)$ has $\delta_\mathrm{i}=1$ but eLC$(mnl,0)$ has $\delta_\mathrm{i}=0$.
The inversion invariant $\delta_\mathrm{i}$ hence depends on the convention which one of the eight inversion centers in the unit cell is chosen in the definition of $\delta_\mathrm{i}$.
However, when we superimpose the two eLCs into an LC that passes \textit{all} eight inversion centers in a unit cell, the value of $\delta_{i}$ is \textit{independent} of the choice of the inversion center, since all eight are occupied in this LC.
We emphasize that only if this is the case, can we hope to observe the physical properties, such as the characteristic boundary states associated with the bulk invariant $\delta_\mathrm{i}$ \cite{Fang2017}, because physical observables should not depend on the conventions.

Moreover, similarly, as detailed in \cref{sec:conv}, the rotation (screw) invariant $\delta_\mathrm{r}=1$ ($\delta_\mathrm{s}=1$) is convention-independent if and only if each rotation (screw) axis in unit cell is occupied by the LC for $n/2\;\mathrm{mod}\;n$ times, where $n=2,4,6$ is the order of the rotation (screw) axis.
For the $S_4$ invariant $\delta_{S_4}=1$ or the hourglass invariant $\delta_\mathrm{h}=1$ to be convention-independent, the LC should occupy each $S_4$ center or glide plane for an odd number of times.
Invariants that are convention-independent are marked blue in \cref{tab:SI2TOP,tab:TOP2}.

\textit{The one-to-many nature of the mapping.}
In the example of space group $P\bar1$, the mappings between indicators and topological invariants are one-to-one.
However, this is in fact the \emph{only} space group where mappings are bijective.
By definition, different sets of indicators must correspond to different sets of invariants, but multiple sets of invariants may correspond to the \emph{same} set of indicators, that is, the mapping from indicators to invariants is one-to-many.

\begin{figure}
\begin{centering}
\includegraphics[width=1\linewidth]{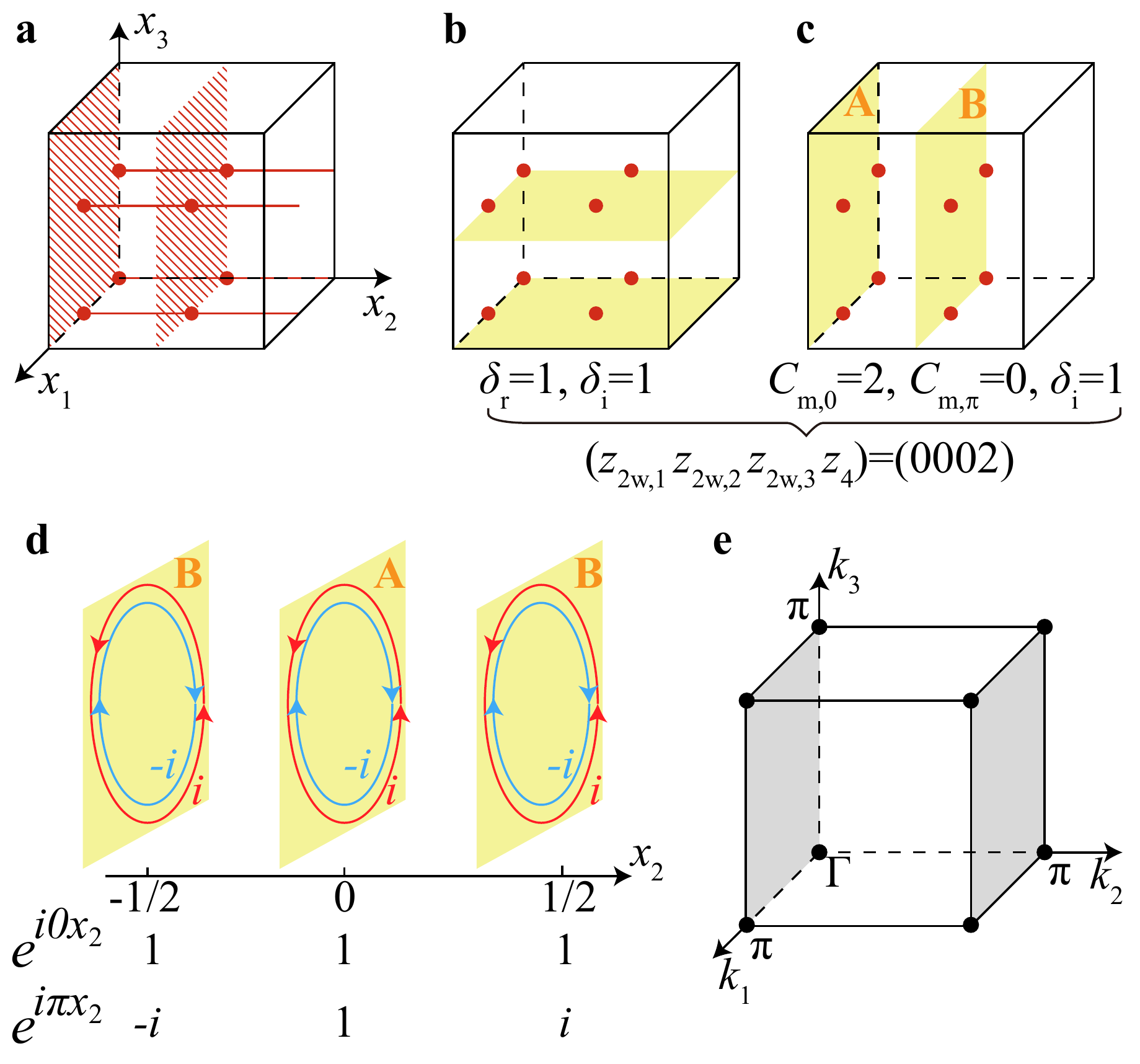}
\par\end{centering}
\protect\caption{\label{fig:3}Two layer constructions for space group $P2/m$, sharing the same set of SI of $(0002)$. {\bf a} All symmetry elements of the space group in one unit cell, including eight inversion centers (red solid circles), four rotation axes (red solid lines), and two mirror planes (shaded planes). {\bf b}, {\bf c} LC1 and LC2 defined in the text, respectively. They have distinct topological invariants but identical indicators. {\bf d} 3D Bl{\"o}ch wave functions in LC2 as superpositions of 2D Bl\"och wave functions with coefficients $e^{ik_2 x_2}$. Here we use red and blue loops to represent the 2D wave functions having mirror eigenvalues $i$ and $-i$ respectively, wherein $i$ wave functions have Chern number 1 and $-i$ wave functions have Chern number -1. For A-eLC the 3D Bl{\"o}ch wave functions at $k_2=0$ and $k_2=\pi$ have the same mirror eigenvalues, leading to identical mirror Chern numbers at $k_2=0$ and $k_2=\pi$. While for B-eLC the Bl{\"o}ch wave functions at $k_2=0$ and $k_2=\pi$ have opposite mirror eigenvalues, leading to opposite mirror Chern numbers at $k_2=0$ and $k_2=\pi$. {\bf e} The two mirror-invariant planes (grey planes) in Brillouin zone. }
\end{figure}

To understand the one-to-many nature of the mapping more concretely, 
 we look at the specific group $P2/m$, containing two mirror planes, 
 four $C_2$-axes and eight inversion centers in each unit cell, all marked in \cref{fig:3}a. 
Now we consider two different LCs illustrated in \cref{fig:3}b,c: in \cref{fig:3}b for LC1, 
 two horizontal planes, each decorated with a 2D TI, occupy all four $C_2$-axes and all eight inversion centers, 
 and in \cref{fig:3}c for LC2, two vertical planes, each decorated with a mirror Chern insulator with $C_\mathrm{m}=1$, occupy the two mirror planes and the eight inversion centers.
%In the following we will show that LC1 and LC2 are topologically distinct (both are nontrivial), but they have identical SI.

Since all inversion centers are occupied in LC1 and LC2, in both cases we have $\delta_\mathrm{i}=1$.
LC1 occupies all four $C_2$-rotation axes, once each, thus having nontrivial rotation invariant $\delta_\mathrm{r}=1$, while LC2 does not occupy any of the rotation axes, having $\delta_\mathrm{r}=0$.
On the other hand, LC2 occupies the two mirror planes, each with 2D TCI having $C_\mathrm{m}=1$. 
According to the calculation in \cref{sec:methods} LC2 has mirror Chern numbers $C_\mathrm{m} = 2$ at $k_z = 0$ plane and $C_\mathrm{m} = 0$ at $k_z=\pi$ plane; and LC1, not occupying any mirror plane, has vanishing mirror Chern number. LC1 and LC2 are therefore topologically distinct states.

Now we turn to the SI of LC1 and LC2.
For space group $P2/m$, the SI have the same group structure $\mathbb{Z}_2\times\mathbb{Z}_2\times\mathbb{Z}_2\times\mathbb{Z}_4$ as that of its subgroup $P\bar1$.
In this case, the value of each indicator remains the same as we break the symmetry down to $P\bar1$.
Viewed as LC in $P\bar1$, both LC1 and LC2 are the superpositions of eLC$(mnl,0)$ and eLC$(mnl,\frac12)$, 
thus having, by the additivity of SI, $z_{2\mathrm{w},i}=0$ and $z_{4}=2$.

The failure in distinguishing LC1 and LC2 by indicators reveals a general ambiguity in the diagnosis of mirror Chern numbers. 
We can add additional even number ($2p$) of A-eLC and even number ($2q$) of B-eLC to LC1 such that the composite state has the same SI and $\delta_\mathrm{r}$ with LC1 but nonzero mirror Chern numbers $C_{\mathrm{m},0}=2p+2q$, $C_{\mathrm{m},\pi}=2p-2q$.
On the other hand, we can also add these additional eLCs to LC2 to get a state having the same SI with LC2 but different mirror Chern numbers $C_{\mathrm{m},0}=2+2p+2q$, $C_{\mathrm{m},\pi}=2p-2q$.
%Therefore the two classes of LCs, wherein the first has invariants $\delta_r=1$, $C_{m,0}-C_{m,\pi}=0\;\mathrm{mod}\;4$ and the second has invariants $\delta_{r}=0$, $C_{m,0}-C_{m,\pi}=2\;\mathrm{mod}\;4$, have the same SI.
The proof here can be generalized to any space group having mirror planes and perpendicular rotation axes, providing that the order of the rotation is even.
In all these space groups, as shown in \cref{tab:SI2TOP}, the TCIs having invariants $\delta_\mathrm{r}=1$ and $C_{\mathrm{m},0}-C_{\mathrm{m},\pi}=0\;\mathrm{mod}\; 2n$ and the TCIs having $\delta_\mathrm{r}=0$ and $C_{\mathrm{m},0}-C_{\mathrm{m},\pi}=n\;\mathrm{mod}\; 2n$ have the same SI, here $n$ is the order of the rotation axis.
The two possible sets of invariants shown in \cref{fig:1}, wherein one is $C_\mathrm{m(001)}=4\;\mathrm{mod}\;8$, $\delta_\mathrm{r(001)}=0$ and the other is $C_\mathrm{m(001)}=0\;\mathrm{mod}\;8$, $\delta_\mathrm{r(001)}=1$, are the examples for $n=4$.

\textit{Indicators for time-reversal topological insulators.}
In the SI group of each space group, except \#174 and \#187-190, 
 there is one special indicator of $\mathbb{Z}_{2,4,8,12}$-type, 
 denoted by $z_\mathrm{t}$, marked red in \cref{tab:formula,tab:Indicator}.
When this special indicator is odd, the system is the well-known 3D time-reversal topological insulator \cite{Po2017} (TI for short).
The essential difference between a TI and a TCI is that the former only requires time-reversal symmetry, such that it remains nontrivial even when all crystalline symmetries are broken.
TI does \textit{not} have layer constructions, so that the method we use does not apply to SI having $z_\mathrm{t}\in\mathrm{odd}$.
To construct states having $z_\mathrm{t}\in\mathrm{odd}$, we first notice that TI is consistent with all space groups, such that for each space group, we have at least one state that is a TI.
Then we can superimpose this TI with all existing LCs obtained, 
and generate gapped states for all nonzero combinations of SI with $z_\mathrm{t}\in\mathrm{odd}$ 
(but with five exceptions discussed in Discussion section).

Regarding $z_\mathrm{t}\in\mathrm{odd}$, we comment that the value of $z_\mathrm{t}$ generally has a convention dependence on the overall signs in the definition of inversion and the rotation operators.
For example, in space group $P\bar1$, the defining properties of the symmetry operators are $\hat{P}^2=1$, $\hat{T}^2=-1$ and $[\hat{P},\hat{T}]=0$.
It is easy to check that the overall sign in front of $\hat{P}$ can be freely chosen without violating any of the above relations.
In other words, without external references, it is unknown \textit{a priori} if, for example, an $s$-orbital should be assigned with positive or negative parity.
Upon redefining $\hat{P}\rightarrow-\hat{P}$, a state having $z_{4}=1$ goes to $z_{4}=3$ and vice versa.
In similar ways, it is proved in \cref{sec:indicator} that states having the $\mathbb{Z}_8$-indicator $z_8=1,3,5,7$ differ only by convention and so do the states have $\mathbb{Z}_{12}$-indicator $z_{12}=1,5,7,11$.
In these cases, the convention refers to the overall sign in front of inversion operator and the sign in front of the rotation operator.
It is difficult to distinguish these states from each other experimentally.
However, here we emphasize that the SI $z_4=1,3$ (so do the SI $z_8=1,3,5,7$ and $z_{12}=1,5,7,11$) have a relevant difference under a fixed convention, which can be detected by the anomalous boundary between the two phases.
For example, suppose we have a spherical sample of $z_4=3$ phase and fill the space outside the sphere with $z_4=1$ phase, then as long as the geometry keeps inversion symmetry, the boundary state on the spherical surface should be identical with the boundary state between  $z_4=2$ and $z_4=0$ phases, which is known as 1D helical mode (See \cref{sec:Layer} for details).
This is because we can deduct a background of $z_4=1$ phase both inside and outside the sphere without changing the boundary state.

The space groups \#174, \#187, \#189 and \#188, \#190,
where one can not diagnose TI from SI, 
have the SI groups $\mathbb{Z}_3\times\mathbb{Z}_3$ and $\mathbb{Z}_3$, respectively, and the corresponding SI $z_{3\mathrm{m},0}$ and $z_{3\mathrm{m},\pi}$ are the mirror Chern numbers (mod 3) in the $k_3=0$ and $k_3=\pi$ planes. 
In \#188 and \#190 $z_{3\mathrm{m},\pi}$ is trivialized by nonsymmorphic symmetry and thus the corresponding SI groups reduce to $\mathbb{Z}_3$.
In these space groups the TI invariant is the parity of $C_{\mathrm{m},0}-C_{\mathrm{m},\pi}$\cite{Hsieh2012} whereas SI have ambiguity for the parities of mirror Chern numbers, thus TI can never be diagnosed from SI.
For example, $z_{3\mathrm{m},0}=1$, $z_{3\mathrm{m},\pi}=0$ can correspond to $C_{\mathrm{m},0}=1$, $C_{\mathrm{m},\pi}=0$ (a TI) or $C_{\mathrm{m},0}=-2$, $C_{\mathrm{m},\pi}=0$ (not a TI).

\section{Discussion}\label{results}

A byproduct of this study is a complete set of TCIs that can be layer-constructed in all 230 space groups (\cref{tab:eLC,tab:eLC2,tab:SI2TOP,tab:TOP2}), even including groups not having SI.
The abundance of the states thus obtained naturally suggests the question: are all TCI states exhausted in these layer-constructions?
We regret to answer it in the \textit{negative}: layer construction cannot give us the weak topological insulator states in five space groups, namely, \#48, \#86, \#134, \#201 and \#224.
In any one of the five, there is a weak indicator $z_{2w}$, but all layer constructed states have $z_{2w}=0$.
A common character of these space groups is that they have three perpendicular glide planes $\{m_{001} |\frac12 \frac12 0\}$ $\{m_{010}|\frac12 0 \frac12\}$ $\{m_{100}| 0 \frac12 \frac12\}$ such that any single layer having weak index $z_{2\mathrm{w},i}=1$ would be doubled along the $i$-th direction and so the generated eLC has vanishing weak index.  
Explicit (non-LC) tight-binding models for the $z_{2\mathrm{w}}=1$ states are given in \cref{sec:weak}, completing the proof that for any nonzero SI there is at least one corresponding gapped topological state.
These corner cases are somewhat surprising as weak TI have so far been considered most akin to stacking of decoupled 2D TI.

Finally, we comment that all LCs can be used to build 3D symmetry protected topological states of bosons and fermions protected by space group $G$ plus a local group $G_\mathrm{L}$. 
To do this one only needs to decorate each layer with a 2D SPT protected by $G_\mathrm{L}$ instead of the 2D TI.

Towards the completion of the work, we have been aware of a similar study \cite{Khalaf2017}. To our knowledge, the results, when overlapping, are consistent with each other.

\section{Methods}\label{sec:methods}
\textit{A short review of symmetry-based indicator.}
For each momentum in the Brillouin zone, there is an associate subgroup, called the little group, of the space group $G$, under the action of which the momentum is invariant up to a reciprocal lattice vector.
A point is a high-symmetry point, denoted $K_j$, if its little group is greater than the little group of any point in the neighborhood.
A fundamental theorem is that each band at momentum $K_i$ or multiplet of degenerate bands corresponds to an irreducible representation of the little group at $K_i$.
The symmetry data of a band structure is defined as the an integer vector $\mathbf{n}$, each element of which, $n(\xi_i^{K_j})$, is the number of appearance of the $i$-th irreducible representation in the valence bands at the $j$-th high-symmetry momentum $K_j$, where $i=1,...,r_j$ labels the irreducible representations of the little group at $K_j$.
One could further define the addition of two symmetry data as the addition of each entry, which corresponds to, physically, the superposition of two band structures.

For a gapped band structure, the elements of its symmetry data cannot take arbitrary integers, and there are constraints on the symmetry data known compatibility relations\cite{Po2017,Bradlyn2017,Kruthoff2016}.
For example, gapped-ness requires that the occupation numbers at each $K_i$ be the same, i.~e., $\sum_in({\xi^{K_j}_i})=\mathrm{const}$.
All compatibility relations are linear equations so that the symmetry data satisfying all these relations again form a smaller linear space, termed the band structure space, denoted $\{\mathrm{BS}\}$.

On the other hand, we consider the symmetry data of atomic insulators. In atomic insulators, the bands are generated by decoupled atomic orbitals placed at certain Wyckoff positions in the unit cell.
By this definition, one finds that the symmetry data of atomic insulators also form a linear space, denoted $\{\mathrm{AI}\}$ (also called the space of band representations \cite{Bradlyn2017}).
Obviously a symmetry data $\mathbf{n}\in\{\mathrm{AI}\}$ satisfies all compatibility relations, so $\{\mathrm{AI}\}\subseteq\{\mathrm{BS}\}$.
One then naturally considers the quotient space $X_\mathrm{BS}=\{\mathrm{BS}\}/\{\mathrm{AI}\}$.
$X_\mathrm{BS}$ is always a finite group generated by several $\mathbb{Z}_{n=2,3,4,6,8,12}$\cite{Po2017}.
Each generator of $X_\mathrm{BS}$ is called a symmetry-based indicator.

The following properties of indicators should be mentioned:
any two gapped band structures having different sets of SI must be topologically distinct, and any two different symmetry data having the same set of SI only differ from each other by the symmetry data of an atomic insulator.

In \onlinecite{Po2017}, the authors calculate the group structure of the indicators for all 230 space groups.
However, it does not give explicit formulae for the generators. 
In order for application, we derive all these formulae in \cref{sec:indicator}.

\textit{Mirror Chern number of layer construction.}
Below we explicitly calculate the mirror Chern numbers of LC1 and LC2 in \cref{fig:2}.
%The topological invariants of 3D mirror TCI are the mirror Chern numbers defined on mirror-invariant planes in momentum space. 
%Here we explicitly calculate these invariants of LC2.
As shown in \cref{fig:3}e, in BZ of space group $P2/m$ there are two mirror-invariant planes, i.e., the $k_2=0$ and $k_2=\pi$ planes, thus we have two mirror Chern numbers $C_{\mathrm{m},0}$ and $C_{\mathrm{m},\pi}$.
We assume there are only two occupied bands in the vertical 2D TCIs in LC2 and denote the corresponding Bl{\"o}ch wave functions as $|\phi_{\pm i}(\mathbf{k}_\mathrm{2D},x_2)\rangle$.
Here $\pm i$ represent the mirror eigenvalues, where $i$ is the imaginary unit, $\mathbf{k}_\mathrm{2D}=(k_1,k_3)$ is the 2D momentum, and $x_2$ is the position along $\mathbf{a}_2$ where the 2D TCIs are attached.
We also assume that the wave functions with the mirror eigenvalue $i$ ($-i$) give a Chern number $1$ ($-1$) such that the 2D mirror Chern number $C_\mathrm{m}$ equal to $1$.
Under the mirror operation $\hat{M}$ the 2D Bl{\"o}ch wave function $|\phi_{\pm i}(\mathbf{k}_\mathrm{2D},x_2)\rangle$ firstly get a mirror eigenvalue $\pm i$ and then move to the mirror position $-x_2$
\begin{equation}
    \hat{M} |\phi_{\pm i}(\mathbf{k}_\mathrm{2D},x_2)\rangle = \pm i |\phi_{\pm i}(\mathbf{k}_\mathrm{2D},-x_2)\rangle \label{eq:mirror}
\end{equation}
To calculate the mirror Chern numbers of LC2 we divide it into two subsystems: the eLC generated from A layer, and the eLC generated from B layer (\cref{fig:3}c).
Since the total mirror Chern numbers are the sum of mirror Chern numbers of the two subsystems, we need only to analyse the two subsystems respectively. 
The 3D Bl{\"o}ch wave functions of A- and B-eLCs can be constructed as
\begin{align}
    |\psi_{\pm i}^\mathrm{A}(\mathbf{k})\rangle & = \sum_{x_2=0,\pm1\cdots} e^{i k_2 x_2} |\phi_{\pm i}(\mathbf{k}_\mathrm{2D},x_2)\rangle \label{eq:psi_A}\\
    |\psi_{\pm i}^\mathrm{B}(\mathbf{k})\rangle & = \sum_{x_2=\pm\frac12,\pm\frac32\cdots} e^{i k_2 x_2} |\phi_{\pm i}(\mathbf{k}_\mathrm{2D},x_2)\rangle \label{eq:psi_B}
\end{align}
Due to \cref{eq:mirror}, it is direct to show that $|\psi^\mathrm{A}_{i}(k_1,0,k_3)\rangle$ and $|\psi^\mathrm{A}_{i}(k_1,\pi,k_3)\rangle$, both of which are superpositions of $|\phi_{i}(\mathbf{k}_\mathrm{2D},x_2)\rangle$ and thus have the Chern number $1$, have the same mirror eigenvalue $i$ (\cref{fig:3}d).
Thus for A-eLC m the mirror Chern numbers at $k_2=0$ and $k_2=\pi$ are all 1.
On the other hand, $|\psi^\mathrm{B}_{i}(k_1,0,k_3)\rangle$ and $|\psi^\mathrm{B}_{i}(k_1,\pi,k_3)\rangle$,  again both of which have the Chern number $1$, have mirror eigenvalues $i$ and $-i$, respectively (\cref{fig:3}d).
Thus for B-eLC the mirror Chern numbers at $k_2=0$ and $k_2=\pi$ are 1 and $-1$, respectively.
Therefore the total mirror Chern numbers in momentum space are $C_{\mathrm{m},0}=2$ and $C_{\mathrm{m},\pi}=0$ for LC2.
It should be noticed that the values of $C_{\mathrm{m},0}$ and $C_{\mathrm{m},\pi}$ do not depend on the two band assumption we take: as long as the 2D TCI has $C_\mathrm{m}=1$ the results remain the same.
On the other hand, the mirror Chern numbers of LC1 should be zero for both $k_2=0$ and $k_2=\pi$ by the following argument. Without breaking mirror symmetry, each vertical plane can bend symmetrically toward the mirror plane until the two halves coincide on mirror invariant planes in real space, due to the $\mathbb{Z}_2$-nature of each half, the folded plane is topologically equivalent to a trivial insulator. Since LC1 can be smoothly trivialized without breaking mirror symmetry, it must have vanishing mirror Chern numbers.

%{\bf Data availability statement.} The data and code that support the findings of this study are available from the corresponding author upon reasonable request.

\acknowledgements{The authors acknowledge support from Ministry of Science and Technology of China under grant numbers 2016YFA0302400 and 2016YFA0300600, National Science Foundation of China under grant number 11674370 and 11421092, and from Chinese Academy of Sciences under grant number XXH13506-202.}

\bibliographystyle{naturemag}
\bibliography{Ref}

\clearpage
\newpage
\appendix

\section{Topology of layer construction} \label{sec:Layer}

\subsection{General consideration}

A general layer consistent with a lattice is described by a set of Miller indices $(mnl)$ and the distance to the origin point $d$. Formally, we denote such a layer as $(mnl;d)$
\begin{align}
\left(mnl;d\right) & =\Big\{\mathbf{r}|\mathbf{r}\cdot\left(m\mathbf{b}_{1}+n\mathbf{b}_{2}+l\mathbf{b}_{3}\right)=2\pi\left(d+q\right)\nonumber \\
 & \qquad\quad q\in\mathbb{Z},\ 0\le d<1\Big\}\label{eq:layer}
\end{align}
 where $\mathbf{b}_{1,2,3}$ are the reciprocal lattices.
Due to the translation symmetry, the layer consists of an infinite number of planes, each of which is represented by an integer $q$. 
%For brevity, in the rest of this paper we sometimes use $L$ to represent the layer
%$\left(m_{L}n_{L}l_{L};d_{L}\right)$. 
For a layer, denoted as $L$, in a particular space group (SG) $\mathcal{G}$, its symmetry property is described by its little group, which is defined as the subgroup of $\mathcal{G}$ that leaves $L$ invariant
\begin{equation}
\mathcal{S}\left(L\right)=\left\{ s\in\mathcal{G}|sL=L\right\} 
\end{equation}
$\mathcal{S}$ must also be a SG containing the full translation subgroup since by definition any lattice vector would translate the layer in \cref{eq:layer}
 to itself. 
Thus the SG can be decomposed as a finite number of cosets of $\mathcal{S}$
\begin{equation}
\mathcal{G}=g_{0}\mathcal{S}+g_{1}\mathcal{S}+\cdots\label{eq:SPG-S}
\end{equation}
%where all the operations in each coset transform the layer to a same
%new layer. 
By applying all the coset representatives on $L$, we get a set of symmetric layers 
\begin{equation}
\left\{ g_{0}L,g_{1}L,\cdots\right\} \label{eq:symL}
\end{equation}
An elementary layer construction (eLC) can be got by decorating the layers in $\{g_0 L, g_1 L, \cdots\}$
 with a set of symmetric 2D topological states, 
 i.e., 2D topological insulators protected by time-reversal symmetry (TIs) 
 or 2D mirror topological crystalline insulators (TCIs) with mirror Chern number 1. 
We denote the eLC generated from $L$ as $\mathrm{eLC}(L)$.
A general layer construction (LC) can be got by stacking a finite number of eLCs together. 
Formally, a LC state can be expressed as the direct product of some eLC states
\begin{equation}
C=E_{1}^{\otimes c_1} \otimes E_{2}^{\otimes c_2}
\otimes \cdots\label{eq:LC}
\end{equation}
Here $C$ represent the LC, 
 $E_{1}$, $E_{2}$$\cdots$ are the eLCs, and $c_{1},c_{2}\cdots$
 are integers representing the multiplicities of these eLCs.

In this work, the topology of a LC $C$ is represented by a set of
 invariants, denoted as $\{\delta(C) \}$.
The invariant set should be properly designed such that (i) it is complete for LC states, i.e., any two topologically different LCs have different invariants,
 and (ii) each invariant in it is additive, i.e.,
\begin{equation}
\delta\left(C\otimes C^{\prime}\right)
=\delta\left(C\right)+\delta\left(C^{\prime}\right)\label{eq:I-additive}
\end{equation}
Here the strong TI invariant, denoted as $\delta_t$, is not included because the strong TI can not be constructed from 2D topological states.
From \cref{eq:LC,eq:I-additive} it is direct to show that the invariants
of a LC are completely determined by the constituent eLCs and the
corresponding expansion coefficients 
\begin{equation}
\delta\left(C\right)=c_{1}\delta\left(E_{1}\right)+c_{2}\delta\left(E_{2}\right)+\cdots \mod N_\delta \label{eq:I-LC}
\end{equation}
Here $N_\delta$ is the order of the additive group formed by $\delta$.
For example, for mirror Chern number $N_\delta=\infty$ and for hourglass invariant $N_\delta=2$.
Hereafter, we will say that two LCs are \emph{equivalent} with each other if
 their invariant sets are same. 
Consequently, although a LC with negative
 coefficients on eLCs seems not physical it can be equivalent with a physical one.
In this sense, we generalize the coefficients in \cref{eq:I-LC}
 from non-negative integers to any integers. 
Therefore, all the possible topologies can be easily found once
 we got all the nonequivalent eLCs. 
In \cref{sec:find-eLC} we will give the systematic method to generate all 
 these eLCs.
For now, let us introduce the invariant set, which consists of four kinds of 
 known invariants and two newly found invariants.
%All of these invariants are additive.
Since for any single spatial operation we have either found its corresponding TCI invariant or proved the nonexistence of TCI invariant, we \emph{conjecture} that the six kinds of invariants are complete for LC states.

\subsection{Four kinds of topological crystalline insulator invariants\label{sub:5ivt}}

\textit{The weak invariants $\delta_{\mathrm{w},i=1,2,3}$.}
These three $\mathbb{Z}_{2}$ numbers are associated with three primitive lattice
 bases $\mathbf{a}_{1,2,3}$, and, 
 a nontrivial invariant implies that gapless modes must exist on
 the surfaces preserving the translation symmetry generated from the corresponding lattice base.
Here we give the method to calculate the weak invariants of an eLC. 
As shown in \figref[a]{fig:5ivt}, for a layer given in \cref{eq:layer}
 we can easily calculate its intersections with $\mathbf{a}_{1,2,3}$
 as
\begin{equation}
\frac{d+q}{m},\;\frac{d+q}{n},\ \frac{d+q}{l}
\end{equation}
respectively. Thus, on an $\mathbf{a}_{1}$ ($\mathbf{a}_{2}$, $\mathbf{a}_{3}$)
preserving surface, the layer cuts $\mathbf{a}_{1}$ ($\mathbf{a}_{2}$,
$\mathbf{a}_{3}$) by $m$ ($n$, $l$) times per surface unit cell
and each cut contributes to a helical mode if the layer is
filled with 2D TI. As each pair of helical modes can be trivialized
without breaking the translation symmetry, the $i$-th weak invariant
of an eLC should be nontrivial only if the layers in it cuts $\mathbf{a}_{i}$
odd times per unit cell. Therefore, we have
\begin{equation}
\delta_{\mathrm{w},1}(E)=\sum_{L\in E}m_{L}\mod2\label{eq:delta2w-1}
\end{equation}
\begin{equation}
\delta_{\mathrm{w},2}(E)=\sum_{L\in E}n_{L}\mod2\label{eq:delta2w-2}
\end{equation}
\begin{equation}
\delta_{\mathrm{w},3}(E)=\sum_{L\in E}l_{L}\mod2\label{eq:delta2w-3}
\end{equation}
Here $E$ is the eLC, and $L$ sums over all the layers in it.
In the example shown in \figref[a]{fig:5ivt}, the layer's Miller
indices are $\left(201\right)$, thus the corresponding weak invariants
are $\delta_{\mathrm{w},1}=0$, $\delta_{\mathrm{w},2}=0$, $\delta_{\mathrm{w},3}=1$.

\begin{figure}
\begin{centering}
\includegraphics[width=1\linewidth]{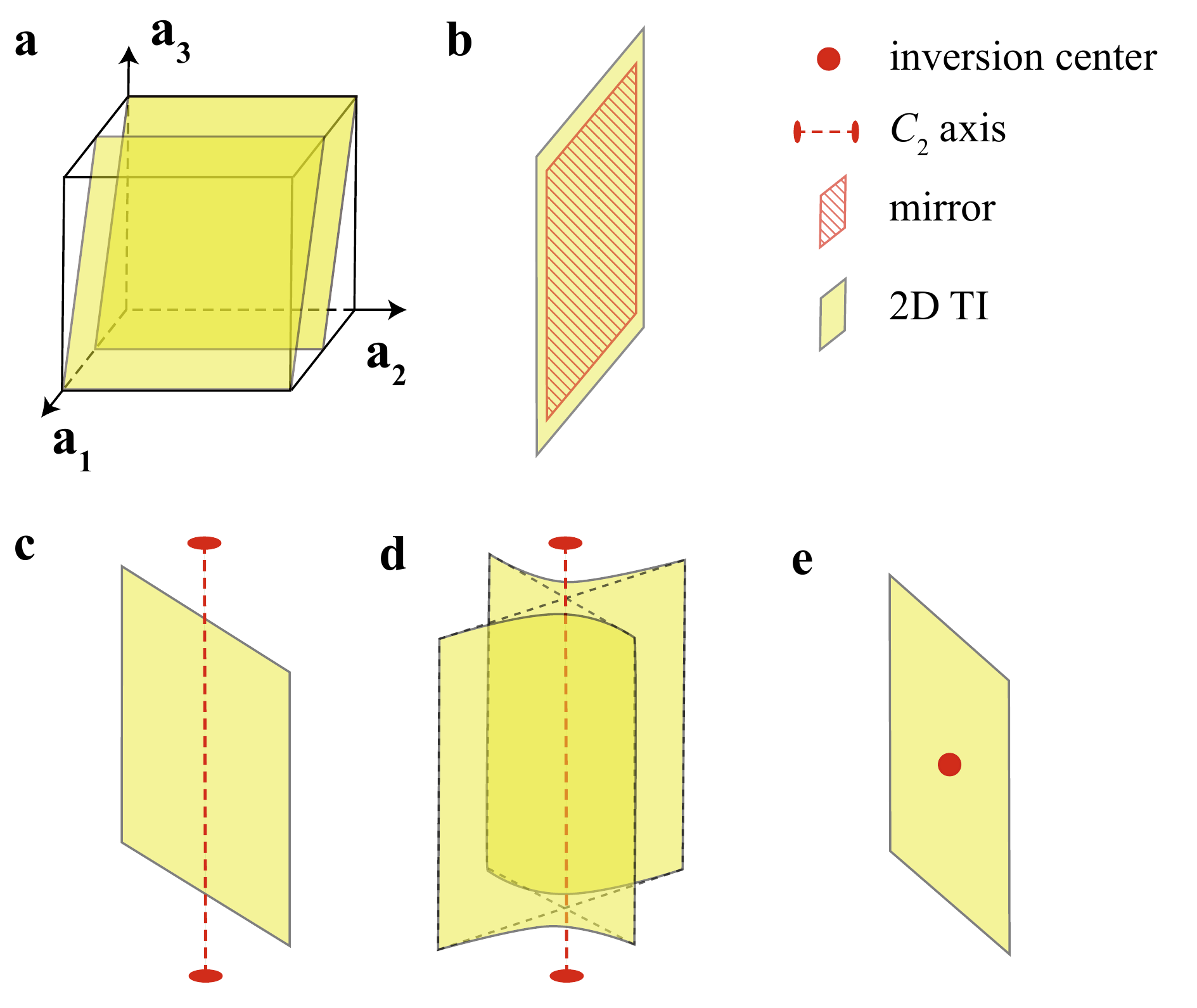}
\par\end{centering}

\protect\caption{
In {\bf a} we illustrate the calculation of weak invariants for a LC. The layer $\left(201;0\right)$ intersects with $\mathbf{a}_{1}$, $\mathbf{a}_{2}$, and $\mathbf{a}_{3}$ by 2, 0, and 1 times per unit cell, and so the weak invariants are $\delta_{\mathrm{w},1}=0$, $\delta_{\mathrm{w},2}=0$, and $\delta_{\mathrm{w},3}=1$.
{\bf b} is the LC of a mirror TCI. 
{\bf c} is the LC with nontrivial $C_2$-rotation invariant. 
{\bf e} is the LC with nontrivial inversion invariant. 
%The common character of the LCs in {\bf c} and {\bf e} is that they can not be dimerized (trivialized) without breaking the corresponding symmetry. 
In {\bf d} we show how a LC occupying the $C_{2}$ axis twice can be dimerized and moved away from the $C_{2}$ axis symmetrically, where the two 2D TIs making up the LC are represented by the blank dashed parallelograms and the dimerized configuration is colored in yellow. \label{fig:5ivt}}
\end{figure}

\textit{The real space mirror Chern number $C_{\mathrm{m}}$.}
In presence of mirror symmetries, 3D mirror TCIs can be got by decorating the
 mirror planes with 2D mirror TCIs.
And, the corresponding mirror Chern numbers (in real space) is simply given by the number of times that 
 the mirror planes are occupied by the 2D mirror TCIs. 
Therefore, at each mirror plane we assign a mirror Chern number,
 which can be calculated as
\begin{equation}
C_{\mathrm{m}}(E)=\sum_{L\in E}N_{{m}}^\mathrm{o}\left(L\right)\label{eq:delta-m}
\end{equation}
\begin{equation}
N_{{m}}^\mathrm{o}\left(L\right)=\begin{cases}
1 & \text{if }m \in L\\
0 & \text{otherwise}
\end{cases}
\end{equation}
for an eLC, where $m\in L$ means the mirror plane is occupied by $L$ and $N_{m}^\mathrm{o}\left(L\right)$ is the number of times that the mirror plane is occupied by $L$.
In the following, we refer such numbers as occupation numbers (ONs).

There should be no worry about the signs of the mirror Chern numbers.
For the 2D state on the mirror plane, the mirror Chern number can be defined as
\begin{equation}
C_\mathrm{m}= \frac{1}{2\pi}\int d^2\mathbf{k} \ 
\mathbf{n}\cdot\mathbf{\Omega}_\mathbf{k}^{(i)}
\end{equation}
where $\mathbf{\Omega}_\mathbf{k}^{(i)}$ is the Berry curvature in the sector with mirror eigenvalue $i$, $\mathbf{n}$ is the normal vector of the mirror,
and $\mathbf{k}$ is integrated in the 2D Brillouin zone on the mirror plane. 
Firstly, this definition does not depend on the choice of the normal vector.
If we choose $\mathbf{n}^\prime=-\mathbf{n}$ as the normal vector, 
then the $\pm i$ mirror eigenvalues interchange with each other and so the 
Berry curvature in $i$ sector becomes $\mathbf{\Omega}^{(i)\prime} = \mathbf{\Omega}^{(-i)} = -\mathbf{\Omega}^{(i)}$, leading to the same $C_\mathrm{m}$.
Secondly, the mirror Chern numbers on all equivalent
 mirror planes in an eLC must equal to each other,
 since under any space group operation both $\mathbf{n}$ and $\mathbf{\Omega}^{(i)}$ transforms as pseudo vectors, i.e., vectors ignoring the inversion, 
 and thus $C_\mathrm{m}$ transforms as a scalar.
In an eLC there is at most one kind of equivalent mirror planes are occupied, 
 therefore we can always choose proper 2D mirror TCIs such that all the corresponding 
 mirror Chern numbers are 1.

%When talking about mirror Chern number, people usually refer to the momentum space mirror Chern number rather than the real space one.
%To meet this convention, we devote the last section in this note to translate $C_{\mathrm{m}}$ into momentum space.

\textit{The hourglass invariant $\delta_{\mathrm{h}}$.}
In presence of glide symmetry, a $\mathbb{Z}_{2}$  bulk invariant can be defined from the Wilson loop operator \cite{Wang2016,aris-HG-2016}.
And the nontrivial topology is manifested by the hourglass fermion on the surface preserving the glide symmetry. 
A fixed-point wavefunction of such a state is a simple stacking of 2D TIs along the glide vector \cite{Ezawa2016}, 
 as shown in \figref[a]{fig:hourglass}.
The two 2D TIs in one unit cell are connected by the glide vector,
 thus they can not be dimerized without breaking the glide symmetry. 
Correspondingly, on the surface preserving the glide symmetry the two helical 
 modes contributed by the two 2D TIs are also connected by the glide vector
 and so also can not be dimerized. 

\begin{figure}
\begin{centering}
\includegraphics[width=1\linewidth]{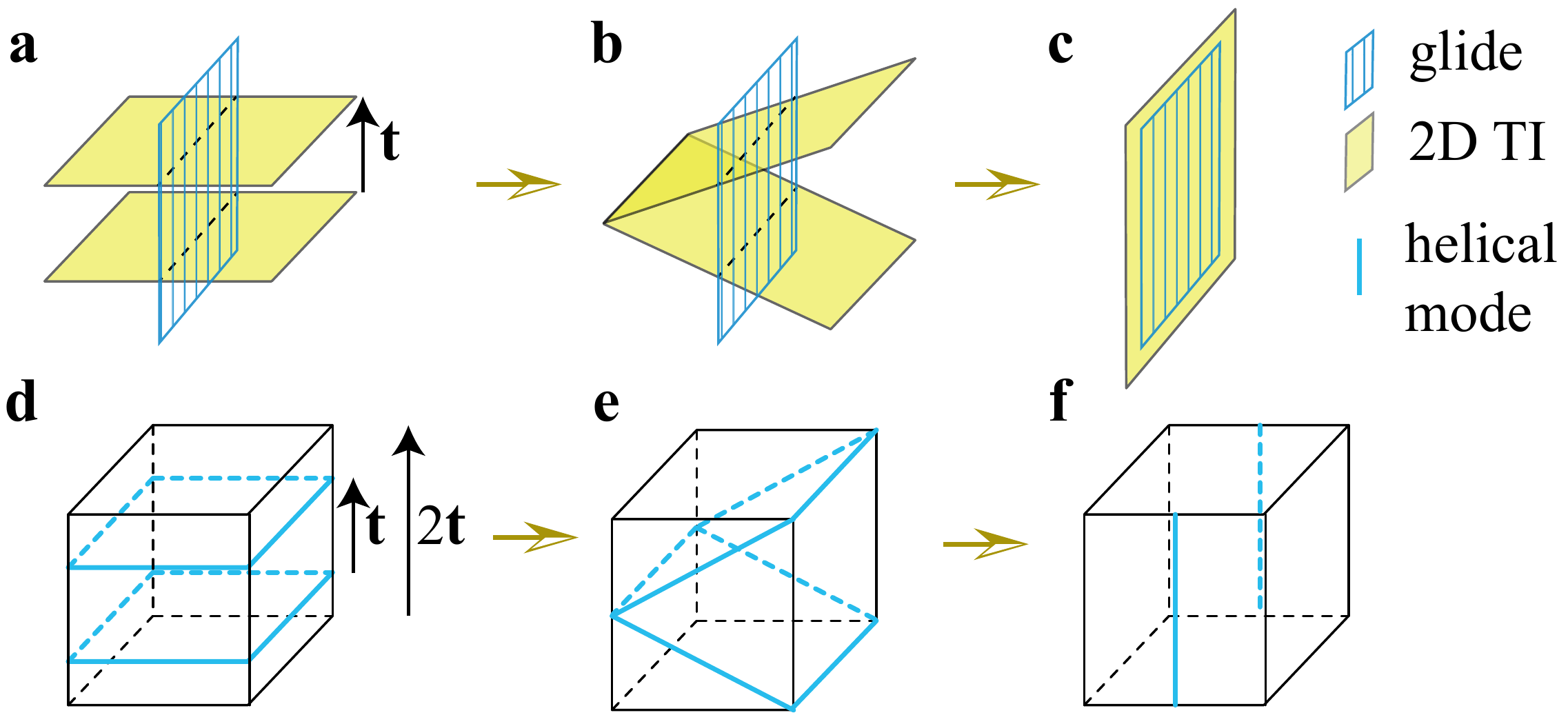}
\par\end{centering}
\protect\caption{
In {\bf a} and {\bf c}, the two LCs having nontrivial hourglass invariant are plotted respectively,
 and in {\bf b} the intermediate state between them is plotted.
The corresponding boundary states on a square cylinder geometry are plotted in {\bf d}-{\bf f}. The translation symmetry on the surface is assumed to be broken.\label{fig:hourglass}}
\end{figure}

Here we propose another glide-protected nontrivial LC configuration
 where only a single 2D TI occupies the glide plane, as shown in \figref[c]{fig:hourglass}.
This LC is nontrivial because the single 2D TI can not be trivialized symmetrically.
The corresponding boundary state on a square cylinder geometry, which is infinitely long to keep the glide symmetry, hosts only two 1D helical modes (\figref[f]{fig:hourglass}) if the other translation symmetry on surface is broken.
As illustrated in \figref[b]{fig:hourglass}, such a state can be symmetrically deformed to the above LC of the hourglass state
 where 2D TIs are stacked along the glide vector.
Therefore, the three LCs plotted in \figref[a-c]{fig:hourglass} are all topologically equivalent.

With the above analysis, we conclude that each pair of layers connected by the glide vector (\figref[a,b]{fig:hourglass}) or each single layer occupying the glide plane (\figref[c]{fig:hourglass}) contributes to one hourglass mode. 
Due to the $\mathbb{Z}_2$ nature of the hourglass invariant, 
 to calculate the $\delta_{\mathrm{h}}$  we should count the number of such configurations in an eLC and take the parity (even or odd) in the end.
Since the minimal lattice translation along the glide vector $\mathbf{t}_\parallel$ is $2\mathbf{t}_\parallel$,
 the layer $L$ in \figref[a,b]{fig:hourglass} contribute $\frac{1}{2\pi}2|\mathbf{t}_\parallel\cdot \mathbf{g}_L|$ helical modes per surface unit cell,
 or $\frac{1}{2\pi}|\mathbf{t}_\parallel\cdot \mathbf{g}_L|$ pairs of helical modes per surface unit cell, 
 where $\mathbf{g}_L = m_L \mathbf{b}_1 + n_L \mathbf{b}_2 + l_L \mathbf{b}_3 $ is the normal vector of $L$.
Therefore, the corresponding invariant can be calculated as
\begin{equation}
\delta_{\mathrm{h}}(E)=\sum_{L\in E} N_{m,\mathbf{t}_\parallel}^\mathrm{o}(L) + N_{m,\mathbf{t}_\parallel}^\mathrm{s}(L)\mod2
\label{eq:delta-2h}
\end{equation}
\begin{equation}
N_{m,\mathbf{t}_\parallel}^\mathrm{o}\left(L\right)=
\begin{cases}
1  & \text{if } m \in L \\
0  & \text{otherwise}
\end{cases}
\label{eq:N-glide1}
\end{equation}
\begin{equation}
N_{m,\mathbf{t}_\parallel}^\mathrm{s}\left(L\right)=
\frac{1}{2\pi}|\mathbf{t}_\parallel\cdot\mathbf{g}_{L}|
\label{eq:N-glide2}
\end{equation}
where  $m$ is the mirror plane, $\mathbf{t}_\parallel$ is glide vector, and $m\in L$ means that the glide plane is occupied by $L$.
$N_{m,\mathbf{t}_\parallel}^\mathrm{o}(L)$ and $N_{m,\mathbf{t}_\parallel}^\mathrm{s}(L)$ are the glide-ON and the glide-stacking-number (glide-SN) contributed by the LC configurations in \figref[c]{fig:hourglass} and \figref[a,b]{fig:hourglass} respectively,

\textit{The rotation invariant $\delta_{\mathrm{r}}$.}
The TCIs protected by $C_{n=2,4,6}$-rotation are proposed
very recently \cite{Fang2017,Song2017}.
%In this state there is no way to fully gap the top and the side boundaries without breaking the corresponding rotation symmetry. 
For a given rotation axis, the corresponding $\mathbb{Z}_2$ invariant $\delta_{\mathrm{r}}$ can be defined as: $1$ if the state can not be adiabatically deformed
 to an atomic insulator symmetrically with respect to the corresponding rotation axis, and $0$ otherwise.
As discussed in \onlinecite{Fang2017}, such states are easy to realize by layer construction.
For example, as illustrated in \figref[c]{fig:5ivt},
 if the $C_{2}$ axis is occupied by only one 2D TI, then this 2D TI can
 not be dimerized symmetrically since the minimal configuration
 that does not occupy the $C_{2}$ axis needs at least two 2D TIs (\figref[d]{fig:5ivt}. 
However, if the $C_{2}$ axis is occupied by two 2D TIs (dashed parallelograms in \figref[d]{fig:5ivt}), the 2D TIs can be dimerized symmetrically. 
Similarly, for the $C_{4}$ and $C_{6}$ axis, the minimal configuration occupying the axis needs
 only 2 and 3 2D TIs, respectively; while the minimal configuration
 that does not occupy the axis needs 4 and 6 2D TIs, respectively.
Therefore the eLC consisting of 2 (3) 2D TIs occupying a $C_{4}$
($C_{6}$) axis is also topologically nontrivial. 
(An exception is the $C_{3}$ axis, for which both the
minimal occupying configuration and the minimal not-occupying configuration
need 3 2D TIs.) 
To calculate the $\mathbb{Z}_{2}$ invariant of a
 $C_n$-rotation axis for a given eLC we need only to count the times that the axis is occupied
\begin{equation}
\delta_{\mathrm{r}}(E)=\sum_{L\in E}\frac{2}{n}N_{C_{n}}^\mathrm{o}\left(L\right)\mod2\label{eq:delta2r-1}
\end{equation}
\begin{equation}
N_{C_{n}}^\mathrm{o}\left(L\right)=\begin{cases}
1 & \text{if }C_{n}\subset L\\
0 & \text{otherwise}
\end{cases}\label{eq:N-rot}
\end{equation}
where $C_n \in L$ means that the $C_n$-rotation axis is occupied by $L$ and $N_{C_{n}}^\mathrm{o}\left(L\right)$ is the rotation-ON of $L$.

\textit{The inversion invariant $\delta_{\mathrm{i}}$.}
The inversion symmetry can also protect a new kind of TCI \cite{Fang2017}.
Such a state can be thought as a double of the centrosymmetric TI. Double of TI is usually considered as a trivial state since the surface Dirac nodes can be gapped, while it is proved that in presence of inversion symmetry even all the surfaces are gapped there must be a 1D helical mode on the inversion preserving
boundary. 
Such a state can be realized by a LC occupying the inversion center (\figref[e]{fig:5ivt}).
Similar with $C_{2}$-rotation-protected TCI, odd number of 2D TIs
occupying the inversion center can not be trivialized without breaking
the inversion while even number can be, leading to the $\mathbb{Z}_{2}$
classification. For each inversion center, we can assign an inversion
invariant $\delta_{\mathrm{i}}$ defined as: $1$ if the state can not be adiabatically
deformed to an atomic limit centrosymmetrically with respect to the
corresponding inversion center, and $0$ otherwise. 
In an eLC state, the invariant $\delta_{\mathrm{i}}$ can be simply counted as the parity of the number of times
 that the inversion center being occupied by the layers.
\begin{equation}
\delta_{\mathrm{i}}(E)=\sum_{L\in E}N_{i}^\mathrm{o}\left(L\right)\mod2
\end{equation}
\begin{equation}
N_{i}^\mathrm{o}\left(L\right)=\begin{cases}
1 & \text{if }i\in L\\
0 & \text{otherwise}
\end{cases}
\end{equation}
where $i\in L$ means that the inversion center $i$ is occupied by $L$ and  $N_{i}^\mathrm{o}\left(L\right)$ is the inversion-ON of $L$.

\subsection{Two new topological crystalline insulator invariants}

\textit{The screw invariant $\delta_\mathrm{s}$.}
In the above we have discussed that the TCI invariant $\delta_\mathrm{r}$ can be defined
 in presence of $C_{n=2,4,6}$-rotation symmetry.
A natural question is whether similar TCI invariant exits in presence of $C_{n=2,4,6}$-screw symmetry.
%The answer is yes.
Suppose that the $C_2$-rotation axis in \figref[c]{fig:5ivt} is replaced by a $C_2$-screw axis, then without breaking the $C_2$-screw symmetry this 2D TI can not be trivialized, whereas double of the LC can be symmetrically trivialized.
Similar statements also hold for $C_4$- and $C_6$-screw symmetries.
Therefore a $\mathbb{Z}_2$ invariant, denoted as $\delta_\mathrm{s}$, can be defined when $C_{n=2,4,6}$-screw presents.
On the other hand, similar with the hourglass state, such a LC can be deformed symmetrically to another LC where 2D TIs are stacked along the screw vector (just like the LC in \figref[a]{fig:hourglass}.
Similar deformations can also happen for $C_{n=4,6}$-screw-protected TCIs.
The key character of the deformed LC is that
 there are odd number of 2D TIs stacked in each screw-vector section,
 i.e., $\sum_{L}\frac{1}{2\pi}|\mathbf{t}_\parallel\cdot \mathbf{g}_L| = 1\ \mathrm{mod}\ 2$.
Here $L$ sums over the layers in the LC,
 $\mathbf{g}_L = m_L \mathbf{b}_1 + n_L\mathbf{b}_2 + l_L \mathbf{b}_3$ 
 is the normal vector of $L$, and $\mathbf{t}_\parallel$ is the screw vector.
Since the two possible LCs are topologically equivalent,
 to calculate $\delta_{\mathrm{r}}$ they should be equally treated.
Therefore, the screw invariant can be calculated as
\begin{equation}
\delta_\mathrm{s} (E)= \sum_{L\in E} \frac{2}{n}N_{C_n,\mathbf{t}_\parallel}^\mathrm{o}(L) + N_{C_n,\mathbf{t}_\parallel}^\mathrm{s}(L) \mod 2
\end{equation}
\begin{equation}
N_{C_n,\mathbf{t}_\parallel}^\mathrm{o}(L) = \begin{cases}
1 & \text{if }C_{n}\subset L\\
0 & \text{otherwise}
\end{cases}
\end{equation}
\begin{equation}
N_{C_n,\mathbf{t}_\parallel}^\mathrm{s}(L) = 
\frac{1}{2\pi}|\mathbf{t}_\parallel\cdot\mathbf{g}_L|
\end{equation}
where $N_{C_n,\mathbf{t}_\parallel}^\mathrm{o}(L)$ and $N_{C_n,\mathbf{t}_\parallel}^\mathrm{s}(L)$ 
 are the screw-ON and the screw-SN contributed by the two possible LC configurations, respectively.

\textit{The $S_{4}$ invariant $\delta_{S_4}$.}
There are three types of symmetry elements, i.e., the point
 like, the line like, and the plane like elements. 
%The line like elements
%include rotation axis and screw axis, and the plane like elements
%include mirror and glide. 
In the above section, all the topologies protected by line like and plane like symmetry elements have been
 discussed. 
However, the discussion about point like elements is incomplete---besides
 the inversion center there is another point like symmetry element,
 i.e., the $S_{4}$ center. 
It should be noticed that, other than $S_{6}$
 which is equivalent with a mirror plus a $C_{3}$ rotation, $S_{4}$
 center is a real point like element which can not be reduced to other
 type symmetry elements.

Now consider the symmetric configurations of 2D TIs occupying a $S_{4}$
center. There are two kinds of them are nontrivial. In the first configuration
(\figref[a]{fig:S4-LC}), only one 2D TI occupies the $S_{4}$ center
and it is perpendicular to the $C_2$-rotation axis given by $S_{4}^{2}$.
This configuration is nontrivial because we can not move or dimerize
the only 2D TI there. In the second configuration (\figref[c]{fig:S4-LC}),
two 2D TIs occupy the $S_{4}$ center and they are parallel with the
 $C_2$-rotation axis given by $S_{4}^{2}$. As shown in \figref[b]{fig:S4-LC},
there is a symmetric deformation connecting it to the first configuration,
thus the second configuration is indeed equivalent with the first
one. From the aspect of boundary state on a sphere geometry, the first configuration contributes one helical mode along the equator (\figref[d]{fig:S4-LC}), while the second configuration contributes two helical modes along two orthogonal meridians (\figref[f]{fig:S4-LC}).
These two boundary states are also connected by a symmetric deformation,
as shown in \figref[e]{fig:S4-LC}. 

Double of the nontrivial configuration must be trivial, because
both the bulk 2D TIs and the corresponding boundary states can be
dimerized without breaking the $S_{4}$ symmetry. Therefore, a $\mathbb{Z}_{2}$
invariant for this new TCI can be defined as: $1$ if the state can
not be deformed to an atomic insulator adiabatically and symmetrically
with respect to the $S_{4}$ symmetry, and $0$ otherwise. 
To calculate this invariant for an eLC, we need only to count the number of 2D
 TIs occupying the $S_{4}$ center
\begin{equation}
\delta_{S_4}(E)=\sum_{L\in E}N_{S_{4}}^\mathrm{o}\left(L\right)+\frac{1}{2}\sum_{L\in E}N_{S_{4}^{2}}^\mathrm{o}\left(L\right)\mod2\label{eq:delta2s}
\end{equation}
\begin{equation}
N_{S_{4}}^\mathrm{o}\left(L\right)=\begin{cases}
1 & \text{if }S_{4}\in L\text{ and }L\perp S_{4}^{2}\\
0 & \text{otherwise}
\end{cases}
\end{equation}
where $N_{S_4}^\mathrm{o}(L)$ is the $S_{4}$-ON of $L$ and $N_{S_{4}^{2}}^\mathrm{o}\left(L\right)$ is the $C_2$-rotation-ON of $L$.
A trick used here is to count the contribution
 of each 2D TI in the second configuration as $\frac{1}{2}$
 such that the two 2D TIs in total contribute $1$.

\begin{figure}
\includegraphics[width=1\linewidth]{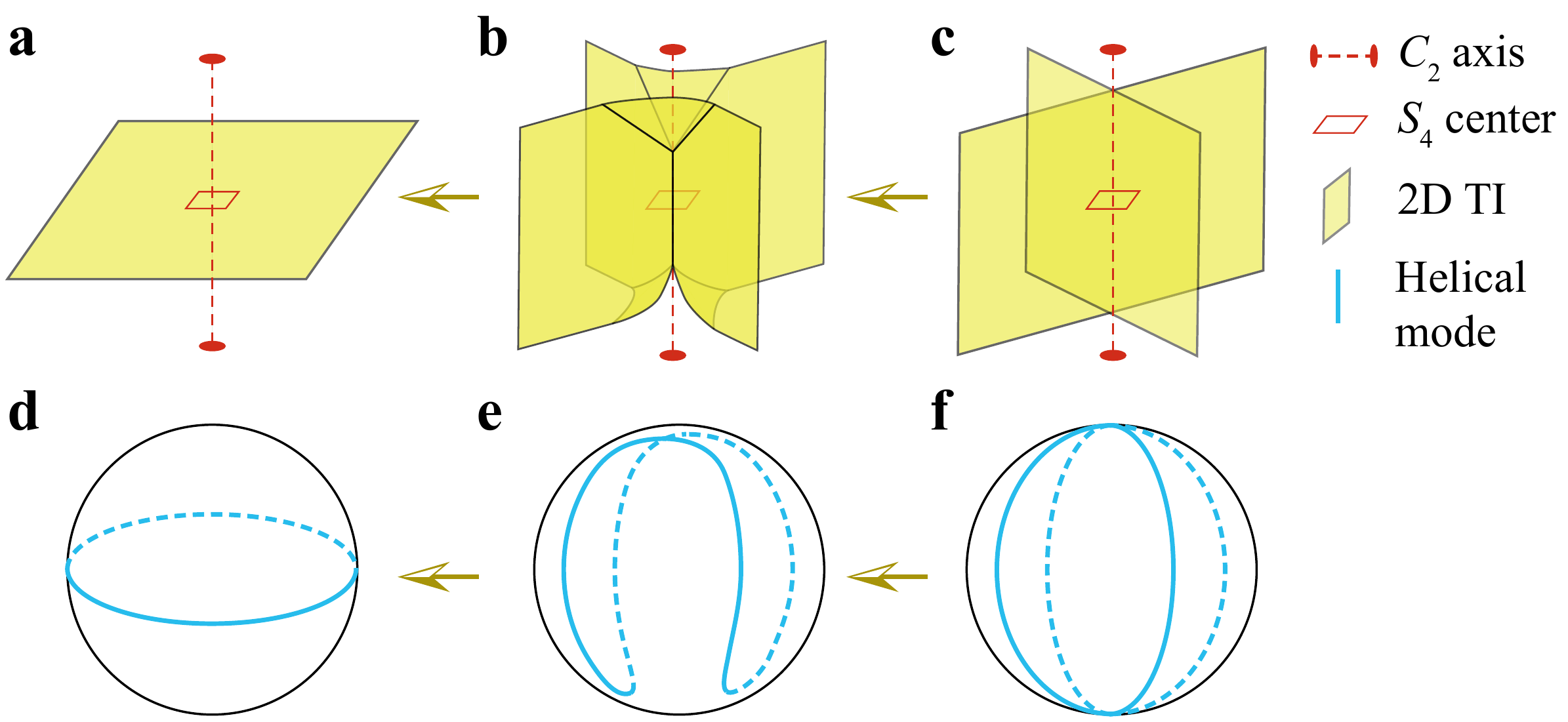}
\protect\caption{$S_{4}$ invariant in LC. In {\bf a} and {\bf c}, we show two
topologically equivalent LCs with nontrivial $S_{4}$ invariant,
and in {\bf b} we show the
symmetric deformation from {\bf c} to {\bf a}. {\bf d}, {\bf f}, {\bf e} are the corresponding
boundary states on a sphere geometry. \label{fig:S4-LC}}
\end{figure}

\subsection{Surface states of topological crystalline insulators}
\begin{figure*}
\includegraphics[width=0.9\linewidth]{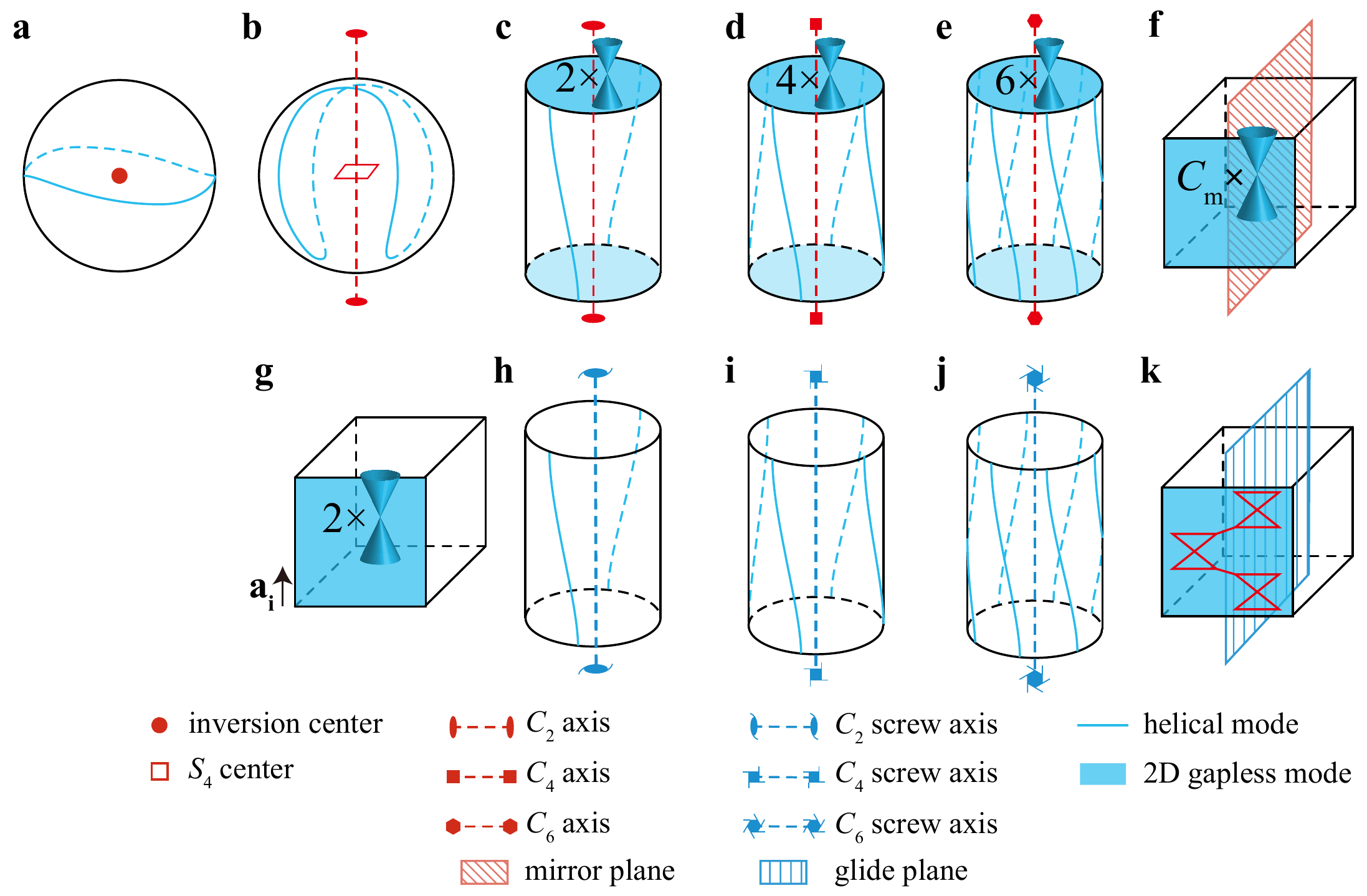}
\protect\caption{\label{fig:surface}Surface states of topological crystalline insulators. {\bf a} The surface state of inversion-protected TCI. {\bf b} The surface state of $S_4$-protected TCI. {\bf c}-{\bf e} The surface states of $C_{n=2,4,6}$-rotation-protected TCI. {\bf f} The surface state of mirror TCI. {\bf g} The surface state of weak TI, here $\mathbf{a}_i$ represents the translation symmetry protecting the weak index. {\bf h}-{\bf j} The surface states of $C_{n=2,4,6}$-screw-protected TCI. {\bf k} The surface state of hourglass TCI. }
\end{figure*}

Here we present a short summary of the topological surface states corresponding to the TCI invariants introduced above (\figref{fig:surface}).

The TCI invariants can be roughly classified into two classes by whether the protecting symmetry is point-group-like or not.
The point-group-like symmetries can be further classified into three cases by the dimensions of symmetry elements.
(i) The point-like symmetries, i.e., the inversion and $S_4$. 
The corresponding invariants are  $\delta_\mathrm{i}$ and $\delta_{S_4}$, and the correspinding anomalous surface states are 1D helical modes, which can be detected by putting the TCIs on finite geometries preserving the point-like symmetries, as shown in \figref[a,b]{fig:surface}.
Under the point-like symmetries the 1D modes transform to themselves such that without breaking the point-like symmetries the 1D modes can not be removed. (See \onlinecite{Fang2017} for more details of surface states of the inversion and $S_4$ invariants, respectively.)
(ii) The line-like symmetries, i.e., the $C_{n=2,4,6}$-rotation.
The correspinding invariant is $\delta_\mathrm{r}$.
And the corresponding anomalous surface state consists of two parts: the $n$ (modulo $2n$) 2D Dirac nodes on surface preserving the rotation symmetry and the $n$ (modulo $2n$) 1D helical modes on finite geometry preserving the rotation symmetry.
Both the two kinds of surface states can be detected by putting the TCI on a cylinder geometry preserving the rotation symmetry, where on the top surface are the 2D Dirac nodes and on the side surface are the 1D helical modes, as shown in  \figref[c-e]{fig:surface}. (See \onlinecite{Fang2017} and \cite{Song2017} for the 2D and 1D surface states, respectively.)
(iii) The plane-like symmetry, i.e., the mirror. 
The correspinding invariant is the mirror Chern number $C_{\mathrm{m}}$, and the correspinding anomalous surface state is the 2D state with $C_\mathrm{m}$ Dirac nodes on surface preserving the mirror symmetry, as shown in \figref[f]{fig:surface} \cite{Hsieh2012,Fulga2016}.

The non-point-group symmetries that can protect TCI invariants can also be classified into three cases.
(i) The translation symmetry.
The correspinding invariant is the weak TI index, i.e., $\delta_\mathrm{w}$, and the corresponding anomalous surface state is the 2D state with even number of Dirac nodes on surface preserving the translation symmetry, as shown in \figref[g]{fig:surface} \cite{Fu2007}.
(ii) The $C_{n=2,4,6}$-screw symmetry. 
The corresponding invariant is $\delta_\mathrm{s}$ and the correspinding anomalous surface states are the 1D helical modes on cylinder geometry preserving the screw symmetry.
As discussed above in this appendix, screw-protected TCI can have identical LCs with the rotation-protected TCI, where the $C_{n=2,4,6}$-screw axis is occupied by $n/2$ times.
Putting such LCs on a cylinder geometry, we get the 1D helical modes shown in \figref[h-j]{fig:surface}.
The only difference with rotation-protected TCI is the absence of 2D gapless surface state because the screw symmetry must be broken on any 2D surface.
(iii) The glide symmetry. The corresponding invariant is $\delta_\mathrm{h}$, and the corresponding anomalous surface state is the 2D hourglass fermion on surface preserving the glide symmetry, as shown in \figref[k]{fig:surface} \cite{Wang2016}.

\subsection{Generating all symmetry elements in a space group}

In this section we give the general strategy to generate all the symmetry elements in a given space group. 
When we talk about the symmetry element associated with a symmetry operation, we mean 
 (i) the center about which the operation takes place, which is given by the invariant
 geometry object under the operation, 
 and (ii) the type of the corresponding operation, i.e., inversion, rotation, screw, etc. 
A general space group operation can be written as $\left\{ p|\mathbf{t}+\mathbf{R}\right\} $ and
 can be interpreted as a point group operation centered at $\mathbf{x}$
 followed by a translation $\mathbf{t}_{\parallel}$
\begin{equation}
p\mathbf{r}+\mathbf{t}+\mathbf{R}=p\left(\mathbf{r}-\mathbf{x}\right)+\mathbf{x}+\mathbf{t}_{\parallel}
\end{equation}
where $\mathbf{x}$ and $\mathbf{t}_{\parallel}$ are determined by
\begin{equation}
\left(1-p\right)\mathbf{x}+\mathbf{t}_{\parallel}=\mathbf{t}+\mathbf{R}\label{eq:elmt-1}
\end{equation}
\begin{equation}
p\mathbf{t}_{\parallel}=\mathbf{t}_{\parallel}\label{eq:elmt-2}
\end{equation}
\begin{comment}
The complete information of the symmetry element associated with $\left\{ p|\mathbf{t}+\mathbf{R}\right\} $
is given by $\mathbf{x}$, $p$, and $\mathbf{t}_{\parallel}$, where
$\mathbf{x}$ gives the center and $p$, $\mathbf{t}_{\parallel}$
together give the type of the operation.
\end{comment}
Beware that $\mathbf{x}$ can be a point, a line, or a plane. 
Symmetry element got from nonzero $\mathbf{R}$ is called ``additional symmetry elements'' \cite{Hahn2002},
 which in general has different location and even different type with the original symmetry element 
 given by $\{p|\mathbf{t}\}$,
 for example, a rotation axis may change to a screw axis, and a mirror plane may change to a glide plane.
Due to the translation symmetry, the TCI invariants defined on symmetry elements locating in different
 cells must be identical with each other.
Therefore, a complete TCI invariant set should include only the TCI invariants defined in the home cell.

\subsection{The momentum space mirror Chern numbers}

As discussed above in this appendix, the real space mirror Chern
 number is easy to calculate for LCs. 
However, people usually understand the mirror Chern numbers in momentum space.
Thus we devote this section to translate the real space mirror Chern numbers to momentum space Mirror Chern numbers.

\begin{figure}
\begin{centering}
\includegraphics[width=0.95\linewidth]{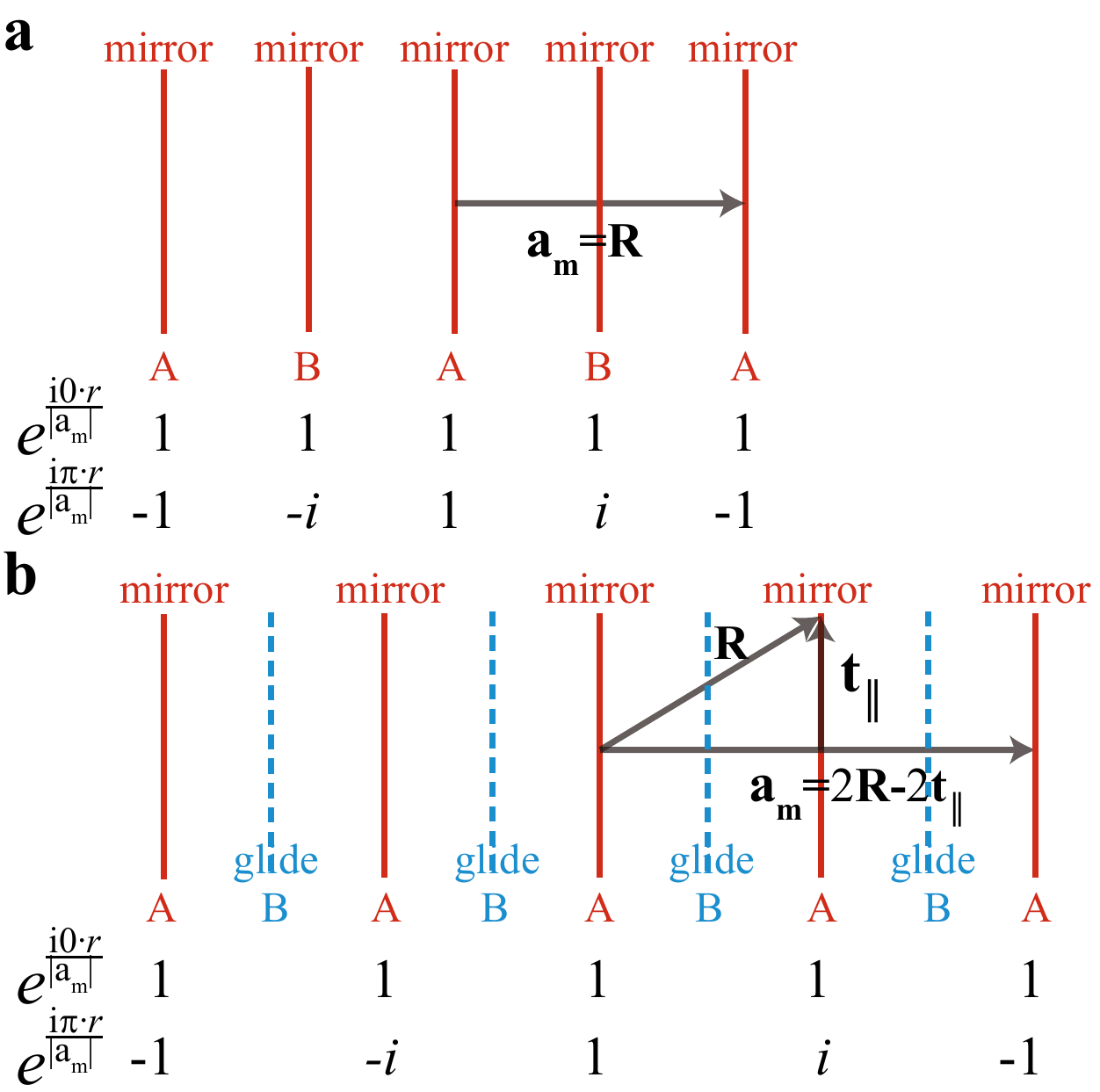}
\par\end{centering}

\protect\caption{The mirror plane generated from $\left\{ m|\mathbf{t}\right\} $
 is denoted as A, and the mirror or glide plane generated from $\left\{ m|\mathbf{t}+\mathbf{R}\right\} $ with $\frac{1}{4\pi}\mathbf{R}\cdot\mathbf{g}_{m}=\frac{1}{2}$ (modulo 1) is denoted as B. 
The $\mathbf{R}$ plotted is the shortest lattice vector giving $\frac{1}{2\pi}\mathbf{R}\cdot\mathbf{g}_m=\frac{1}{2}$.
In {\bf a} B is a also a mirror plane, and the minimal lattice vector perpendicular with A is $\mathbf{a}_{m}=\mathbf{R}$.
In {\bf b} B is a glide plane, and the minimal lattice vector perpendicular with A is $\mathbf{a}_{m}=2\mathbf{R}-2\mathbf{t}_{\parallel}$, where $\mathbf{t}_{\parallel}$ is the glide vector of B.
In the bottom of {\bf a} and {\bf b}, we list the phase factors in the Fourier transformations along $\mathbf{a}_{m}$ for $k=0$ and $k=\pi$. \label{fig:Cm}}
\end{figure}

Consider a mirror operation $\left\{ m|\mathbf{t}\right\} $, where
$\mathbf{t}$ is perpendicular with the mirror. By applying \cref{eq:elmt-1,eq:elmt-2}, we get that the mirror or glide plane generated
from $\left\{ m|\mathbf{t}+\mathbf{R}\right\} $ satisfies
\begin{align}
d & =\frac{1}{2\pi}\mathbf{x}\cdot\mathbf{g}_{m}=\frac{1}{4\pi}\mathbf{t}\cdot\mathbf{g}_{m}+\frac{1}{4\pi}\mathbf{R}\cdot\mathbf{g}_{m}
\end{align}
\begin{equation}
\mathbf{t}_{\parallel}=\mathbf{R}-\left(\mathbf{R}\cdot\hat{\mathbf{g}}_{m}\right)\hat{\mathbf{g}}_{m}
\end{equation}
Here $\mathbf{g}_{m}$ is the normal vector of the mirror, i.e., the
 minimal reciprocal vector perpendicular with the mirror, $\hat{\mathbf{g}}_{m}$
 is the unit vector along $\mathbf{g}_{m}$, $d$ is the position of
 the mirror or glide plane, and $\mathbf{t}_{\parallel}$ is the translation
 parallel with the glide plane (zero for mirror). 
Since $\frac{1}{2\pi}\mathbf{R}\cdot\mathbf{g}_m$ must be an integer, 
 $\frac{1}{4\pi}\mathbf{R}\cdot\mathbf{g}_{m}$ is either an integer or a half-integer.
Due to the translation symmetry,
 all $\mathbf{R}$ giving integer $\frac{1}{4\pi}\mathbf{R}\cdot\mathbf{g}_{m}$ 
 correspond to the original mirror plane A,
 and all $\mathbf{R}$ giving half-integer $\frac{1}{4\pi}\mathbf{R}\cdot\mathbf{g}_{m}$
 correspond to the new mirror or glide plane B.
The two cases that B is a mirror plane or a glide plane, which depend on the value of $\mathbf{t}_\parallel$, are very different. 
For the first case, we have two real space mirror Chern numbers for A and B,
 respectively. 
While, for the second case, we have only one real space mirror Chern number
 for A. 
For the first case, as shown in \figref[a]{fig:Cm}, 
 the minimal lattice vector perpendicular to the mirror, denoted as $\mathbf{a}_{m}$,
 is just the shortest $\mathbf{R}$ giving $\frac{1}{4\pi}\mathbf{R}\cdot\mathbf{g}_{m}=\frac{1}{2}$,
 and in one period there are two distinct mirror planes. 
Here we choose the mirror operation on A to define the mirror eigenvalue. 
Applying the Fourier transformation along $\mathbf{a}_{m}$, we find that for the two mirror-invariant
 momenta $k=0,\pi$ A subsystem has the same mirror eigenvalues
 while B subsystem has two converse mirror eigenvalues. 
Therefore, the mirror Chern number at $k=0$ and $k=\pi$ is given by
\begin{equation}
C_{\mathrm{m},0}=C_\mathrm{{m},A}+C_\mathrm{{m},B}
\end{equation}
\begin{equation}
C_{\mathrm{m},\pi}=C_\mathrm{{m},A}-C_\mathrm{{m},B}
\end{equation}
 respectively, where $C_\mathrm{{m},A}$ and $C_\mathrm{{m},B}$ are the two
 real space mirror Chern numbers. 
For the second case, as shown in \figref[b]{fig:Cm}, 
 the minimal lattice vector perpendicular to the mirror is given by $2\mathbf{R}-2\mathbf{t}_{\parallel}$,
 where $\mathbf{R}$ is the shortest lattice vector giving $\frac{1}{4\pi}\mathbf{R}\cdot\mathbf{g}_{m}=\frac{1}{2}$,
 and in one period there are two same mirror planes. 
Applying the Fourier transformation, we get
\begin{equation}
C_{\mathrm{m},0}=2C_\mathrm{{m},A}
\end{equation}
\begin{equation}
C_{\mathrm{m},\pi}=C_\mathrm{{m},A}-C_\mathrm{{m},A}=0
\end{equation}
where $C_{\mathrm{m},\pi}$ is always zero.

\clearpage
\newpage
\section{Convention dependence of topological invariants} \label{sec:conv}

There may be more than one symmetry elements that are identical with each other but locate at different positions.
For example, for any centrosymmetric SG there are eight inversion centers in a unit cell.
In the following we call such symmetry elements as noncoincident identical symmetry elements.
In general case, noncoincident identical symmetry elements are given by $\{p|\mathbf{t+R}\}$,
 where $p$ and $\mathbf{t}$ are fixed and $\mathbf{R}$ goes over lattice vectors
 that merely change the position of the symmetry operation (\cref{sec:Layer}). 
On each of the noncoincident symmetry elements we should assign a TCI invariant,
 e.g., eight inversion invariants for the eight inversion centers.
However, after an exhaustive enumeration over 230 SGs, we find that the TCI invariants defined on 
 noncoincident identical symmetry elements are not independent with each other.
For glide, rotation, inversion, screw, and $S_{4}$ symmetries, 
 one of the invariants can be uniquely determined from another
 and the three weak invariants. 
Therefore, among each kind of noncoincide identical symmetry elements
 it is enough to choose one to define the invariant. 
The convention for this choice in given in \cref{sec:table}. 

Here we take SG \#2 ($P\bar{1}$) as an example to show this convention dependence.
As shown in \figref[a]{fig:LCind}, if we choose the inversion center at origin
 to define the inversion invariant, then $E_1$, $E_2$, $E_3$ have nontrivial
 invariants, and $E_4$ has trivial invariant.
While, if we choose the inversion center at $(0,0,\frac{1}{2})$ to define the 
 inversion invariant, then $E_1$, $E_2$, $E_4$ have nontrivial invariants, and $E_3$ has 
 trivial invariant.

It should be emphasized that physically nontrivial topology should not depend on this
 convention.
In other words, the topology is physical as long as the invariants defined on every of the noncoincident
 symmetry elements are all nontrivial.
Taking SG $P\bar{1}$ as example, none of $E_1$, $E_2$, $E_3$, $E_4$ is physically nontrivial, because in any of
 them there are four empty inversion centers around which the 2D TIs can
 be dimerized symmetrically by breaking the inversion symmetry on the other four centers.
On the other hand, a physical nontrivial state can be realized as $E_3 \otimes E_4$, 
 where all the eight inversion centers are occupied and so $\delta_{\mathrm{i}}=1$ does not depend on the choice of inversion center.
In the ultimate presentation of our results (\cref{tab:SI2TOP}), 
 convention-independent $\mathbb{Z}_2$ invariants will be marked.

\clearpage
\newpage
\section{Generating all nonequivalent elementary layer constructions} \label{sec:find-eLC}

\subsection{Nonequivalent layers}

The topology of an eLC is completely determined by the little group,
 the weak invariants, the ONs, and the SNs of \emph{any}
 layer in it, because by applying the coset representatives of the
 little group, we can reproduce weak invariants, ONs, and SNs
 of all the other layers in the eLC and so reproduce the topology
 of the eLC. 
Let us denote a layer in an eLC as $L$,
 its little group as $\mathcal{S}\left(L\right)$, and the corresponding
 coset representatives as $g_{0}$, $g_{1}$, $\cdots$.
The eLC consists of $\left\{ g_{0}L,g_{1}L,\cdots\right\} $. 
Firstly, according to \cref{eq:delta2w-1,eq:delta2w-2,eq:delta2w-3},
 the weak invariants of $g_{i}L$ are just the parities of $g_i L$'s Miller indices
 and so are determined by the rotation matrix of $g_i$ (an integer matrix
 in the basis of lattice vectors) and the parities of $L$'s Miller
 indices ($L$'s weak invariants). 
Secondly, the ONs (SNs) on any symmetry element $e$ of $g_{i}L$ can be calculated
 as $N_{e}^\mathrm{o}\left(g_{i}L\right)=N_{g_{i}^{-1}eg_i}^\mathrm{o}\left(L\right)$
 ($N_{e}^\mathrm{s}\left(g_{i}L\right)=N_{g_{i}^{-1}eg_i}^\mathrm{s}\left(L\right)$).
In the following, we will say two layers are \emph{equivalent} with each other if the eLCs generated 
 from them are topologically equivalent. 

Here we give a sufficient condition for equivalence:
 two layers $L$ and $L^\prime$ must be equivalent with each other if 
 (i) $\mathcal{S}\left(L\right)=\mathcal{S}\left(L^{\prime}\right)$,
 (ii) $\delta_{\mathrm{w},i=1,2,3}\left(L\right)=\delta_{\mathrm{w},i=1,2,3}\left(L^{\prime}\right)$,
 (iii) $N_{e}^\mathrm{o}\left(L\right)=N_{e}^\mathrm{o}\left(L^{\prime}\right)$ for $e$
  as mirror (glide) planes, $C_2$-rotation (screw) axes, inversion centers, and $S_4$ centers,
 and (iv) $n_{\mathbf{t}_\parallel} N_e^\mathrm{s}(L)=n_{\mathbf{t}_\parallel} N_e^\mathrm{s}(L^\prime) \mod 2$ 
 for $e$ as glide planes and $C_{n=2,4,6}$-screw axes,
 where $n_{\mathbf{t}_\parallel}$ is the minimal multiplier that makes $\mathbf{t}_\parallel$ a lattice vector,
 and $\mathbf{t}_\parallel$ is the glide or screw vector.
For example, for glide and $C_2$-screw $n_{\mathbf{t}_\parallel}=2$.
Now let us prove the sufficient condition.
Firstly, according to (i)-(iii) and the discussion in last paragraph,
 the mirror, $C_2$-rotation, inversion, $S_4$ invariants, and weak invariants of $\mathrm{eLC}(L)$
 and $\mathrm{eLC}(L^\prime)$ must equal to each other.
Secondly, since a single layer's ON on a $C_{4}$-rotation ($C_6$-rotation) axis equals to its ON
 on the $C_2$-rotation axis given by $C_{4}^{2}$ ($C_{6}^{3}$),
 according to (i) and (iii) the $C_4$-rotation ($C_6$-rotation) invariants of $\mathrm{eLC}(L)$ and
 $\mathrm{eLC}(L^\prime)$ must equal to each other.
Thirdly, for $e$ as glide planes two cases should be discussed separately 
 and in both cases the hourglass invariants of $\mathrm{eLC}(L)$ and $\mathrm{eLC}(L^\prime)$ equal to each other.
In the first case we assume $N_g^\mathrm{o}(L)=N_g^\mathrm{o}(L^\prime)\neq 0$, where $g$ is the glide operation.
As both $L$ and $L^\prime$ occupy the glide plane, we have $N_g^\mathrm{s}(L)=N_g^\mathrm{s}(L^\prime)=0$.
According to (i) the hourglass invariants of $\mathrm{eLC}(L)$ and $\mathrm{eLC}(L^\prime)$ must equal to each other.
In the second case we assume $N_g^\mathrm{o}(L)=N_g^\mathrm{o}(L^\prime)=0$, which implies both $L$ and $L^\prime$ are not invariant under the glide operation.
However, we can introduce glide-invariant ``composite layers'' as $\tilde{L}=L\otimes g L$ and $\tilde{L}^\prime=L^\prime \otimes g L^\prime$,
 whose little group is given by $\mathcal{S}(L)\oplus g\mathcal{S}(L)=\mathcal{S}(L^\prime)\oplus g\mathcal{S}(L^\prime)$.
On one hand, the discussion in last paragraph also applies for these ``composite layers''
 and so the topologies of the eLCs are uniquely determined by the little groups, weak invariants, ONs, and SNs 
 of the ``composite layers''.
On the other hand, we have 
\begin{align}
N_g^\mathrm{s}(\tilde{L}) &= \frac{1}{2\pi}(|\mathbf{t}_\parallel\cdot\mathbf{g}_L| + |\mathbf{t}_\parallel\cdot g \mathbf{g}_L|)
                  = 2\frac{1}{2\pi}|\mathbf{t}_\parallel\cdot\mathbf{g}_L| \nonumber \\
                 &= 2 N_g^\mathrm{s}(L)
\end{align}
and $N_g^\mathrm{s}(\tilde{L}^\prime) = 2 N_g^\mathrm{s}(L^\prime)$ according to  \cref{eq:N-glide2}.
Therefore, the hourglass invariants of $\mathrm{eLC}(L)$ and $\mathrm{eLC}(L^\prime)$ on $g$ 
 should equal to each other as long as (i) and (iv) are satisfied.
Fourthly, the proof for $C_{n=2,4,6}$-screw axes is parallel with the proof for glide plane,
 where a ``composite layer'' should be introduced if the layer does not occupy the screw axis.

To exhaust all nonequivalent eLCs, we need only to exhaust
 all nonequivalent layers, which is done in two steps---in the following two sections we first give the strategy to generate all nonequivalent layers with same Miller indices, and then the strategy to exhaust all Miller indices.

\subsection{Nonequivalent layers with same Miller index}

In this section, we will prove that for a layer with given Miller indices 
 $\left(mnl;d\right)$ 
 there are only finite cases of $d$ to give nonequivalent layers, 
 and give the algorithm to find these cases. 
Since the weak invariants and SNs do not depend on $d$,
 in the following two $d$'s will be identified as equivalent if and only if
 the layer given by them have same little group and ONs on mirror (glide) planes, $C_2$-rotation (screw) axes,
 inversion centers, and $S_4$ centers.

Let us start with the discussion of little group. For convenience,
we set $\mathbf{g}=m\mathbf{b}_{1}+n\mathbf{b}_{2}+l\mathbf{b}_{3}$.
Then the little group $\mathcal{S}\left(mnl;d\right)$ consists of two
kinds of operations, (i) the operations that leave $\mathbf{g}$ invariant
\begin{align}
\mathcal{S}_{\mathrm{I}}\left(mnl\right) & =\Big\{\left\{ p|\mathbf{t}+\mathbf{R}\right\} \in\mathcal{G}\big|p\mathbf{g}=\mathbf{g}\nonumber \\
 & \qquad\frac{1}{2\pi}\mathbf{t}\cdot\mathbf{g}=0\ \mathrm{mod}\ 1\Big\}
\end{align}
and (ii) the operations that reverse the direction of $\mathbf{g}$
\begin{align}
\mathcal{S}_{\mathrm{II}}\left(mnl;d\right) & =\Big\{\left\{ p|\mathbf{t}+\mathbf{R}\right\} \in\mathcal{G}\big|p\mathbf{g}=-\mathbf{g}\nonumber \\
 & \qquad\frac{1}{4\pi}\left(\mathbf{t}+\mathbf{R}\right)\cdot\mathbf{g}=d\ \mathrm{mod}\ \frac{1}{2}\Big\}
\label{eq:SII}
\end{align}
The additional conditions in the braces come from the simultaneous
layer equations ( \cref{eq:layer}) before and after the operation,
i.e.,
\begin{equation}
\mathbf{r}\cdot\mathbf{g}=2\pi d\mod2\pi
\end{equation}
\begin{equation}
\left(p\mathbf{r}+\mathbf{t}+\mathbf{R}\right)\cdot\mathbf{g}=2\pi d\mod2\pi
\end{equation}
Apparently, the dependence of $\mathcal{S}\left(mnl;d\right)$ on $d$
comes from the conditions in $\mathcal{S}_{\mathrm{II}}$, which require
the layer to locate at some special positions.
%\begin{equation}
%d=\frac{1}{4\pi}\left(\mathbf{t}+\mathbf{R}\right)\cdot\mathbf{g}\mod\frac{1}{2}
%\end{equation}
To make these conditions more explicit, we simply list all the special
positions as 
\begin{align}
\mathcal{D}_{mnl}^{\prime} & =\bigg\{d\ \mathrm{mod}\ 1\bigg|d=\frac{\mathbf{t}\cdot\mathbf{g}}{4\pi},\left(\frac{\mathbf{t}\cdot\mathbf{g}}{4\pi}+\frac{1}{2}\right)\nonumber \\
 & \qquad\qquad\qquad\forall\left\{ p|\mathbf{t}\right\} \in\mathcal{G}\quad p\mathbf{g}=-\mathbf{g}\bigg\}\label{eq:d-points}
\end{align}
where we only take one of the points that coincide, and $\frac{1}{4\pi}\mathbf{R}\cdot\mathbf{g}$
is omitted because it is either an integer or a half-integer. 
Then, for any $d\in\mathcal{D}^{\prime}$, 
 $\mathcal{S}_\mathrm{II}(mnl;d)$ can be uniquely determined,  
 and for any $d\notin\mathcal{D}^{\prime}$
 there must be $\mathcal{S}_{\mathrm{II}}\left(mnl;d\right)=\emptyset$.
Use $d_{0}$ to represent the general point not in $\mathcal{D}^{\prime}$,
the full classification of $d$ from $0$ to $1$ can be formally
written as 
\begin{equation}
\mathcal{D}_{mnl}=\left\{ d_{0}\right\} \cup\mathcal{D}_{m_{L}n_{L}l_{L}}^{\prime}\label{eq:d-class}
\end{equation}
%where each entry is associated with a little group.
Here each entry of $\mathcal{D}_{mnl}$ can be thought
 as a Wyckoff position of the 1D SG projected from the 3D SG along the direction specified by the Miller indices,
 and,
 each entry of $\mathcal{D}_{mnl}$ is associated with a little group $\mathcal{S}(mnl;d)$.

Now we prove that, not only the little group but also the ONs
 for each $d\in\mathcal{D}_{mnl}$ can be determined.
They include ONs for (i) inversion centers,
 (ii) $C_2$-rotation (screw) axes perpendicular with $\mathbf{g}$,
 (iii) mirror (glide) planes perpendicular with $\mathbf{g}$, and (iv)
 $S_{4}$ centers with $S_{4}^{2}$ parallel with $\mathbf{g}$.
Notice that operations that contribute to \cref{eq:d-points}
 can only be (i) inversion, (ii) $C_2$-rotation (screw) perpendicular
 with $\mathbf{g}_{L}$, (iii) mirror (glide) perpendicular with $\mathbf{g}_{L}$,
 (iv) $S_{4}$ with $S_{4}^{2}$ parallel with $\mathbf{g}_{L}$, 
 corresponding to the the four kinds of ONs.
According to \cref{eq:elmt-1,eq:elmt-2}, $d$ in \cref{eq:d-points}
 is nothing but the possible positions of the corresponding symmetry
 elements. 
To be specific, for the symmetry element given by $\left\{ p|\mathbf{t}+\mathbf{R}\right\} $,
project both sides of \cref{eq:elmt-1} on $\mathbf{g}$,
we get its position as 
\begin{equation}
\frac{1}{2\pi}\mathbf{x}\cdot\mathbf{g}=\frac{1}{4\pi}\left(\mathbf{t}+\mathbf{R}\right)\cdot\mathbf{g}
\label{eq:pos-proj}
\end{equation}
which equals to either $\frac{1}{4\pi}\mathbf{t}\cdot\mathbf{g}$
or $\frac{1}{4\pi}\mathbf{t}\cdot\mathbf{g}+\frac{1}{2}$ (modulo
1). 
Therefore, all the possible positions of these symmetry elements have been included in 
 $\mathcal{D}_{mnl}$,
 and for each $d\in\mathcal{D}_{mnl}$,
 the four kinds of ONs can be calculated by checking
 whether the corresponding position (\cref{eq:pos-proj}) coincides with $d$.
The trivial case where all the ONs are zero is given by $d=d_{0}$.

\subsection{Exhausting Miller indices}

In this section, we will prove that to generate all nonequivalent
eLCs only a finite number of Miller indices around $(0,0,0)$ need
to be considered, and give the concrete algorithm to find these Miller
indices.

\begin{figure}
\begin{centering}
\includegraphics[width=1\linewidth]{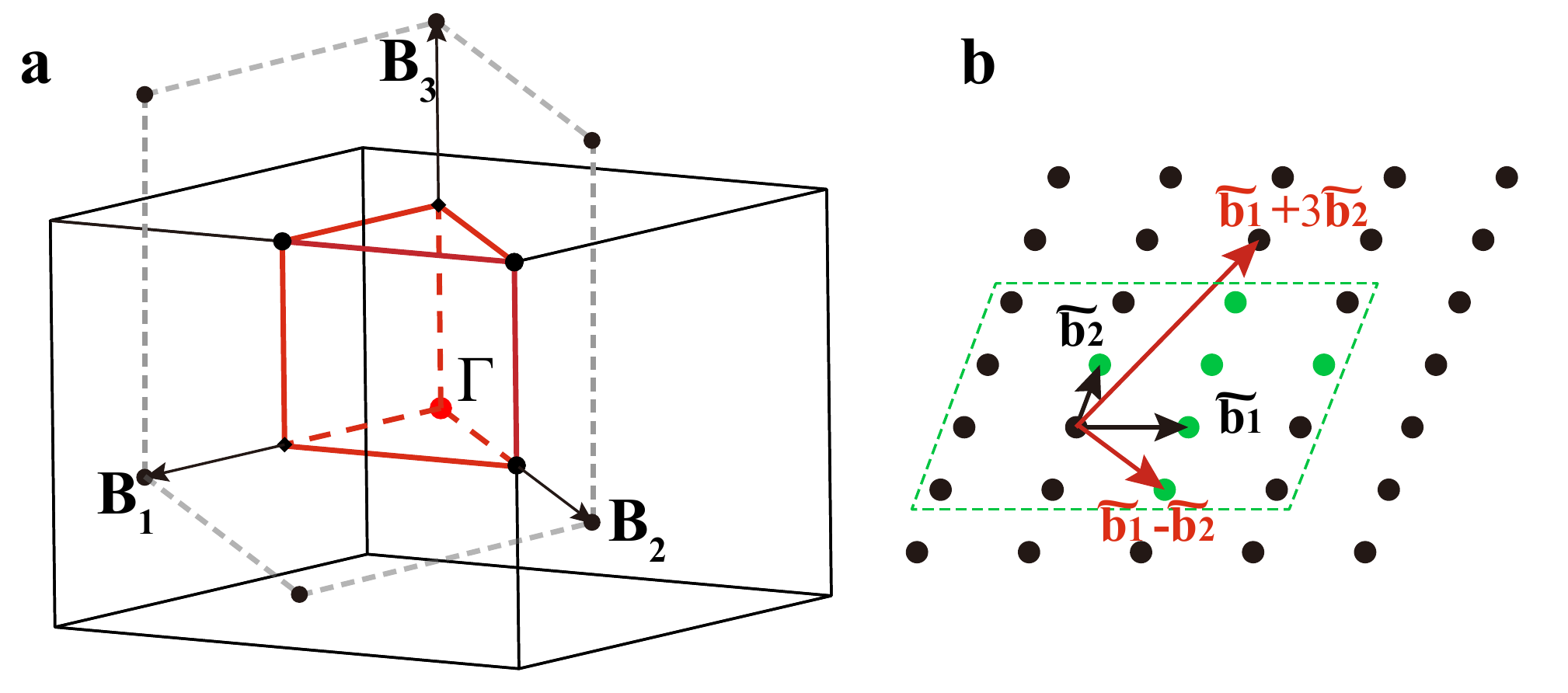}
\par\end{centering}

\protect\caption{\label{fig:Miller}In {\bf a} is plotted the irreducible domain, which
is spanned by the three reciprocal vectors $\mathbf{B}_{1}$, $\mathbf{B}_{2}$,
and $\mathbf{B}_{3}$ and can be thought as the extension of the irreducible
Brillouin zone (plotted by the red lines) to infinity.
In general case, a reciprocal lattice in the irreducible domain can be expanded by $\mathbf{B}_{1/2/3}$ with {\it positive} integers or fractions. 
%In this irreducible domain, there are three high symmetry
%directions along $\mathbf{B}_{1}$, $\mathbf{B}_{2}$, and $\mathbf{B}_{3}$
%respectively, and three high symmetry planes spanned by $\mathbf{B}_{1}$
%$\mathbf{B}_{2}$, $\mathbf{B}_{2}$ $\mathbf{B}_{3}$, and $\mathbf{B}_{3}$
%$\mathbf{B}_{1}$ respectively. 
In {\bf b} we show the 6 reciprocal lattices
(green circles) that need to be considered on a high symmetry plane
for the purpose of generating nonequivalent eLCs, where $\tilde{\mathbf{b}}_{1}$
and $\tilde{\mathbf{b}}_{2}$ are the primitive bases on this plane.
For example, all the eLCs generated from $\tilde{\mathbf{b}}_{1}+3\tilde{\mathbf{b}}_{2}$
can also be generated from $\tilde{\mathbf{b}}_{1}-\tilde{\mathbf{b}}_{2}$.}
\end{figure}

Two sets of Miller indices $\left(mnl\right)$ $\left(m^{\prime}n^{\prime}l^{\prime}\right)$
 are called to be \emph{equivalent} with each other if one of them
 can be transformed from another by a SG operation, i.e.,
\begin{equation}
\exists\left\{ p|\mathbf{t}\right\} \in\mathcal{G}\quad s.t.\quad\mathbf{g}=\pm p\mathbf{g}^{\prime}
\label{eq:MillerEQ}
\end{equation}
 where $\mathbf{g}=m\mathbf{b}_{1}+n\mathbf{b}_{2}+l\mathbf{b}_{3}$
 and $\mathbf{g}^{\prime}=m^{\prime}\mathbf{b}_{1}+n^{\prime}\mathbf{b}_{2}+l^{\prime}\mathbf{b}_{3}$.
%Then by definition an eLC must consist of layers with all the equivalent
%Miller indices. 
Based on the equivalent relation we introduce the concept of irreducible domain of
 reciprocal lattice---the minimal subset of reciprocal lattice from which the full reciprocal
 lattice can be reproduced by equivalence relation (\cref{eq:MillerEQ}).
The irreducible domain can be thought as the extension of the irreducible
 Brillouin zone to infinity, as shown in \figref[a]{fig:Miller}.
The irreducible domain consists of high symmetry directions (edges),
 high symmetry planes (surfaces), and general points (bulk). 
Similar with the irreducible Brillouin zone, any reciprocal vector outside
 the irreducible domain can be transformed from a vector in it by a SG operation. 
%Analysis in \ref{sub:eq-layer} shows that the topology
% of an eLC can be inferred from any layer in it, which of course can be chosen as the one with its Miller indices
% in the irreducible domain.
Thus to cover all the nonequivalent eLCs, we need only to enumerate
 all the acceptable Miller indices in the irreducible domain. 
Due to the coprime condition of Miller indices, along an edge only the shortest
 vector give acceptable Miller indices. 
While in a surface or the bulk of the irreducible domain there are infinite acceptable
 Miller indices. 
Here we will prove that even in the latter case only finite Miller indices around $(000)$ need to be considered. 

We firstly give the proof and algorithm for Miller indices in a surface
 of the irreducible domain. 
As shown in \figref[b]{fig:Miller}, $\tilde{\mathbf{b}}_{1}$ and $\tilde{\mathbf{b}}_{2}$ are chosen
 as the primitive bases in the surface such that any reciprocal vector in
 the surface can be expanded on them with integer coefficients
\begin{equation}
\mathbf{g}=\tilde{m}\tilde{\mathbf{b}}_{1}+\tilde{n}\tilde{\mathbf{b}}_{2}\label{eq:gtilde}
\end{equation}
A layer on the surface $(\tilde{m}\tilde{n};d)$ is given by a pair of coprime integers $\tilde{m},\tilde{n}$
 and a position $d$. 
As discussed in the above two sections, the relevant properties of a layer  includes (i) the little group, (ii) the weak invariants, (iii) the ONs on inversion centers, $C_2$-rotation (screw) axes, mirror (glide) planes, and $S_4$ centers, 
 and (iv) the multiples of glide-SNs and screw-SNs, i.e,
\begin{equation}
n_{\mathbf{t}_\parallel} \frac{1}{2\pi}\mathbf{t}_\parallel\cdot\left(\tilde{m}\tilde{\mathbf{b}}_{1}+\tilde{n}\tilde{\mathbf{b}}_{2}\right)\mod2\label{eq:Ng-surf}
\end{equation}
 where $\mathbf{t}_\parallel$ is the glide or screw vector 
 and $n_{\mathbf{t}_\parallel}$ is the minimal multiplier that makes $\mathbf{t}_\parallel$ a lattice vector.
Now we prove that, for any integers $r$ and $s$ if $\tilde{m}-4r$ and $\tilde{n}-4s$
 are coprime the four relevant properties of the layer $\left(\tilde{m}-4r,\tilde{n}-4s;d\right)$
 are identical with the $\left(\tilde{m}\tilde{n};d\right)$'s.
Firstly, as we count $\tilde{m}\tilde{\mathbf{b}}_{1}+\tilde{n}\tilde{\mathbf{b}}_{2}$
as a general vector in the Miller surface, the little group should only consist
of operations that leave both $\tilde{\mathbf{b}}_{1}$ and $\tilde{\mathbf{b}}_{2}$
invariant or reverse, i.e.,
\begin{equation}
\mathcal{S}\left(\tilde{m}\tilde{n};d\right)=\mathcal{S}_{\mathrm{I}}\left(\tilde{m}\tilde{n}\right)\cup\mathcal{S}_{\mathrm{II}}\left(\tilde{m}\tilde{n};d\right)
\end{equation}
\begin{align}
\mathcal{S}_{\mathrm{I}}\left(\tilde{m}\tilde{n}\right) & =\Big\{\left\{ p|\mathbf{t}+\mathbf{R}\right\} \in\mathcal{G}\big|p\tilde{\mathbf{b}}_{1}=\tilde{\mathbf{b}}_{1},\ p\tilde{\mathbf{b}}_{2}=\tilde{\mathbf{b}}_{2}\nonumber \\
 & \qquad\frac{1}{2\pi}\mathbf{t}\cdot\left(\tilde{m}\tilde{\mathbf{b}}_{1}+\tilde{n}\tilde{\mathbf{b}}_{2}\right)=0\ \mathrm{mod}\ 1\Big\}\label{eq:SI-surf}
\end{align}
\begin{align}
\mathcal{S}_{\mathrm{II}}\left(\tilde{m}\tilde{n};d\right) & =\Big\{\left\{ p|\mathbf{t}+\mathbf{R}\right\} \in\mathcal{G}\big|p\tilde{\mathbf{b}}_{1}=-\tilde{\mathbf{b}}_{1},\ p\tilde{\mathbf{b}}_{2}=-\tilde{\mathbf{b}}_{2}\nonumber \\
 & \qquad\frac{1}{4\pi}\mathbf{t}\cdot\left(\tilde{m}\tilde{\mathbf{b}}_{1}+\tilde{n}\tilde{\mathbf{b}}_{2}\right)=d\ \mathrm{mod}\ \frac{1}{2}\Big\}\label{eq:SII-surf}
\end{align}
Operation in $\mathcal{S}_{\mathrm{I}}$ can only be mirror or glide, 
 and operation in $\mathcal{S}_{\mathrm{II}}$ can only be inversion, $C_2$-rotation, or $C_2$-screw. 
Thus $\mathbf{t}$ in \cref{eq:SI-surf,eq:SII-surf} must be either a lattice vector or half of a lattice vector,
 which implies 
\begin{equation}
\frac{1}{2\pi}\mathbf{t}\cdot\left(4r\tilde{\mathbf{b}}_{1}+4s\tilde{\mathbf{b}}_{2}\right)=0\mod1
\end{equation}
\begin{equation}
\frac{1}{4\pi}\mathbf{t}\cdot\left(4r\tilde{\mathbf{b}}_{1}+4s\tilde{\mathbf{b}}_{2}\right)=0\mod\frac{1}{2}
\end{equation}
Then it follows that $\mathcal{S}\left(\tilde{m}\tilde{n};d\right)=\mathcal{S}\left(\tilde{m}-4r,\tilde{n}-4s;d\right)$.
Secondly, as $4r$ and $4s$ does not change any parities of the Miller indices, there must be 
 $\delta_{\mathrm{w},i=1,2,3}\left(\tilde{m}\tilde{n};d\right)=\delta_{\mathrm{w},i=1,2,3}\left(\tilde{m}-4r,\tilde{n}-4s;d\right)$.
Thirdly, since $\tilde{m}\tilde{\mathbf{b}}_{1}+\tilde{n}\tilde{\mathbf{b}}_{2}$ is
 a general vector in the Miller surface, the symmetry elements occupied by $(\tilde{m}\tilde{n};d)$ can only be inversion centers,
 $C_2$-rotation axes, or $C_2$-screw axes, as given by
\begin{align}
\mathcal{E}\left(\tilde{m}\tilde{n};d\right) & =\Big\{\left\{ p|\mathbf{t}+\mathbf{R}\right\} \in\mathcal{G}\big|p\tilde{\mathbf{b}}_{1}=-\tilde{\mathbf{b}}_{1},\ p\tilde{\mathbf{b}}_{2}=-\tilde{\mathbf{b}}_{2},\nonumber \\
 & \qquad\frac{1}{4\pi}\left(\mathbf{t}+\mathbf{R}\right)\cdot\left(\tilde{m}\tilde{\mathbf{b}}_{1}+\tilde{n}\tilde{\mathbf{b}}_{2}\right)=d\ \mathrm{mod}\ 1\Big\}\label{eq:Eset-surf}
\end{align}
 where $\mathbf{t}$ is either a lattice vector or half of a lattice vector.
ONs on these occupied symmetry elements are 1, and the ONs on all other symmetry elements are 0.
Then it follows that $\mathcal{E}\left(\tilde{m}\tilde{n};d\right)=\mathcal{E}\left(\tilde{m}-4r,\tilde{n}-4s;d\right)$
 due to the relation
\begin{equation}
\frac{1}{4\pi}\left(\mathbf{t}+\mathbf{R}\right)\cdot\left(4r\tilde{\mathbf{b}}_{1}+4s\tilde{\mathbf{b}}_{2}\right)=0\mod1
\end{equation}
Fourthly, the multiples of glide-SNs or screw-SNs (\cref{eq:Ng-surf}) of $\left(\tilde{m}\tilde{n};d\right)$
 and $\left(\tilde{m}-4r,\tilde{n}-4s;d\right)$ must also equal to
 each other for similar reason. 
In the end, we show that, for any coprime $\tilde{m},\tilde{n}$ their
correspondences $\tilde{m}^{\prime},\tilde{n}^{\prime}$ in the interval
$\left[-1,2\right]$ got by $\tilde{m}^{\prime}=\tilde{m}-4r$ and
$\tilde{n}^{\prime}=\tilde{n}-4s$ must also be coprime. 
All the possible non-coprime pairs of $\tilde{m}^{\prime},\tilde{n}^{\prime}$
 can be $\left(0,2\right)$, $\left(2,0\right)$, $\left(2,2\right)$,
 however, if $\tilde{m}^{\prime},\tilde{n}^{\prime}$
 have the common divisor $2$ then $\tilde{m},\tilde{n}$ must also
 have the common divisor $2$, which contradicts with the presumption.
Therefore, the conclusion is that to generate all nonequivalent eLCs
from the Miller indices within a surface of the irreducible domain
we need only to consider the reciprocal vectors $\tilde{m}\tilde{\mathbf{b}}_{1}+\tilde{n}\tilde{\mathbf{b}}_{2}$
with $\tilde{m},\tilde{n}=-1,0,1,2$, as shown by the dashed green
box in \figref[b]{fig:Miller}. 
After removing the parallel vectors, only the six vectors marked as green in \figref[b]{fig:Miller} need to be considered.

The discussion for the general Miller indices in the bulk of the irreducible domain
 is much more simple.
In this case the only possible operation in the little group is inversion and the only possible occupied symmetry
 element is inversion center. 
As we take the convention where the inversion center is put at the origin, 
 the two layers $\left(mnl;d\right)$ $\left(m-2r,n-2s,l-2t;d\right)$ have same little group and
 same inversion-ONs. 
As $2r$, $2s$, $2t$ are even numbers, they also have same weak invariants, same glide-SNs,
 and same screw-SNs (\cref{eq:Ng-surf}).
Therefore, in the bulk of the irreducible domain, we need only to consider $m\mathbf{b}_{1}+n\mathbf{b}_{2}+l\mathbf{b}_{3}$ with
$m,n,l=0,1$.

\clearpage
\newpage
\section{Symmetry-based indicators\label{sec:indicator}}

In \onlinecite{Po2017,Bradlyn2017}, a complete theory of symmetry-based indicator (SI) has been developped, 
 which in principle is all we can know about band topology from symmetry eigenvalues.
However, the explicit expressions of these SI are absent in the original work.
Here we give explicit formulae for \emph{all} the SI.
In the rest of this paper, these formulae are referred as Fu-Kane-like formulae.
%These formulae consist of the Fu-Kane formulae \cite{Fu2007} for strong and weak
%TIs, the mirror Chern number formulae, the $P4/m$-based $\mathbb{Z}_{8}$ formula \cite{Song2017}, 
% and a newly found $S_{4}$-based $\mathbb{Z}_{2}$ formula. 
%Particularly, the nonzero $S_{4}$-based SI corresponds to strong TI for a fully gapped state.

\subsection{Symmetry-based indicators}

For a system with given SG, its band representation (BR) for a group
of selected bands (the occupied bands for example) can be written
as an integer vector 
\begin{equation}
\mathrm{BR}=\left[n\left(\xi_{1}^{K_{1}}\right),n\left(\xi_{2}^{K_{1}}\right)\cdots,n\left(\xi_{1}^{K_{2}}\right)\cdots\right]^{T}
\end{equation}
where each entry gives the number of a irreducible representation
(irrep) $\xi$ at a momentum $K$. On one hand, all the compatibility
relation allowed BRs form a linear space, whose bases can be written
as
\begin{equation}
\left\{ B_{1},B_{2}\cdots B_{D}\right\} 
\end{equation}
On the other hand, all the BRs that can be generated from symmetric
Wannier functions, i.e., the BRs of atomic insulators, also form a linear space,
with bases
\begin{equation}
\left\{ A_{1},A_{2}\cdots A_{D^{\prime}}\right\} 
\end{equation}
By an exhaustive enumeration over 230 SGs, the author of \onlinecite{Po2017}
 find that the dimensions of these two linear spaces are always same
 ($D=D^{\prime}$) and all $B_{i}$ can be expanded by $\left\{ A_{i}\right\} $
 with integral or fractional coefficients. 
$B_{i}$ with fractional coefficients on $\left\{ A_{i}\right\} $ must be topologically nontrivial
 because it mismatches with all possible atomic BRs. 
Then the SI of a given BR are defined as its expansion coefficients on the nontrivial $B_{i}$'s.
In this work we call each coefficient as a indicator or a SI, 
 and the coefficient set as a SI set.

An observation follows, if $\mathcal{G}$, $\mathcal{G}^{\prime}$
 are two SGs and $\mathcal{G}^{\prime}\subset\mathcal{G}$, then we
 can build a many-to-one mapping from the SI sets in $\mathcal{G}$
 to the SI sets in $\mathcal{G}^{\prime}$ by breaking the additional
 symmetries.
If this mapping is one to one, i.e., any two different SI sets in
 $\mathcal{G}$ map to two different SI sets in $\mathcal{G}^{\prime}$,
 we say that the SI sets in $\mathcal{G}$ are \emph{induced} from
 the SI sets in $\mathcal{G}^{\prime}$. 
%Here we take $\mathcal{G}=C2/m$ and $\mathcal{G}^{\prime}=P\bar{1}$ for example, whose SI sets
% form the group $\mathbb{Z}_{2}\times\mathbb{Z}_{2}\times\mathbb{Z}_{4}$
% and $\mathbb{Z}_{2}\times\mathbb{Z}_{2}\times\mathbb{Z}_{2}\times\mathbb{Z}_{4}$
% respectively. 
%By a careful study on their BRs, we find the mapping as 
%\begin{equation}
%\left(001\right)\to\left(0001\right)
%\end{equation}
%\begin{equation}
%\left(010\right)\to\left(0010\right)
%\end{equation}
%\begin{equation}
%\left(100\right)\to\left(1101\right)
%\end{equation}
%which implies that the SI sets of $C2/m$ can be indeed induced
% from the SI sets of $P\bar{1}$. 
After an exhaustive study, we find that the SI in \emph{all} SGs can be induced from the
following six SGs
\[
\#2\ (P\bar{1}),\;\#81\ (P\bar{4}),\;\#83\ (P4/m),\ \#174\ (P\bar{6})
\]
\[
\#175\ (P6/m),\ \#176\ \left(P6_{3}/m\right)
\]
Therefore, in order for the complete Fu-Kane-like formulae, we need
only to derive the Fu-Kane-like formulae for this six SGs.

\subsection{Five kinds of Fu-Kane-like formulae}

\textit{$P\bar{1}$ formulae.}
SI of SG \#2 ($P\bar{1}$) form the group $\mathbb{Z}_{2}\times\mathbb{Z}_{2}\times\mathbb{Z}_{2}\times\mathbb{Z}_{4}$.
As discussed in \onlinecite{Po2017}, the 
$\mathbb{Z}_{4}$ and $\mathbb{Z}_{2}$ generators are the strong and
weak TIs, respectively. 
The $\mathbb{Z}_{2}$ indicators, denoted as $z_{2\mathrm{w},i=1,2,3}$ in the following, can be calculated by the
 Fu-Kane formula. 
And, by a detailed analysis on the BR, we find that the $\mathbb{Z}_{4}$ indicator can be calculated as
\begin{equation}
z_{4}=\sum_{\mathbf{K}\in\mathrm{TRIM}}\frac{n_{\mathbf{K}}^{-}-n_{\mathbf{K}}^{+}}{2}\mod4\label{eq:z4i}
\end{equation}
Here $n_{\mathbf{K}}^{-}$ is the number of occupied odd-parity Kramer
pairs at $\mathbf{K}$, $n_{\mathbf{K}}^{+}$ is the number of occupied even-parity
Kramer pairs at $\mathbf{K}$, and $\mathbf{K}$ is summed over all
the eight time-reversal invariant momenta (TRIMs). According to the Fu-Kane criterion, $z_{4}=1,3$
corresponds to strong TI. 
%(As $\sum_{\mathbf{K}}n_{\mathbf{K}}^{-}+\sum_{\mathbf{K}}n_{\mathbf{K}}^{+}$
%must be even, we have $\sum_{\mathbf{K}}n_{\mathbf{K}}^{-}=1\mod2$
%for $z_{4}=1,3$.) 
And, according to the discussion in \cref{sec:LCind}, $z_{4}=2$ corresponds to the inversion-protected TCI.

\textit{$P4/m$ formulae.}
SI of SG \#83 ($P4/m$) form the group $\mathbb{Z}_{2}\times\mathbb{Z}_{4}\times\mathbb{Z}_{8}$,
whose complete Fu-Kane-like formulae have been derived in \onlinecite{Song2017}.
Here we only summarize the main results. The $\mathbb{Z}_{2}$ indicator 
is the weak TI indicator $z_{2\mathrm{w},1}$, which equals to $z_{2\mathrm{w},2}$
according to the $C_{4}$ symmetry. The $\mathbb{Z}_{4}$ indicator is
the mirror Chern number (modulo 4) of the $k_{z}=\pi$ plane and will
be denoted as $z_{4\mathrm{m},\pi}$ in the following. 
And the $\mathbb{Z}_{8}$ indicator is given by
\begin{equation}
z_{8}=\left[3n_{\frac{3}{2}}^{+}-3n_{\frac{3}{2}}^{-}-n_{\frac{1}{2}}^{+}+n_{\frac{1}{2}}^{-}\right]/2\mod8
\label{eq:z8}
\end{equation}
where the definitions for $n_{\frac{3}{2}}^{+}$, $n_{\frac{3}{2}}^{-}$,
$n_{\frac{1}{2}}^{+}$, $n_{\frac{1}{2}}^{-}$ are given in \cref{tab:z8}. 
According to \onlinecite{Song2017}, if $z_{2\mathrm{w},1}=0$ and $z_{4\mathrm{m},\pi}=0$,
 odd $z_{8}$ corresponds to strong TI, $z_{8}=2,6$ corresponds to mirror TCI with
 mirror Chern number 2 (modulo 4) in the $k_{z}=0$ plane, and $z_{8}=4$
 corresponds to either mirror TCI with mirror Chern number 4 (modulo
 8) in the $k_{z}=0$ plane or TCI with nontrivial $C_{4}$-rotation invariant. 

\renewcommand\arraystretch{1.5}
\begin{table*}
\begin{centering}
\begin{tabular}{|>{\centering}p{2cm}|>{\centering}p{2cm}|c|c|}
\hline 
Lattice & SGs & $n$ & Definitions for $n_{\frac{3}{2}}^{+}$, $n_{\frac{3}{2}}^{-}$, $n_{\frac{1}{2}}^{+}$,
$n_{\frac{1}{2}}^{-}$ \\
\hline 
\hline 
\multirow{4}{2cm}{\centering Tetragonal primitive} & \multirow{4}{2cm}{\centering 83 (123, 124, 127, 128)\footnotemark[1]} & $n_{\frac{1}{2}}^{+}$ & $n(E_{\frac{1}{2}\mathrm{g}}^{\Gamma})+n(E_{\frac{1}{2}\mathrm{g}}^\mathrm{M})+n(E_{\frac{1}{2}\mathrm{g}}^\mathrm{Z})+n(E_{\frac{1}{2}\mathrm{g}}^\mathrm{A})+n(E_{\frac{1}{2}\mathrm{g}}^\mathrm{X})+n(E_{\frac{1}{2}\mathrm{g}}^\mathrm{R})$\\
\cline{3-4} 
 &  & $n_{\frac{1}{2}}^{-}$ & $n(E_{\frac{1}{2}\mathrm{u}}^{\Gamma})+n(E_{\frac{1}{2}\mathrm{u}}^\mathrm{M})+n(E_{\frac{1}{2}\mathrm{u}}^\mathrm{Z})+n(E_{\frac{1}{2}\mathrm{u}}^\mathrm{A})+n(E_{\frac{1}{2}\mathrm{u}}^\mathrm{X})+n(E_{\frac{1}{2}\mathrm{u}}^\mathrm{R})$\\
\cline{3-4} 
 &  & $n_{\frac{3}{2}}^{+}$ & $n(E_{\frac{3}{2}\mathrm{g}}^{\Gamma})+n(E_{\frac{3}{2}\mathrm{g}}^\mathrm{M})+n(E_{\frac{3}{2}\mathrm{g}}^\mathrm{Z})+n(E_{\frac{3}{2}\mathrm{g}}^\mathrm{A})+n(E_{\frac{1}{2}\mathrm{g}}^\mathrm{X})+n(E_{\frac{1}{2}\mathrm{g}}^\mathrm{R})$\\
\cline{3-4} 
 &  & $n_{\frac{3}{2}}^{-}$ & $n(E_{\frac{3}{2}\mathrm{u}}^{\Gamma})+n(E_{\frac{3}{2}\mathrm{u}}^\mathrm{M})+n(E_{\frac{3}{2}\mathrm{u}}^\mathrm{Z})+n(E_{\frac{3}{2}\mathrm{u}}^\mathrm{A})+n(E_{\frac{1}{2}\mathrm{u}}^\mathrm{X})+n(E_{\frac{1}{2}\mathrm{u}}^\mathrm{R})$\\
\hline 
\multirow{4}{2cm}{\centering Tetragonal body-centred} & \multirow{4}{*}{87 (139, 140)\footnotemark[1]} & $n_{\frac{1}{2}}^{+}$ & $n(E_{\frac{1}{2}\mathrm{g}}^{\Gamma})+n(E_{\frac{1}{2}\mathrm{g}}^\mathrm{M})+n(E_{\frac{1}{2}\mathrm{g}}^\mathrm{X})+2n(E_{\frac{1}{2}\mathrm{g}}^\mathrm{N})+n(E_{\frac{1}{2}}^\mathrm{P})$\footnotemark[2]\\
\cline{3-4} 
 &  & $n_{\frac{1}{2}}^{-}$ & $n(E_{\frac{1}{2}\mathrm{u}}^{\Gamma})+n(E_{\frac{1}{2}\mathrm{u}}^\mathrm{M})+n(E_{\frac{1}{2}\mathrm{u}}^\mathrm{X})+2n(E_{\frac{1}{2}\mathrm{u}}^\mathrm{N})+n(E_{\frac{3}{2}}^\mathrm{P})$\\
\cline{3-4} 
 &  & $n_{\frac{3}{2}}^{+}$ & $n(E_{\frac{3}{2}\mathrm{g}}^{\Gamma})+n(E_{\frac{3}{2}\mathrm{g}}^\mathrm{M})+n(E_{\frac{1}{2}\mathrm{g}}^\mathrm{X})+2n(E_{\frac{1}{2}\mathrm{g}}^\mathrm{N})+n(E_{\frac{3}{2}}^\mathrm{P})$\\
\cline{3-4} 
 &  & $n_{\frac{3}{2}}^{-}$ & $n(E_{\frac{3}{2}\mathrm{u}}^{\Gamma})+n(E_{\frac{3}{2}\mathrm{u}}^\mathrm{M})+n(E_{\frac{1}{2}\mathrm{u}}^\mathrm{X})+2n(E_{\frac{1}{2}\mathrm{u}}^\mathrm{N})+n(E_{\frac{1}{2}}^\mathrm{P})$\\
\hline 
\multirow{4}{2cm}{\centering Cubic primitive} & \multirow{4}{*}{221} & $n_{\frac{1}{2}}^{+}$ & $n(E_{\frac{1}{2}\mathrm{g}}^{\Gamma})+n(F_{\frac{3}{2}\mathrm{g}}^{\Gamma})+n(E_{\frac{1}{2}\mathrm{g}}^\mathrm{R})+n(F_{\frac{3}{2}\mathrm{g}}^\mathrm{R})+2n(E_{\frac{1}{2}\mathrm{g}}^\mathrm{M})+n(E_{\frac{3}{2}\mathrm{g}}^\mathrm{M})+2n(E_{\frac{1}{2}\mathrm{g}}^\mathrm{X})+n(E_{\frac{3}{2}\mathrm{g}}^\mathrm{X})$\\
\cline{3-4} 
 &  & $n_{\frac{1}{2}}^{-}$ & $n(E_{\frac{1}{2}\mathrm{u}}^{\Gamma})+n(F_{\frac{3}{2}\mathrm{u}}^{\Gamma})+n(E_{\frac{1}{2}\mathrm{u}}^\mathrm{R})+n(F_{\frac{3}{2}\mathrm{u}}^\mathrm{R})+2n(E_{\frac{1}{2}\mathrm{u}}^\mathrm{M})+n(E_{\frac{3}{2}\mathrm{u}}^\mathrm{M})+2n(E_{\frac{1}{2}\mathrm{u}}^\mathrm{X})+n(E_{\frac{3}{2}\mathrm{u}}^\mathrm{X})$\\
\cline{3-4} 
 &  & $n_{\frac{3}{2}}^{+}$ & $n(F_{\frac{3}{2}\mathrm{g}}^{\Gamma})+n(E_{\frac{5}{2}\mathrm{g}}^{\Gamma})+n(F_{\frac{3}{2}\mathrm{g}}^\mathrm{R})+n(E_{\frac{5}{2}\mathrm{g}}^\mathrm{R})+2n(E_{\frac{3}{2}\mathrm{g}}^\mathrm{M})+n(E_{\frac{1}{2}\mathrm{g}}^\mathrm{M})+2n(E_{\frac{3}{2}\mathrm{g}}^\mathrm{X})+n(E_{\frac{1}{2}\mathrm{g}}^\mathrm{X})$\\
\cline{3-4} 
 &  & $n_{\frac{3}{2}}^{-}$ & $n(F_{\frac{3}{2}\mathrm{u}}^{\Gamma})+n(E_{\frac{5}{2}\mathrm{u}}^{\Gamma})+n(F_{\frac{3}{2}\mathrm{u}}^\mathrm{R})+n(E_{\frac{5}{2}\mathrm{u}}^\mathrm{R})+2n(E_{\frac{3}{2}\mathrm{u}}^\mathrm{M})+n(E_{\frac{1}{2}\mathrm{u}}^\mathrm{M})+2n(E_{\frac{3}{2}\mathrm{u}}^\mathrm{X})+n(E_{\frac{1}{2}\mathrm{u}}^\mathrm{X})$\\
\hline 
\multirow{8}{2cm}{\centering Cubic face-centred} & \multirow{4}{*}{225} & $n_{\frac{1}{2}}^{+}$ & $n(E_{\frac{1}{2}\mathrm{g}}^{\Gamma})+n(F_{\frac{3}{2}\mathrm{g}}^{\Gamma})+2n(E_{\frac{1}{2}\mathrm{g}}^\mathrm{X})+n(E_{\frac{3}{2}\mathrm{g}}^\mathrm{X})+2n(E_{\frac{1}{2}\mathrm{g}}^\mathrm{L})+2n(E_{\frac{3}{2}\mathrm{g}}^\mathrm{L})+n(E_{\frac{1}{2}}^\mathrm{W})$\\
\cline{3-4} 
 &  & $n_{\frac{1}{2}}^{-}$ & $n(E_{\frac{1}{2}\mathrm{u}}^{\Gamma})+n(F_{\frac{3}{2}\mathrm{u}}^{\Gamma})+2n(E_{\frac{1}{2}\mathrm{u}}^\mathrm{X})+n(E_{\frac{3}{2}\mathrm{u}}^\mathrm{X})+2n(E_{\frac{1}{2}\mathrm{u}}^\mathrm{L})+2n(E_{\frac{3}{2}\mathrm{u}}^\mathrm{L})+n(E_{\frac{3}{2}}^\mathrm{W})$\\
\cline{3-4} 
 &  & $n_{\frac{3}{2}}^{+}$ & $n(F_{\frac{3}{2}\mathrm{g}}^{\Gamma})+n(E_{\frac{5}{2}\mathrm{g}}^{\Gamma})+2n(E_{\frac{3}{2}\mathrm{g}}^\mathrm{X})+n(E_{\frac{1}{2}\mathrm{g}}^\mathrm{X})+2n(E_{\frac{1}{2}\mathrm{g}}^\mathrm{L})+2n(E_{\frac{3}{2}\mathrm{g}}^\mathrm{L})+n(E_{\frac{3}{2}}^\mathrm{W})$\\
\cline{3-4} 
 &  & $n_{\frac{3}{2}}^{-}$ & $n(F_{\frac{3}{2}\mathrm{u}}^{\Gamma})+n(E_{\frac{5}{2}\mathrm{u}}^{\Gamma})+2n(E_{\frac{3}{2}\mathrm{u}}^\mathrm{X})+n(E_{\frac{1}{2}\mathrm{u}}^\mathrm{X})+2n(E_{\frac{1}{2}\mathrm{u}}^\mathrm{L})+2n(E_{\frac{3}{2}\mathrm{u}}^\mathrm{L})+n(E_{\frac{1}{2}}^\mathrm{W})$\\
\cline{2-4} 
 & \multirow{4}{*}{226} & $n_{\frac{1}{2}}^{+}$ & $n(E_{\frac{1}{2}\mathrm{g}}^{\Gamma})+n(F_{\frac{3}{2}\mathrm{g}}^{\Gamma})+n(E_{\frac{3}{2}\mathrm{g}}^\mathrm{X})+n(E_{\frac{1}{2}\mathrm{u}}^\mathrm{X})+n(E_{\frac{3}{2}\mathrm{u}}^\mathrm{X})$\\
\cline{3-4} 
 &  & $n_{\frac{1}{2}}^{-}$ & $n(E_{\frac{1}{2}\mathrm{u}}^{\Gamma})+n(F_{\frac{3}{2}\mathrm{u}}^{\Gamma})+n(E_{\frac{3}{2}\mathrm{u}}^\mathrm{X})+n(E_{\frac{1}{2}\mathrm{g}}^\mathrm{X})+n(E_{\frac{3}{2}\mathrm{g}}^\mathrm{X})$\\
\cline{3-4} 
 &  & $n_{\frac{3}{2}}^{+}$ & $n(F_{\frac{3}{2}\mathrm{g}}^{\Gamma})+n(E_{\frac{5}{2}\mathrm{g}}^{\Gamma})+n(E_{\frac{1}{2}\mathrm{g}}^\mathrm{X})+n(E_{\frac{1}{2}\mathrm{u}}^\mathrm{X})+n(E_{\frac{3}{2}\mathrm{u}}^\mathrm{X})$\\
\cline{3-4} 
 &  & $n_{\frac{3}{2}}^{-}$ & $n(F_{\frac{3}{2}\mathrm{u}}^{\Gamma})+n(E_{\frac{5}{2}\mathrm{u}}^{\Gamma})+n(E_{\frac{1}{2}\mathrm{u}}^\mathrm{X})+n(E_{\frac{1}{2}\mathrm{g}}^\mathrm{X})+n(E_{\frac{3}{2}\mathrm{g}}^\mathrm{X})$\\
\hline 
\multirow{4}{2cm}{\centering Cubic body-centred} & \multirow{4}{*}{229} & $n_{\frac{1}{2}}^{+}$ & $n(E_{\frac{1}{2}\mathrm{g}}^{\Gamma})+n(F_{\frac{3}{2}\mathrm{g}}^{\Gamma})+n(E_{\frac{1}{2}\mathrm{g}}^\mathrm{H})+n(F_{\frac{3}{2}\mathrm{g}}^\mathrm{H})+3n(E_{\frac{1}{2}\mathrm{g}}^\mathrm{N})+n(E_{\frac{1}{2}}^\mathrm{P})+n(F_{\frac{3}{2}}^\mathrm{P})$\\
\cline{3-4} 
 &  & $n_{\frac{1}{2}}^{-}$ & $n(E_{\frac{1}{2}\mathrm{u}}^{\Gamma})+n(F_{\frac{3}{2}\mathrm{u}}^{\Gamma})+n(E_{\frac{1}{2}\mathrm{u}}^\mathrm{H})+n(F_{\frac{3}{2}\mathrm{u}}^\mathrm{H})+3n(E_{\frac{1}{2}\mathrm{u}}^\mathrm{N})+n(F_{\frac{3}{2}}^\mathrm{P})+n(E_{\frac{5}{2}}^\mathrm{P})$\\
\cline{3-4} 
 &  & $n_{\frac{3}{2}}^{+}$ & $n(F_{\frac{3}{2}\mathrm{g}}^{\Gamma})+n(E_{\frac{5}{2}\mathrm{g}}^{\Gamma})+n(F_{\frac{3}{2}\mathrm{g}}^\mathrm{H})+n(E_{\frac{5}{2}\mathrm{g}}^\mathrm{H})+3n(E_{\frac{1}{2}\mathrm{g}}^\mathrm{N})+n(F_{\frac{3}{2}}^\mathrm{P})+n(E_{\frac{5}{2}}^\mathrm{P})$\\
\cline{3-4} 
 &  & $n_{\frac{3}{2}}^{-}$ & $n(F_{\frac{3}{2}\mathrm{u}}^{\Gamma})+n(E_{\frac{5}{2}\mathrm{u}}^{\Gamma})+n(F_{\frac{3}{2}\mathrm{u}}^\mathrm{H})+n(E_{\frac{5}{2}\mathrm{u}}^\mathrm{H})+3n(E_{\frac{1}{2}\mathrm{u}}^\mathrm{N})+n(E_{\frac{1}{2}}^\mathrm{P})+n(F_{\frac{3}{2}}^\mathrm{P})$\\
\hline 
\end{tabular}
\par\end{centering}

\footnotetext[1]{The equations here are derived for SG \#83 (\#87) but also applicable to the SGs in the bracket,
 which are supergroups of SG \#83 (\#87).
To apply these equations for these supergroups, one should omit the additional symmetries and count them as SG \#83 (\#87).}

\footnotetext[2]{In SG \#87, the little group at $N$ is $C_i$ and the irrep notations 
 in \onlinecite{point-group_1994} is $A_{\frac{1}{2}\mathrm{g}}$ and  $A_{\frac{1}{2}\mathrm{u}}$, 
 both of which are one dimensional. 
However, due to the Kramer's theorem, 
 the irreps at $N$ should be double degenerate, 
 thus we adopt the two dimensional notations $E_{\frac{1}{2}\mathrm{g}}$ and  $E_{\frac{1}{2}\mathrm{u}}$.}

\raggedright{}\protect\caption{\label{tab:z8}The concrete expressions for $n_{\frac{3}{2}}^{+}$,
$n_{\frac{3}{2}}^{-}$, $n_{\frac{1}{2}}^{+}$, $n_{\frac{1}{2}}^{-}$
in the $z_{8}$ Fu-Kane-like formulae in all applicable SGs. The notations
of high symmetry momenta follow the standard convention \cite{aroyo_brillouin-zone_2014},
and the notations of point group irreps follow \onlinecite{point-group_1994}.}
\end{table*}
\renewcommand\arraystretch{1.0}

\textit{$P\bar{6}$ formulae.}
SI of SG \#174 ($P\bar{6}$) form the group $\mathbb{Z}_{3}\times\mathbb{Z}_{3}$.
This SG has a horizontal mirror and a vertical $C_3$-rotation.
The $\mathbb{Z}_{3}$ indicators are
 the mirror Chern numbers (modulo 3) of the $k_{z}=0$ and the $k_{z}=\pi$
 planes. 
This two indicators, denoted as $z_{3\mathrm{m},0}$ and $z_{3\mathrm{m},\pi}$, 
 can be derived by applying the Chern number formula for $C_{3}$-invariant
 insulator \cite{Fang2012} in each mirror eigenvalue sector. 
In \cref{tab:formula} we give the concrete expressions and their
 generalizations in other SGs with $C_{3}$ and mirror symmetries.

\textit{$P6/m$ formulae.}
SI of SG \#175 ($P6/m$) form the group $\mathbb{Z}_{6}\times\mathbb{Z}_{12}$.
As this SG has a vertical $C_6$-rotation symmetry, a horizontal
mirror symmetry, and an inversion symmetry, we can at least calculate
two mirror Chern numbers (modulo 6) in the $k_{z}=0$ and $k_{z}=\pi$
planes, denoted as $z_{6\mathrm{m},0}$ and $z_{6\mathrm{m},\pi}$,
by applying the Chern number formula \cite{Fang2012}
in each mirror eigenvalue sector, and a $z_{4}$ indicator,
as tabulated in \cref{tab:formula}. 
It seems that the SI group should be $\mathbb{Z}_6\times\mathbb{Z}_6\times\mathbb{Z}_4$.
However, the three indicators are not independent---$z_{6\mathrm{m},0}+z_{6\mathrm{m},\pi}$ and $z_4$ must have
 same parity because odd values of both indicate a strong TI, leading to the $\mathbb{Z}_6\times\mathbb{Z}_{12}$ group.
Choosing proper BR bases, we find that the $\mathbb{Z}_{6}$ generator corresponds to $z_{6\mathrm{m},0}=5$,
 $z_{6\mathrm{m},\pi}=1$, $z_{4}=0$, and the $\mathbb{Z}_{12}$ generator
 corresponds to $z_{6\mathrm{m},0}=1$, $z_{6\mathrm{m},\pi}=0$, $z_{4}=1$. 
In this convention, the relations between the $\mathbb{Z}_6$, $\mathbb{Z}_{12}$ indicators,
 denoted as $z_6$ and $z_{12}$ respectively,
 and the $z_{6\mathrm{m},0}$, $z_{6\mathrm{m},\pi}$,  $z_{4}$ indicators can be derived as
\begin{equation}
    z_{6\mathrm{m},0} = -z_{6} + z_{12} \mod 6
\end{equation}
\begin{equation}
    z_{6\mathrm{m},\pi} = z_{6} \mod 6
\end{equation}
\begin{equation}
    z_{4} = z_{12} \mod 4
\end{equation}
Therefore, the $z_6$ indicator is directly given by $z_{6\mathrm{m},\pi}$, and the $z_{12}$ indicator 
 is determined by the following two equations
\begin{align}
    z_{12} \mod 6 &= z_{6\mathrm{m},0} + z_{6\mathrm{m},\pi}  \\
    z_{12} \mod 4 &= z_{4}
\end{align}
%In table
%\ref{tab:z12} we list the relation between the $z_{6\mathrm{m},0}+z_{6\mathrm{m},\pi}$,
%$z_{4i}$ and the $\mathbb{Z}_{12}$ factor, 
which is equivalent to
\begin{equation}
z_{12} =\big\{\bar{z}_{6\mathrm{m}}+3\big[\left(\bar{z}_{6\mathrm{m}}-z_{4}\right)\ \mathrm{mod}\ 4\big]\big\}\mod12\label{eq:z12}
\end{equation}
where 
\begin{equation}
    \bar{z}_{6\mathrm{m}} = z_{6\mathrm{m},0} + z_{6\mathrm{m},\pi} \mod 6
\end{equation}

\renewcommand\arraystretch{1.6}
\begin{table*}
\begin{centering}
\begin{tabular}{|c|>{\centering}p{4.5cm}|p{10cm}<{\centering}|}
\hline 
Indicator & SGs &  Formula\\
\hline 
\hline 
$z_{2\mathrm{w},j=1,2,3}$ & All SGs with inversion & $\sum_{\mathbf{K}}^{\prime}n_{\mathbf{K}}^{-}\mod2$ \footnotemark[1] \\
\hline 
\textcolor{red}{$z_{4}$} & All SGs with inversion & $\sum_{\mathbf{K}\in\mathrm{TRIM}}\frac{1}{2}n_{\mathbf{K}}^{-}-\frac{1}{2}n_{\mathbf{K}}^{+}\mod4$\\
\hline 
\textcolor{red}{$z_{2}$} & All SGs with $S_{4}$ & $\sum_{\mathbf{K}}\frac{1}{2}n_{\mathbf{K}}^{\frac{3}{2}}-\frac{1}{2}n_{\mathbf{K}}^{\frac{1}{2}}\mod2$ \footnotemark[2]\\
\hline 
\multirow{5}{*}{$z_{4\mathrm{m},\pi}$} & \multirow{2}{*}{83 (123, 127)\footnotemark[3]} & $\frac{3}{2}n(E_{\frac{3}{2}\mathrm{g}}^\mathrm{Z})-\frac{3}{2}n(E_{\frac{3}{2}\mathrm{u}}^\mathrm{Z})-\frac{1}{2}n(E_{\frac{1}{2}\mathrm{g}}^\mathrm{Z})+\frac{1}{2}n(E_{\frac{1}{2}\mathrm{u}}^\mathrm{Z})+\frac{3}{2}n(E_{\frac{3}{2}\mathrm{g}}^\mathrm{A})-\frac{3}{2}n(E_{\frac{3}{2}\mathrm{u}}^\mathrm{A})$ \\
& & $-\frac{1}{2}n(E_{\frac{1}{2}\mathrm{g}}^\mathrm{A})+\frac{1}{2}n(E_{\frac{1}{2}\mathrm{u}}^\mathrm{A})+n(E_{\frac{1}{2}\mathrm{g}}^\mathrm{R})-n(E_{\frac{1}{2}\mathrm{u}}^\mathrm{R})\mod4$\\
\cline{2-3} 
 & \multirow{3}{*}{221} & $\frac{3}{2}n(E_{\frac{3}{2}\mathrm{g}}^\mathrm{X})-\frac{3}{2}n(E_{\frac{3}{2}\mathrm{u}}^\mathrm{X})-\frac{1}{2}n(E_{\frac{1}{2}\mathrm{g}}^\mathrm{X})+\frac{1}{2}n(E_{\frac{1}{2}\mathrm{u}}^\mathrm{X})$ \\
& & $+\frac{3}{2}n(E_{\frac{5}{2}\mathrm{g}}^\mathrm{R})-\frac{3}{2}n(E_{\frac{5}{2}\mathrm{u}}^\mathrm{R})+n(F_{\frac{3}{2}\mathrm{g}}^\mathrm{R})-n(F_{\frac{3}{2}\mathrm{u}}^\mathrm{R})-\frac{1}{2}n(E_{\frac{1}{2}\mathrm{g}}^\mathrm{R})+\frac{1}{2}n(E_{\frac{1}{2}\mathrm{u}}^\mathrm{R})$ \\
& & $+n(E_{\frac{1}{2}\mathrm{g}}^\mathrm{M})+n(E_{\frac{3}{2}\mathrm{g}}^\mathrm{M})-n(E_{\frac{1}{2}\mathrm{u}}^\mathrm{M})-n(E_{\frac{3}{2}\mathrm{u}}^\mathrm{M})\mod4$\\
\hline 
\textcolor{red}{$z_{8}$} & 83, 87, 123, 124, 127, 128, 139, 140, 221, 225, 226, 229 & $\frac{3}{2}n_{\frac{3}{2}}^{+}-\frac{3}{2}n_{\frac{3}{2}}^{-}-\frac{1}{2}n_{\frac{1}{2}}^{+}+\frac{1}{2}n_{\frac{1}{2}}^{-}\mod8$ \footnotemark[4]\\
\hline 
\multirow{2}{*}{$z_{3\mathrm{m},0}$} & \multirow{2}{*}{174 (187, 188, 189, 190)\footnotemark[3]} & 
 $-\frac{1}{2}n(^1E^K_{\frac{1}{2}}) +\frac{3}{2}n(^1E^K_\frac{3}{2}) +\frac{1}{2}n(^1E^K_\frac{5}{2})
 -\frac{1}{2}n(^2E^K_{\frac{1}{2}}) +\frac{3}{2}n(^2E^K_\frac{3}{2}) +\frac{1}{2}n(^2E^K_\frac{5}{2})$ \\
 & & $+n(E_{\frac{1}{2}}^{\Gamma})-n(E_{\frac{5}{2}}^{\Gamma}) \mod 3$ \\
\hline 
\multirow{2}{*}{$z_{3\mathrm{m},\pi}$} & \multirow{2}{*}{174 (187, 189) \footnotemark[3]} & 
 $-\frac{1}{2}n(^1E^\mathrm{H}_{\frac{1}{2}}) +\frac{3}{2}n(^1E^\mathrm{H}_\frac{3}{2}) +\frac{1}{2}n(^1E^\mathrm{H}_\frac{5}{2})
 -\frac{1}{2}n(^2E^\mathrm{H}_{\frac{1}{2}}) +\frac{3}{2}n(^2E^\mathrm{H}_\frac{3}{2}) +\frac{1}{2}n(^2E^\mathrm{H}_\frac{5}{2})$ \\
& & $+n(E_{\frac{1}{2}}^\mathrm{A})-n(E_{\frac{5}{2}}^\mathrm{A}) \mod 3$ \\
\hline 
\multirow{2}{*}{$z_{6\mathrm{m},0}$} & \multirow{2}{*}{175 (191, 192), 176 (193, 194)\footnotemark[3]} & 
$\frac{3}{2}n(E_{\frac{3}{2}\mathrm{g}}^{\Gamma})-\frac{5}{2}n(E_{\frac{5}{2}\mathrm{g}}^{\Gamma})-\frac{1}{2}n(E_{\frac{1}{2}\mathrm{g}}^{\Gamma})-\frac{3}{2}n(E_{\frac{3}{2}\mathrm{u}}^{\Gamma})+\frac{5}{2}n(E_{\frac{5}{2}\mathrm{u}}^{\Gamma})+\frac{1}{2}n(E_{\frac{1}{2}\mathrm{u}}^{\Gamma})$ \\
& & $+3n(E_{\frac{3}{2}}^{K})-5n(E_{\frac{5}{2}}^{K})-n(E_{\frac{1}{2}}^{K})+\frac{3}{2}n(E_{\frac{1}{2}\mathrm{g}}^\mathrm{M})-\frac{3}{2}n(E_{\frac{1}{2}\mathrm{u}}^\mathrm{M})\mod6$\\
\hline 
\multirow{2}{*}{$z_{6\mathrm{m},\pi}$} & \multirow{2}{*}{175 (191, 192) \footnotemark[3]} & 
$\frac{3}{2}n(E_{\frac{3}{2}\mathrm{g}}^\mathrm{A})-\frac{5}{2}n(E_{\frac{5}{2}\mathrm{g}}^\mathrm{A})-\frac{1}{2}n(E_{\frac{1}{2}\mathrm{g}}^\mathrm{A})-\frac{3}{2}n(E_{\frac{3}{2}\mathrm{u}}^\mathrm{A})+\frac{5}{2}n(E_{\frac{5}{2}\mathrm{u}}^\mathrm{A})+\frac{1}{2}n(E_{\frac{1}{2}\mathrm{u}}^\mathrm{A})$ \\
& & $+3n(E_{\frac{3}{2}}^\mathrm{H})-5n(E_{\frac{5}{2}}^\mathrm{H})-n(E_{\frac{1}{2}}^\mathrm{H})+\frac{3}{2}n(E_{\frac{1}{2}\mathrm{g}}^\mathrm{L})-\frac{3}{2}n(E_{\frac{1}{2}\mathrm{u}}^\mathrm{L})\mod6$\\
\hline 
\textcolor{red}{$z_{12}$} & 175, 191, 192 & $\big\{ \bar{z}_{6\mathrm{m}} +3\big[\left( \bar{z}_{6\mathrm{m}}-z_{4}\right)\ \mathrm{mod}\ 4\big]\big\} \mod12$ \footnotemark[5] \\
\hline
\textcolor{red}{$z_{12}^\prime$} & 176, 193, 194 & $\big\{ z_{6\mathrm{m},0}+3\big[\left(z_{6\mathrm{m},0}-z_{4}\right)\ \mathrm{mod}\ 4\big]\big\} \mod12$\\
\hline 
\end{tabular}
\par\end{centering}

\footnotetext[1]{$\mathbf{K}$ is summed over the four TRIMs with $k_j=\pi$.}

\footnotetext[2]{$\mathbf{K}$ is summed over the four $S_4$ invariant TRIMs,  $n_{\mathbf{K}}^{\frac{1}{2}}$ is the number of Kramer pairs at $\mathbf{K}$ with $\mathrm{tr}\left[D\left(S_{4}\right)\right]=\sqrt{2}$,  $n_{\mathbf{K}}^{\frac{3}{2}}$ is the number of Kramer pairs at $\mathbf{K}$ with $\mathrm{tr}\left[D\left(S_{4}\right)\right]=-\sqrt{2}$, and $D(S_4)$ is the  representation matrix on the corresponding Kramer pair.}

\footnotetext[3]{The equation is derived for the SG in front of the bracket but also applicable to the SGs in the bracket,
 which are supergroups of the SG in front of the bracket.
To apply the equation for these supergroups, one should omit the additional symmetries and count them as the corresponding subgroup.}

\footnotetext[4]{The concrete definitions for $n_{\frac{3}{2}}^{+}$, $n_{\frac{3}{2}}^{-}$, $n_{\frac{1}{2}}^{+}$, and $n_{\frac{1}{2}}^{-}$ is given in \cref{tab:z8}.}
\footnotetext[5]{Here $\bar{z}_{6\mathrm{m}} = z_{6\mathrm{m},0} + z_{6\mathrm{m},\pi} \mod 6$ }

\protect\caption{\label{tab:formula}Fu-Kane-like formulae for all SI.
The indicators whose odd values correspond to strong TI are printed in red.
The notations of high symmetry momenta follow the standard convention \cite{aroyo_brillouin-zone_2014},
and the notations of point group irreps follow \onlinecite{point-group_1994}.}
\end{table*}
\renewcommand\arraystretch{1.0}

%\begin{table}
%\begin{centering}
%\begin{tabular}{|>{\centering}p{2cm}|c|c|c|c|c|}
%\hline 
%\multicolumn{3}{|c|}{175, 191, 192} & \multicolumn{3}{c|}{176, 193, 194}\\
%\hline 
%\hline 
%$z_{6\mathrm{m},0}+z_{6\mathrm{m},\pi}$ mod 6 & $z_{4i}$ & $z_{12}$ & $z_{3\mathrm{m},0}$ & $z_{4i}$ & $z_{12}^{\prime}$\\
%\hline 
%0 & 0 & 0 & 0 & 0 & 0\\
%\hline 
%1 & 1 & 1 & 1 & 1 & 1\\
%\hline 
%2 & 2 & 2 & 2 & 2 & 2\\
%\hline 
%3 & 3 & 3 & 0 & 3 & 3\\
%\hline 
%4 & 0 & 4 & 1 & 0 & 4\\
%\hline 
%5 & 1 & 5 & 2 & 1 & 5\\
%\hline 
%0 & 2 & 6 & 0 & 2 & 6\\
%\hline 
%1 & 3 & 7 & 1 & 3 & 7\\
%\hline 
%2 & 0 & 8 & 2 & 0 & 8\\
%\hline 
%3 & 1 & 9 & 0 & 1 & 9\\
%\hline 
%4 & 2 & 10 & 1 & 2 & 10\\
%\hline 
%5 & 3 & 11 & 2 & 3 & 11\\
%\hline 
%\end{tabular}
%\par\end{centering}

%\protect\caption{\label{tab:z12}The relations between mirror Chern numbers (modulo
%3 or 6), inversion index $z_{4i}$, and the $\mathbb{Z}_{12}$ factors
%$z_{12}$ in SGs 175, 176, 191, 192, 193, 194.}
%
%
%\end{table}

\textit{$P6_{3}/m$ formulae.}
SI of SG \#176 ($P6_{3}/m$) form the group $\mathbb{Z}_{12}$.
Due to the horizontal mirror symmetry a mirror Chern number for the $k_z=0$ plane can be defined,
 and, its value (modulo 6) can be calculated from symmetry eigenvalues at high symmetry momenta in the $k_z=0$ plane,
 where the $C_6$-screw is equivalent with $C_6$-rotation,
 by applying the Chern number formula for $C_6$-invariant insulator \cite{Fang2012}
(The mirror Chern number for $k_z=\pi$ plane always, however, equals to zero due to the $C_6$-screw symmetry.)
Here we denote this the mirror Chern number (modulo 6) as indicator $z_{6\mathrm{m},0}$.
On the other hand, due to the inversion symmetry the $z_4$ indicator can be also defined.
Similar with indicators in SG $P6/m$, $z_{6\mathrm{m},0}$ and $z_{4}$ have same parity, 
 leading to the SI group $\mathbb{Z}_{12}$.
Therefore, the $\mathbb{Z}_{12}$ indicator can be calculated as
\begin{equation}
z_{12}^{\prime}=\left\{ z_{6\mathrm{m},0}+3\left[\left(z_{6\mathrm{m},0}-z_{4}\right)\ \mathrm{mod}\ 4\right]\right\} \mod12\label{eq:z12p}
\end{equation}
Here we use the prime notation to distinguish it from the $\mathbb{Z}_{12}$
 indicator in SG $P6/m$ (\cref{eq:z12}).

\begin{table*}
\begin{centering}
\begin{tabular}{|c|c|p{8cm}<{\centering}|}
\hline 
SI set & SI group & SGs\\
\hline 
\hline 
\multirow{4}{*}{$z_{2\mathrm{w},1}$ $z_{2\mathrm{w},2}$ $z_{2\mathrm{w},3}$ \textcolor{red}{$z_{4}$} } & $\mathbb{Z}_{2}\times\mathbb{Z}_{2}\times\mathbb{Z}_{2}\times\mathbb{Z}_{4}$ & 2, 10, 47\\
\cline{2-3} 
 & $\mathbb{Z}_{2}\times\mathbb{Z}_{2}\times\mathbb{Z}_{4}$ & 11, 12, 13, 49, 51, 65, 67, 69\\
\cline{2-3} 
 & $\mathbb{Z}_{2}\times\mathbb{Z}_{4}$ & 14, 15, 48, 50, 53, 54, 55, 57, 59, 63, 64, 66, 68, 71, 72, 73, 74,
84, 85, 86, 125, 129, 131, 132, 134, 147, 148, 162, 164, 166, 200,
201, 204, 206, 224\\
\cline{2-3} 
 & $\mathbb{Z}_{4}$ & 52, 56, 58, 60, 61, 62, 70, 88, 126, 130, 133, 135, 136, 137, 138,
141, 142, 163, 165, 167, 202, 203, 205, 222, 223, 227, 228, 230\\
\hline 
\textcolor{red}{$z_{2}$} & $\mathbb{Z}_{2}$ & 81, 82, 111, 112, 113, 114, 115, 116, 117, 118, 119, 120, 121, 122,
215, 216, 217, 218, 219, 220\\
\hline 
$z_{2\mathrm{w},1}$ $z_{4\mathrm{m},\pi}$ \textcolor{red}{$z_{8}$}  & $\mathbb{Z}_{2}\times\mathbb{Z}_{4}\times\mathbb{Z}_{8}$ & 83, 123\\
\hline 
$z_{2\mathrm{w},1}$ \textcolor{red}{$z_{8}$}  & $\mathbb{Z}_{2}\times\mathbb{Z}_{8}$ & 87, 124, 139, 140, 229\\
\hline 
$z_{4\mathrm{m},\pi}$ \textcolor{red}{$z_{8}$}  & $\mathbb{Z}_{4}\times\mathbb{Z}_{8}$ & 127, 221\\
\hline 
\textcolor{red}{$z_{8}$}  & $\mathbb{Z}_{8}$ & 128, 225, 226\\
\hline 
$z_{3\mathrm{m},0}$ $z_{3\mathrm{m},\pi}$ & $\mathbb{Z}_{3}\times\mathbb{Z}_{3}$ & 174, 187, 189\\
\hline 
$z_{3\mathrm{m},0}$ & $\mathbb{Z}_{3}$ & 188, 190\\
\hline 
$z_{6\mathrm{m},\pi}$ \textcolor{red}{$z_{12}$}  & $\mathbb{Z}_{6}\times\mathbb{Z}_{12}$ & 175, 191\\
\hline 
\textcolor{red}{$z_{12}$} & $\mathbb{Z}_{12}$ & 192\\
\hline 
\textcolor{red}{$z_{12}^\prime$} & $\mathbb{Z}_{12}$ & 176, 193, 194\\
\hline 
\end{tabular}
\par\end{centering}

\protect\caption{SI in all SGs.
The indicators whose odd values correspond to strong TI are printed in red.
\label{tab:Indicator}}
\end{table*}

\subsection{The $S_{4}$ Fu-Kane-like formula}

The SI of SG \#81 ($P\bar{4}$) form the group $\mathbb{Z}_{2}$,
 and we find that the $\mathbb{Z}_2$ indicator can be calculated as
\begin{equation}
z_{2}=\sum_{\mathbf{K}}\frac{n_{\mathbf{K}}^{\frac{3}{2}}-n_{\mathbf{K}}^{\frac{1}{2}}}{2}\mod2\label{eq:z2s}
\end{equation}
where $\mathbf{K}$ is summed over all the four $S_{4}$-invariant
TRIMs, $n_{\mathbf{K}}^{\frac{3}{2}}$ is the number of Kramer pairs
at $\mathbf{K}$ with $\mathrm{tr}\left[D\left(S_{4}\right)\right]=-\sqrt{2}$,
$n_{\mathbf{K}}^{\frac{1}{2}}$ is the number of Kramer pairs at $\mathbf{K}$
with $\mathrm{tr}\left[D\left(S_{4}\right)\right]=\sqrt{2}$, and
$D\left(S_{4}\right)$ is the $S_{4}$ representation matrix on the
corresponding Kramer pair. 
%This formula is also induced to other SGs
% with $S_{4}$ symmetry, as shown in table \ref{tab:Indicator}. 

The $S_{4}$ indicator $z_{2}$ has nothing to do with the $S_{4}$-invariant $\delta_{S_4}$ defined in \cref{sec:Layer}. 
By the models following, we find that $z_{2}=1$ corresponds to a strong TI or a Weyl semimetal.
Since for $z_{2}=1$ there are two topologically distinct phases, i.e.,  $\delta_{S_4}=0$ and $\delta_{S_4}=1$ phases,
 we conjecture that the Weyl semimetal phase is the intermediate state between them.
Correspondingly, $z_{2}=0$ corresponds to an insulator with $\delta_{t}=0$ and $\delta_{S_4}=0$ or $1$, or a Weyl semimetal.

\textit{Strong TI.}
Consider the model
\begin{align}
\hat{H}(\mathbf{k}) &= \left[\Delta-\sum_{i}\cos k_i\right]\tau_z\sigma_0 + \sum_i\sin k_i \tau_x \sigma_i
\end{align}
where $\tau_{x,y,z}$ and $\sigma_{x,y,z}$ are Pauli matrices,
 and $\sigma_{0}$ is two by two identity matrix,
 the bases is set as $|\frac{1}{2}\rangle$, $|\bar{\frac{1}{2}}\rangle$, $|\frac{3}{2}\rangle$, $|\bar{\frac{3}{2}}\rangle$,
 i.e., the bases of $E_\frac{1}{2}$ and $E_\frac{3}{2}$ irreps.
This model has rotoreflection symmetry $\hat{S}_4=\tau_z e^{i\frac{\pi}{4}\sigma_z}$ 
 and time-reversal symmetry $\hat{T}=-i\sigma_y K$.
On one hand, the $z_{2}$ indicator can be calculated as: 1 for $1<|\Delta|<3$ and 0 for $|\Delta|<1$ or $|\Delta|>3$.
On the other hand, according to \onlinecite{shen_book}, this model is indeed a strong TI when $1<|\Delta|<3$.
Therefore, in this model $z_{2}=1$ corresponds to strong TI.

Now, by the k$\cdot$p analysis below, we argue that this correspondence is universal.
We assume $\Delta=3+m$ and expand the model around $(000)$ to linear terms in $k$.
Then a Dirac Hamiltonian with mass $m$ can be got as
\begin{equation}
    H(\mathbf{k}) \approx m\tau_z \sigma_0 + \sum_i k_i \tau_x \sigma_i
\end{equation}
It should be noticed that, although this k$\cdot$p model is derived from the above tight binding model,
 it is the universal Hamiltonian around the phase transition point from strong TI to trivial insulator \cite{Chiu_RMP_2016}.
On one hand, the topological phase transition is given by the change of $\mathrm{sgn}(m)$.
On the other hand, the change of $\mathrm{sgn}(m)$ must lead to a change of $z_{2}$ due to the following reason.
If the model is $S_4$ invariant, 
 the $S_4$ operator has to be $\hat{S}_4=\tau_z \sigma^{i\frac{\pi}{4}\sigma_z}$ (up to a $\pm 1$ sign),
 thus the change of $\mathrm{sgn}(m)$ inverts the $E_\frac{1}{2}$ and $E_\frac{3}{2}$ irreps at $(000)$,
 leading to a change of $z_{2}$.
Therefore, the topological phase transition is always accompanied by the change of $z_{2}$ indicator,
 and the correspondence between $z_{2}=1$ and strong TI should be universal.

\begin{figure}
\begin{centering}
\includegraphics[width=1\linewidth]{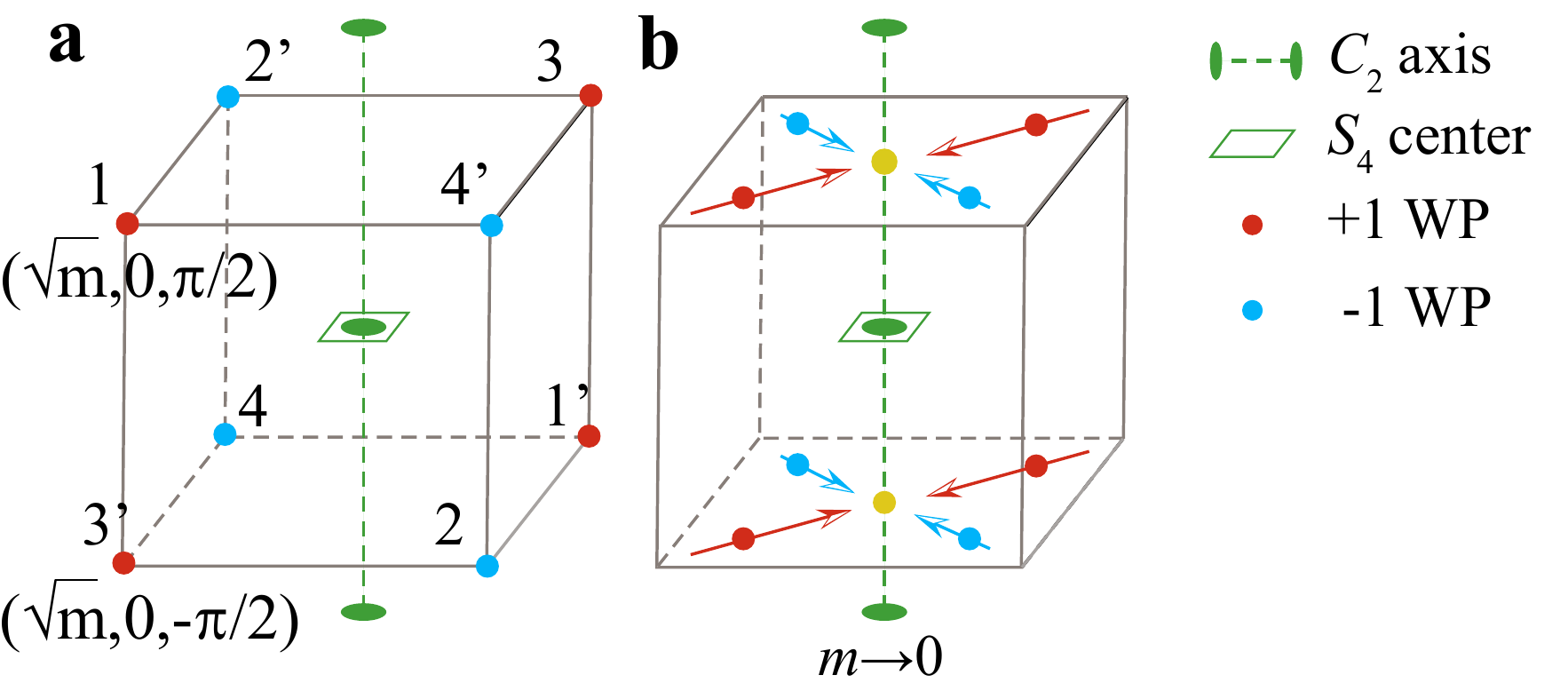}
\par\end{centering}

\protect\caption{The $S_{4}$ Weyl semimetal. In {\bf a} we plot
the eight Weyl points in model \cref{eq:Weyl}, where the red
circles represent +1-charge Weyl points and the blue circles represent
-1-charge Weyl points. The Weyl points labeled with 1, 2, 3, 4 transform
to each other in turn under the $S_{4}$ operation, and transform
to the Weyl points labeled with 1', 2', 3', 4' respectively under the
time reversal. In {\bf b}, we plot the motion of these Weyl points as
$m$ approaches zero. When $m=0$ four Weyl points merge to a
double Dirac point. \label{fig:S4-Weyl}}

\end{figure}

\textit{Weyl semimetal with $z_{2}=1$.}
Consider the model
\begin{align}
\hat{H}\left(\mathbf{k}\right) & =\left(2-\sum_{i}\cos k_{i}\right)\tau_{z}\sigma_{0}+m\sin k_{z}\tau_{x}\sigma_{z}\nonumber \\
 & +\left(\cos k_{x}-\cos k_{y}\right)\tau_{x}\sigma_{0}+\sin k_{x}\sin k_{y}\tau_{y}\sigma_{z}\label{eq:Weyl}
\end{align}
This Hamiltonian has time-reversal
 symmetry $\hat{T}=-i\sigma_{y}$ and rotoreflection symmetry $\hat{S}_{4}=\tau_{z}e^{i\frac{\pi}{4}\sigma_{z}}$,
 and the occupied states at the four $S_{4}$-invariant TRIMs give $z_{2}=1$ (at half-filling). 
Expand this model around $\left(00\frac{\pi}{2}\right)$
 for the spin up ($\sigma_{z}=1$) and spin down ($\sigma_{z}=-1$)
 components respectively, we get
\begin{equation}
\hat{H}^{\uparrow}(\mathbf{k})\approx\delta k_{z}\tau_{z}+\left(k_{y}^{2}-k_{x}^{2}+m\right)\tau_{x}+k_{x}k_{y}\tau_{y}
\end{equation}
and
\begin{equation}
\hat{H}^{\downarrow}(\mathbf{k})\approx\delta k_{z}\tau_{z}+\left(k_{y}^{2}-k_{x}^{2}-m\right)\tau_{x}-k_{x}k_{y}\tau_{y}
\end{equation}
where $\delta k_z = k_z - \frac{\pi}{2}$.
Two +1-change Weyl points and two -1-change Weyl points are found
 to locate at $\left(\pm\sqrt{m},0,\frac{\pi}{2}\right)$ and $\left(0,\pm\sqrt{m},\frac{\pi}{2}\right)$,
 respectively. 
(Here we assume $m$ is a positive small quantity.)
Similarly, by expanding this model around $\left(0,0,-\frac{\pi}{2}\right)$,
we get two other +1-change Weyl points at $\left(\pm\sqrt{m},0,-\frac{\pi}{2}\right)$
and two other -1-change Weyl points at $\left(0,\pm\sqrt{m},-\frac{\pi}{2}\right)$.
The eight Weyl points at generic momenta form the minimal configuration satisfying both
time-reversal and $S_{4}$ symmetries, as shown in \figref[a]{fig:S4-Weyl}.

Without a band inversion at $S_{4}$-invariant TRIMs, there is no way to annihilate these Weyl points.
For example, as shown in \figref[b]{fig:S4-Weyl}, by tuning $m$ to zero the two +1-charge
Weyl points and the two -1-charge Weyl points at the $k_{z}=\frac{\pi}{2}$
plane will merge at $\left(00\frac{\pi}{2}\right)$ but can not annihilate
each other. 
Instead, they form a double Dirac point, which is protected
 by the $S_{4}$ and time-reversal symmetries. 
(The terms anti-commuting with the $\tau_z \sigma_0$, $\tau_x \sigma_0$, and $\sigma_y \sigma_z$
 terms in \cref{eq:Weyl} can only be $\tau_y \sigma_x$ and $\tau_y \sigma_y$.
 The corresponding coefficients, denoted as $I(\mathbf{k})$ and $J(\mathbf{k})$ respectively, 
 should satisfy $I(\mathbf{k})=I(-\mathbf{k})$, $J(\mathbf{k})=J(-\mathbf{k})$,
 $I(-k_y,k_x,-k_z)=-J(\mathbf{k})$, $J(S_4\mathbf{k})=I(\mathbf{k})$ 
 and vanish at the Dirac points.)
To annihilate the two double Dirac points at $(0,0,\pm\pi/2)$, one need to move
 them to $\Gamma$, which causes a band inversion.
In this particular model the band inversion causes a change of $z_{2}$,
 however, in general case it does not.
For instance, here we consider a product state consisting of a Weyl semimetal model with $z_{2}=1$ and two strong TI models.
The $z_{2}$ indicator of this product state is given by $z_{2}=1+1+1\ \mathrm{mod}\ 2=1$.
Then the process annihilating the Weyl points without changing $z_{2}$ can be constructed as:
 (i) move the eight Weyl points to $(000)$ to cause a band inversion,
 (ii) at the same time close and reopen the gap of one of the two TI models.
After this process, all the Weyl points are gapped and the $z_{2}$ indicator remains unchanged,
 i.e. $z_{2}=0+0+1=1$.

It should be emphasized that in all the other non-centrosymmetric SGs with nontrivial SI groups,
 i.e., \#174 and \#187-190,  Weyl points can be annihilated without band inversion at high symmetry momentum.
In all these five SGs $k_z=0$-plane is mirror-invariant, and Weyl points at generic momenta should
 constitute symmetric pairs about the mirror, where each pair consists of two Weyl points with opposite charges.
Therefore, each pair of Weyl points can move toward each other until they annihilate each other at the $k_z=0$-plane.
To be specific, let us assume the k$\cdot$p model of a pair of Weyl points moved to the $k_z=0$-plane as
\begin{equation}
    \hat{H}(\delta\mathbf{k}) = \sum_i \delta k_i \tau_z\sigma_i
\end{equation}
where $\delta \mathbf{k} = \mathbf{k} - \mathbf{k}_c$ and $\mathbf{k}_c$ is the position of the two Weyl points.
The mirror operator can be chosen as $\hat{m}_z=i \tau_x \sigma_z$.
Apparently, the mass term $m\tau_x$ is symmetry-allowed.
Thus the Weyl points can pairwise annihilate at generic point in the $k_z=0$ plane without any level crossing 
 at high symmetry momentum.

%$P\bar{4}$ is a special SG which  has representation enforced Weyl semimetal.
%On the contrary, in other non-centrosymmetric SGs with nontrivial SI group, i.e.,
% SGs \#174 and \#187-190 according to table \ref{tab:Indicator}, 
% the Weyl semimetal phase can always be deformed to a gapped state without changing the band irreps.
%In these groups, there is a horizontal mirror, thus Weyl points at generic momenta
% occur in pairs as $(k_x,k_y,\pm k_z)$, and the two Weyl points there have opposite charges.
%Then we can move such a pair of Weyl points getting close and annihilate them at the horizontal
% mirror, which does not change the irreps at high symmetry momenta.

%\subsubsection{Weyl semimetal with $z_{2s}=0$}

%...

\subsection{Convention dependence of $z_{4}$, $z_{8}$, $z_{12}$, $z_{12}^\prime$ indicator}
In the text we have discussed that by redefining $\hat{P}\to -\hat{P}$
 $z_{4}=1,3$ turn into each other,
 and without external reference $z_{4}=1,3$ should be physically identical.
Such a convention dependence also exists for the $z_8$, $z_{12}$, and $z_{12}^\prime$ indicator (\cref{tab:formula}).

For the $z_8$ indicator, according to {\cref{tab:z8}}, 
 redefining $\hat{P}\to -\hat{P}$ interchanges $n_{\frac{3}{2}/\frac{1}{2}}^+$
 and $n_{\frac{3}{2}/\frac{1}{2}}^-$, and redefining $\hat{C}_4 \to -\hat{C}_4$ interchanges
 $n_\frac{3}{2}^{+/-}$ and  $n_\frac{1}{2}^{+/-}$.
%Here we take a $\mathbb{Z}_8$ generator as
% $n_\frac{1}{2}^{+}=-1$, $n_\frac{1}{2}^{-}=1$, $n_\frac{3}{2}^+=n_\frac{3}{2}^-=0$.
%Then we get $P\to -P$ and $C_4 \to -C_4$ makes $z_8=1\to z_8=7$ and $z_8=1\to z_8=5$, respectively. 
%Multiplying this generator with integers, we conclude that
% the odd values $z_8=1,3,5,7$ differ from each other only by a convention,
% and the even values $z_8=2,6$ also differ from each other by a convention.
Substitute this relation into \cref{eq:z8}, we find that there are only four kinds of physically distinct cases:
 (i) $z_8=0$, (ii) $z_8=1,3,5,7$ corresponding to the strong TI,
 (iii) $z_8=2,6$, and (iv) $z_8=4$.

The $z_{12}$ indicator is completely determined by the 
 mirror Chern number indicator $z_{6\mathrm{m},0}$, $z_{6\mathrm{m},\pi}$ and the $z_{4}$ indicator,
 wherein $z_{4}=1,3$ turn into each other when redefining $\hat{P}\to -\hat{P}$,
 and $z_{6\mathrm{m},0/\pi}=1,5$ turn into each other  when redefining $\hat{C}_6\to -\hat{C}_6$.
%Substitute this into Eq. (\ref{eq:z12}) we get that $z_{12}=1$ turns into $z_{12}=7$
% and $z_{12}=5$ under the redefining $P\to-P$ and $C_6\to -C_6$ respectively.
Substitute this into \cref{eq:z12} we conclude that there are only six kinds of physically distinct cases:
 (i) $z_{12}=0$, (ii) $z_{12}=1,5,7,11$, (iii) $z_{12}=2,10$, (iv) $z_{12}=3,9$,
 (v) $z_{12}=4,8$, and (vi) $z_{12}=6$,
 wherein (ii) and (iv) correspond to strong TI.
The difference between (ii) and (iv) is the magnitude of mirror Chern numbers:
 in (ii) $C_{\mathrm{m},0}+C_{\mathrm{m},\pi} = \pm 1$ (modulo 6),
 while in (iv) $C_{\mathrm{m},0}+C_{\mathrm{m},\pi} = 3$ (modulo 6).

The discussion for the $z_{12}^\prime$ indicator is similar with the discussion for the $z_{12}$ indicator.
And the conclusion is same---there are only six kinds of physically distinct cases, i.e.,
 (i) $z_{12}^\prime=0$, (ii) $z_{12}^\prime=1,5,7,11$, (iii) $z_{12}^\prime=2,10$, 
 (iv) $z_{12}^\prime=3,9$, (v) $z_{12}^\prime=4,8$, and (vi) $z_{12}^\prime=6$,
 wherein (ii) and (iv) corresponds to strong TI.
And, the difference between (ii) and (iv) is also the magnitude of mirror Chern numbers:
 in (ii) $C_{\mathrm{m},0} = \pm 1$ (modulo 6),
 while in (iv) $C_{\mathrm{m},0} = 3$ (modulo 6).

\subsection{Fu-Kane-like formulae for all symmetry-based indicators}

To summarize, we tabulate all the Fu-Kane-like formulae 
 in \cref{tab:formula} and their relation with SI groups in \cref{tab:Indicator}.
Two comments are made here.
The first is that, in the second to fourth rows of \cref{tab:Indicator},
the Fu-Kane-like formulae seem to ``mismatch'' with the SI
groups. This is simply because some weak
indicators are trivialized by some SG operations. For example, the SI
group of SG \#11 ($P2_{1}/m$) is $\mathbb{Z}_{2}\times\mathbb{Z}_{2}\times\mathbb{Z}_{4}$
instead of $\mathbb{Z}_{2}\times\mathbb{Z}_{2}\times\mathbb{Z}_{2}\times\mathbb{Z}_{4}$
because the $C_2$-screw axis (along $\mathbf{a}_{2}$) enforce
the four TRIMs at $k_{2}=\pi$ to have same parities and so enforce
$z_{2\mathrm{w},2}=0$. Thus, there should be no worry about this ``mismatch''.
The second is that, the SI itself can not tell us whether a
state is fully gapped. 
Therefore, to apply the Fu-Kane-like formulae one should firstly check whether the band is fully gapped.

\clearpage
\newpage
\section{Symmetry-based indicator of layer construction}\label{sec:LCind}

%In sections \ref{sec:Layer}-\ref{sec:find-eLC}, we build a systematic
% framework to use eLCs as real space building blocks to generate TCIs.
%In section \ref{sec:indicator}, we derive the complete Fu-Kane-like
% formulae for all SGs with nontrivial SI groups.
In \cref{sec:Layer,sec:conv,sec:find-eLC}, we developped systematic methods 
 to (i) diagnose the TCI invariants of an eLC according to its geometry configuration
 and (ii) enumerate all nonequivalent eLCs.
With all the nonequivalent eLCs at hand it is direct to enumerate all the TCI states consistent with LCs
 by stacking the eLCs together.
On the other hand, if the SI of eLCs can be calculated then the SI of all these TCI states 
 can also be calculated due to the additive property of SI.
In other words, we can use eLC as intermedia to complete the mapping from SI to TCI invariants.
However, the inputs of Fu-Kane-like formulae derived in \cref{sec:indicator} are the numbers of irreps at high symmetry momenta, which are not explicit for eLC.
Therefore, to finish the mapping in this appendix we give the method to calculate the SI of eLCs.

Firstly, one should notice that the SI can be uniquely determined from a complete set of topological invariants,
 because by definition two different SI corresponds to two different topologies. 
Here we take the \emph{assumption} that the the seven kinds of invariants introduced in \cref{sec:Layer} are complete for LC states. 
Therefore the SI of a LC is uniquely determined by these invariants. 
%Secondly, since both the topological invariants and the symmetry SI are additive, in
% order to calculate the SI of all LCs, we need only to calculate the SI of eLCs. 
Secondly, as shown in \cref{sec:indicator} SI in all SGs can be induced from only six SGs, 
 thus to calculate the SI of an eLC $E$ in any SG $\mathcal{G}$ we can (i) reduce
 $\mathcal{G}$ to one of the six SGs, say $\mathcal{G}^{\prime}$,
 (ii) calculate the topological invariants of $E$ in $\mathcal{G}^{\prime}$,
 and (iii) calculate the SI in $\mathcal{G}^{\prime}$ due to
 the invariants. 
Therefore, we need only to derive the SI of eLCs in these six SGs.

\subsection{$P\bar{1}$}

According to the discussion in \cref{sec:Layer}, the TCI invariants
 of SG \#2 ($P\bar{1}$) should include three weak invariants $\delta_{\mathrm{w},i=1,2,3}$
 and an inversion invariant $\delta_{\mathrm{i}}$, which is defined on the inversion
 center at the origin (see \cref{sec:table} for the convention).
There are only four independent eLCs---$E_{1}=\mathrm{eLC}\left(100;0\right)$,
 $E_{2}=\mathrm{eLC}\left(010;0\right)$, $E_{3}=\mathrm{eLC}\left(001;0\right)$,
 $E_{4}=\mathrm{eLC}\left(001;\frac{1}{2}\right)$, whose invariants
 are
\begin{equation}
\left\{ \delta_{\mathrm{w},i=1,2,3},\delta_{\mathrm{i}}\right\} \left(E_{1}\right)=\left\{ 100,1\right\} 
\end{equation}
\begin{equation}
\left\{ \delta_{\mathrm{w},i=1,2,3},\delta_{\mathrm{i}}\right\} \left(E_{2}\right)=\left\{ 010,1\right\} 
\end{equation}
\begin{equation}
\left\{ \delta_{\mathrm{w},i=1,2,3},\delta_{\mathrm{i}}\right\} \left(E_{3}\right)=\left\{ 001,1\right\} 
\end{equation}
\begin{equation}
\left\{ \delta_{\mathrm{w},i=1,2,3},\delta_{\mathrm{i}}\right\} \left(E_{4}\right)=\left\{ 001,0\right\} 
\end{equation}
 as shown in \figref[a]{fig:LCind} and tabulated in \cref{tab:eLC}.
The SI set includes three weak TI indicators $z_{2\mathrm{w},i=1,2,3}$
 and one $\mathbb{Z}_{4}$ indicator $z_{4}$, where $z_{2\mathrm{w},i=1,2,3}$
 equals to $\delta_{\mathrm{w},i=1,2,3}$ according to the Fu-Kane criterion.
To calculate the $z_{4}$ indicator, let us assume that the 2D TI in
 $E_{3}$ (\figref[a]{fig:LCind}) gives the BR
 \begin{equation}
 E_{\frac{1}{2}\mathrm{u}}^{\Gamma}\; E_{\frac{1}{2}\mathrm{g}}^\mathrm{X}\; E_{\frac{1}{2}\mathrm{g}}^\mathrm{Y}\; E_{\frac{1}{2}\mathrm{g}}^\mathrm{V}
 \end{equation}
 in the $k_{z}=0$ plane. 
Taking this 2D TI as building block, we get the 3D BRs for the four eLCs, as shown in \figref[a]{fig:LCind}.
It should be noticed that, for $E_{4}$, the $k_{3}=\pi$ states have opposite parities with $k_{3}=0$ states; while for $E_{1,2,3}$, the $k_{1,2,3}=\pi$ states have same parities with $k_{1,2,3}=0$ states.
(Analysis on parities resembles the analysis on mirror eigenvalues, as discussed in \cref{sec:Layer} and sketched in \figref[a]{fig:Cm}.) 
Substitute these BRs into \cref{tab:formula}, we get
\begin{equation}
\left\{ z_{2\mathrm{w},i=1,2,3},z_{4}\right\} \left(E_{1}\right)=\left\{ 100,2\right\} 
\end{equation}
\begin{equation}
\left\{ z_{2\mathrm{w},i=1,2,3},z_{4}\right\} \left(E_{2}\right)=\left\{ 010,2\right\} 
\end{equation}
\begin{equation}
\left\{ z_{2\mathrm{w},i=1,2,3},z_{4}\right\} \left(E_{3}\right)=\left\{ 001,2\right\} 
\end{equation}
\begin{equation}
\left\{ z_{2\mathrm{w},i=1,2,3},z_{4}\right\} \left(E_{4}\right)=\left\{ 001,0\right\} 
\end{equation}

\begin{figure}
\begin{centering}
\includegraphics[width=1\linewidth]{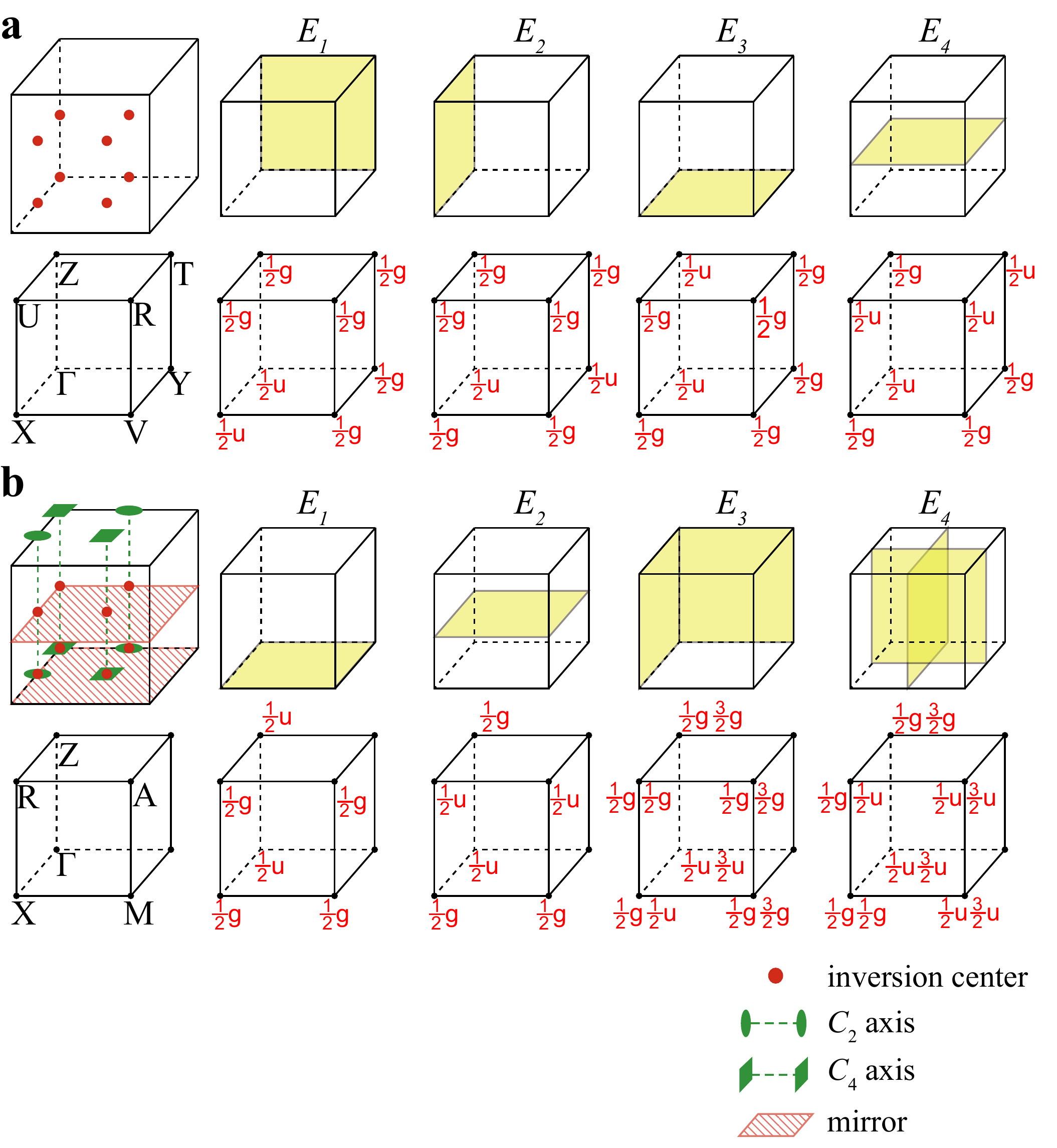}
\par\end{centering}

\protect\caption{{\bf a} are the band representations of the four independent eLCs
in SG $P\bar{1}$, and {\bf b} are the band representations of the four
independent eLCs in SG $P4/m$. In the top panel of each, we plot
the symmetry elements and the real space eLCs, 
in the bottom panel of each, we plot the irreducible
Brillouin zone and the irreps of the corresponding eLCs. \label{fig:LCind}}

\end{figure}

\subsection{$P4/m$}

The invariants in SG \#83 ($P4/m$) include three weak invariants $\delta_{\mathrm{w},i=1,2,3}$,
 two mirror Chern numbers $C_{\mathrm{m},0/\pi}$, a $C_{4}$-rotation invariant
 $\delta_\mathrm{r}$, an inversion invariant $\delta_{\mathrm{i}}$, and a $S_4$ invariant $\delta_{S_4}$.
The independent eLCs are $E_{1}=\mathrm{eLC}\left(001;0\right)$,
 $E_{2}=\mathrm{eLC}\left(001;\frac{1}{2}\right)$, $E_{3}=\left(100;0\right)$,
 $E_{4}=\left(100;\frac{1}{2}\right)$, and their invariants are
\begin{align}
 & \left\{ \delta_{\mathrm{w},i=1,2,3},C_{\mathrm{m},k=0,\pi},\delta_{\mathrm{r}},\delta_\mathrm{i},\delta_{S_4}\right\} \left(E_{1}\right)\nonumber \\
= & \left\{ 001,11,0,1,1\right\} \label{eq:P4/m-E1ivt}
\end{align}
\begin{align}
 & \left\{ \delta_{\mathrm{w},i=1,2,3},C_{\mathrm{m},k=0,\pi},\delta_{\mathrm{r}},\delta_\mathrm{i},\delta_{S_4}\right\} \left(E_{2}\right)\nonumber \\
= & \left\{ 001,1\bar{1},0,0,0\right\} 
\end{align}
\begin{align}
 & \left\{ \delta_{\mathrm{w},i=1,2,3},C_{\mathrm{m},k=0,\pi},\delta_{\mathrm{r}},\delta_\mathrm{i},\delta_{S_4}\right\} \left(E_{3}\right)\nonumber \\
= & \left\{ 110,00,1,0,1\right\} 
\end{align}
\begin{align}
 & \left\{ \delta_{\mathrm{w},i=1,2,3},C_{\mathrm{m},k=0,\pi},\delta_{\mathrm{r}},\delta_\mathrm{i},\delta_{S_4}\right\} \left(E_{4}\right)\nonumber \\
= & \left\{ 110,00,0,0,0\right\} 
\end{align}
, as shown in \figref[b]{fig:LCind}. We assume the 2D TI in $E_{1}$
has the BR
\begin{equation}
E_{\frac{1}{2}\mathrm{u}}^{\Gamma}\; E_{\frac{1}{2}\mathrm{g}}^\mathrm{X}\; E_{\frac{1}{2}\mathrm{g}}^\mathrm{M}
\end{equation}
in the $k_{3}=0$ plane, using which as building Blocks we can derive
the BRs of $E_{2}$, $E_{3}$, $E_{4}$. In $E_{1}$ the $k_{3}=\pi$
states have same parities with $k_{3}=0$ states; while in $E_{2}$
the $k_{3}=\pi$ states have opposite parities with the $k_{3}=0$
states. In $E_{3}$ and $E_{4}$, the parities at TRIMs can be derived
in a similar way, and the irreps at $C_{4}$-invariant TRIMs must
consist of a pair of $E_{\frac{1}{2},g/u}$ and $E_{\frac{3}{2},g/u}$
because the $C_{4}$ operation transforms a layer to another such
that the $C_{4}$ representation matrix must be traceless. 
The BRs of the four eLCs are summarized in \figref[b]{fig:LCind}, 
 substitute which into equations in \cref{tab:formula} we get
\begin{equation}
\left\{ z_{2\mathrm{w},1},z_{4\mathrm{m},\pi},z_{8}\right\} \left(E_{1}\right)=\left\{ 0,1,2\right\} 
\end{equation}
\begin{equation}
\left\{ z_{2\mathrm{w},1},z_{4\mathrm{m},\pi},z_{8}\right\} \left(E_{2}\right)=\left\{ 0,3,0\right\} 
\end{equation}
\begin{equation}
\left\{ z_{2\mathrm{w},1},z_{4\mathrm{m},\pi},z_{8}\right\} \left(E_{3}\right)=\left\{ 1,0,4\right\} 
\end{equation}
\begin{equation}
\left\{ z_{2\mathrm{w},1},z_{4\mathrm{m},\pi},z_{8}\right\} \left(E_{4}\right)=\left\{ 1,0,0\right\} \label{eq:P4/m-E4ind}
\end{equation}
\cref{eq:P4/m-E1ivt} to \cref{eq:P4/m-E4ind} are consistent with
the results in \onlinecite{Song2017}.

\subsection{$P\bar{6}$}

SG \#174 ($P\bar{6}$) has five invariants, i.e., three weak invariants
$\delta_{\mathrm{w},i=1,2,3}$ and two mirror Chern numbers $C_{\mathrm{m},k_z=0,\pi}$.
And, the two independent eLCs $E_{1}=\mathrm{eLC}\left(001;0\right)$,
$E_{2}=\mathrm{eLC}\left(001;\frac{1}{2}\right)$ correspond to
\begin{equation}
\left\{ \delta_{\mathrm{w},i=1,2,3},C_{\mathrm{m},k=0,\pi}\right\} \left(E_{1}\right)=\left\{ 001,11\right\} 
\end{equation}
\begin{equation}
\left\{ \delta_{\mathrm{w},i=1,2,3},C_{\mathrm{m},k=0,\pi}\right\} \left(E_{2}\right)=\left\{ 001,1\bar{1}\right\} 
\end{equation}
Correspondingly, the SI (Chern number modulo 3) can be directly
got as
\begin{equation}
\left\{ z_{3\mathrm{m},k=0,\pi}\right\} \left(E_{1}\right)=\left\{ 11\right\} 
\end{equation}
\begin{equation}
\left\{ z_{3\mathrm{m},k=0,\pi}\right\} \left(E_{2}\right)=\left\{ 12\right\} 
\end{equation}

\subsection{$P6/m$}

SG \#175 ($P6/m$) has eight TCI invariants, i.e., three weak invariants $\delta_{\mathrm{w},i=1,2,3}$,
two mirror Chern numbers $C_{\mathrm{m},k=0,\pi}$,
a $C_{6}$-rotation invariant $\delta_{\mathrm{r}}$, and  an inversion invariant $\delta_\mathrm{i}$
And, the three independent eLCs $E_{1}=\mathrm{eLC}\left(001;0\right)$,
 $E_{2}=\mathrm{eLC}\left(001;\frac{1}{2}\right)$, $E_{3}=\mathrm{eLC}\left(100;0\right)$ correspond to
\begin{align}
 & \left\{ \delta_{\mathrm{w},i=1,2,3},C_{\mathrm{m},k=0,\pi},\delta_\mathrm{r},\delta_\mathrm{i}\right\} \left(E_{1}\right)\nonumber \\
= & \left\{ 001,11,0,1\right\} 
\end{align}
\begin{align}
 & \left\{ \delta_{\mathrm{w},i=1,2,3},C_{\mathrm{m},k=0,\pi},\delta_\mathrm{r},\delta_\mathrm{i}\right\} \left(E_{2}\right)\nonumber \\
= & \left\{ 001,1\bar{1},0,0\right\} 
\end{align}
\begin{align}
 & \left\{ \delta_{\mathrm{w},i=1,2,3},C_{\mathrm{m},k=0,\pi},\delta_\mathrm{r},\delta_\mathrm{i}\right\} \left(E_{3}\right)\nonumber \\
= & \left\{ 000,00,1,1\right\} 
\end{align}
According to \cref{sec:indicator}, SG $P6/m$ has the SI set $\left\{ z_{6\mathrm{m},\pi},z_{12}\right\} $.
The $z_{6\mathrm{m},\pi}$ indicator can be got as $C_{\mathrm{m},k=\pi}\ \mathrm{mod}\ 6$,
 and the $z_{12}$ indicator can be determined from $z_{6\mathrm{m},k=0}+z_{6\mathrm{m},k=\pi}$ and $z_{4}$ (\cref{tab:formula}),
 wherein $z_{6\mathrm{m},k=0}+z_{6\mathrm{m},k=\pi}$ can be calculated as $ C_{\mathrm{m},k=0}+C_{\mathrm{m},k=\pi}\ \mathrm{mod}\ 6$ and
 the $z_4$ indicator can be calculated by group reduction to $P\bar{1}$. 
We get
\begin{equation}
\left\{ z_{6\mathrm{m},\pi},z_{12}\right\} \left(E_{1}\right)=\left\{ 1,2\right\} 
\end{equation}
\begin{equation}
\left\{ z_{6\mathrm{m},\pi},z_{12}\right\} \left(E_{2}\right)=\left\{ 5,0\right\} 
\end{equation}
\begin{equation}
\left\{ z_{6\mathrm{m},\pi},z_{12}\right\} \left(E_{3}\right)=\left\{ 0,6\right\} 
\end{equation}

\subsection{$P6_{3}/m$}

The invariants in SG \#176 ($P6_{3}/m$) include three weak invariants 
 $\delta_{\mathrm{w},i=1,2,3}$, two mirror Chern numbers $C_{\mathrm{m},k=0,\pi}$,
 an inversion invariant $\delta_{\mathrm{i}}$, 
 and a $C_{6}$-screw invariant $\delta_\mathrm{s}$. 
There are two independent eLCs, $E_{1}=\mathrm{eLC}\left(001;0\right)$,
 $E_{2}=\mathrm{eLC}\left(001;\frac{1}{4}\right)$, 
 whose invariants are
\begin{align}
 & \left\{ \delta_{\mathrm{w},i=1,2,3},C_{\mathrm{m},k=0,\pi},\delta_{\mathrm{i}},\delta_\mathrm{s}\right\} \left(E_{1}\right)\nonumber \\
= & \left\{ 000,00,1,1\right\} 
\end{align}
\begin{align}
 & \left\{ \delta_{\mathrm{w},i=1,2,3},C_{\mathrm{m},k=0,\pi},\delta_{\mathrm{i}},\delta_\mathrm{s}\right\} \left(E_{2}\right)\nonumber \\
= & \left\{ 000,20,0,1\right\} 
\end{align}
Similar with SG \#175 ($P6/m$), their SI can be calculated as
\begin{equation}
z_{12}^{\prime}(E_{1})=6, \qquad z_{12}^\prime(E_2)=8
\end{equation}

\subsection{$P\bar{4}$}

$z_{2}=1$ corresponds to strong TI or Weyl semimetal, neither of which can be realized by layer construction.
Thus all the eLCs in SG $P\bar{4}$ have a trivial indicator.

\clearpage
\newpage
\section{Weak topological insulators beyond layer constructions}\label{sec:weak}

Using the layer construction method introduced above we success to obtain all the SI in all SGs except 
 five (and only five) corner cases,
 where we find that the compatibility-relation allowed weak TIs can not be realized in any LC.
These corner cases are SGs \#48 ($Pnnn$), \#86 ($P4_2/n$), \#134 ($P4_2/nnm$), \#201 ($Pn\bar{3}$), and \#224 ($Pn\bar{3}m$),
 all of which are centrosymmetric and have the SI group $\mathbb{Z}_2\times\mathbb{Z}_4$.
The $\mathbb{Z}_4$ indicator is $z_{4}$ and the $\mathbb{Z}_2$ indicator is the weak TI indicator $z_{2\mathrm{w},1}$,
 which equals to the other two weak TI indicators, i.e., $z_{2\mathrm{w},1}=z_{2\mathrm{w},2}=z_{2\mathrm{w},3}$,
 due to the compatibility relation.

In this appendix, we will construct a tight-binding model for the weak TI in SG $Pn\bar{3}m$.
Since all the other four SGs are subgroups of $Pn\bar{3}m$ and the corresponding symmetry breaking from 
 $Pn\bar{3}m$ does not expand the cells, 
 the weak TI models in the other four SGs can be got 
 from this model by a slight breaking of additional crystalline symmetries.
Therefore, by this model we show that 
 (i) compatibility-relation allowed weak TIs in these SGs can indeed be realized in tight-binding models 
 and (ii) these weak TIs are beyond the scope of layer construction.

\begin{figure}
\begin{centering}
\includegraphics[width=1\linewidth]{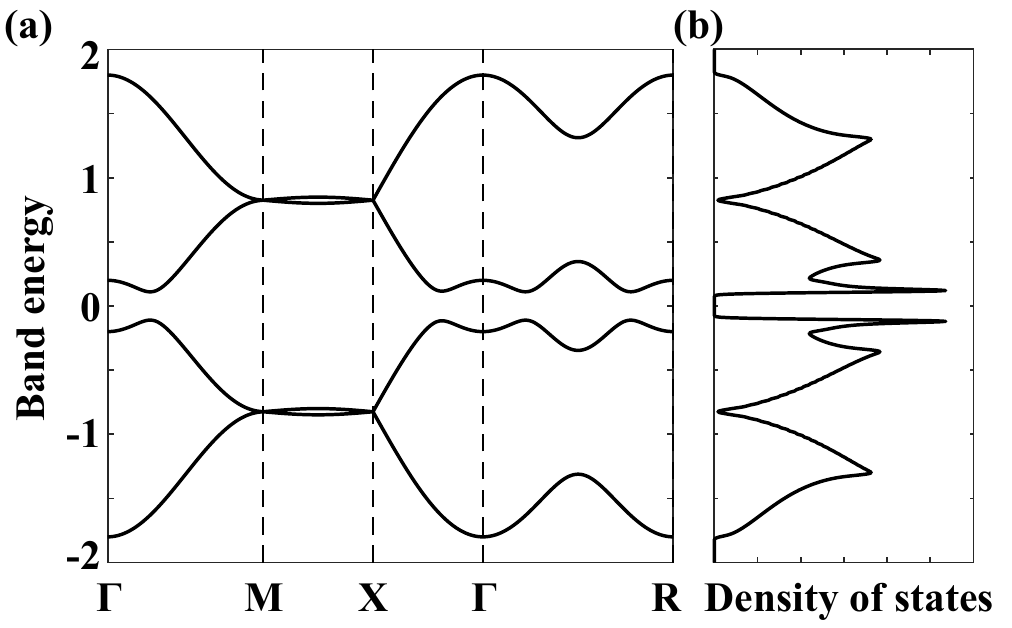}
\par\end{centering}
\protect\caption{
In {\bf a} and {\bf b} the band structure and density of states of the weak TI model for SG $Pn\bar{3}m$ 
 are ploted, respectively.
The parameters are set as $t_1=-t_2=1$, $\Delta=0.8$, $\lambda_1=\lambda_2=0.2$.
\label{fig:224WTI}}
\end{figure}

Now let us construct the model. 
$Pn\bar{3}m$ has a primitive cubic lattice, and its SG generators include inversion $P=\{-1|000\}$, glide $n=\{m_{001}|\frac{1}{2}\frac{1}{2}0\}$,
 rotation $C_3=\{3_{111}|000\}$, and mirror $m=\{m_{1\bar{1}0}|000\}$ \cite{aroyo_bilbao_2009}.
Here we consider the Wyckoff position $2a$, which include two positions $(\frac{1}{4}\frac{1}{4}\frac{1}{4})$ 
 and $(\bar{\frac{1}{4}}\bar{\frac{1}{4}}\bar{\frac{1}{4}})$ and has the site-symmetry group $T_d$.
We choose the bases of $E_\frac{1}{2}$ and $E_\frac{5}{2}$ irreps on $2a$ as the tight-binding model bases.
Then the SG operator on Bloch bases can be derived as
\begin{equation}
    \hat{P} = \mu_0 \tau_x \sigma_0 \label{eq:224-P}
\end{equation}
\begin{equation}
    \hat{n} = -i e^{-i\frac{k_x+k_y}{2}}\mu_0 \left[ \cos\frac{k_x+k_y}{2} - \sin\frac{k_x+k_y}{2}\right] \sigma_z \label{eq:224-n}
\end{equation}
\begin{equation}
    \hat{C_3} = \frac{1}{2}\mu_0 \tau_0[\sigma_0-i\sigma_x-i\sigma_y-i\sigma_z] \label{eq:224-C3}
\end{equation}
\begin{equation}
    \hat{m} = \frac{1}{\sqrt{2}} \mu_z \tau_0[-i\sigma_x+i\sigma_y] \label{eq:224-m}
\end{equation}
where $\mu_i$, $\tau_i$, $\sigma_i$ are pauli matrices representing the $E_\frac{1}{2}$ and $E_\frac{5}{2}$ irreps,
 the $(\frac{1}{4}\frac{1}{4}\frac{1}{4})$ and $(\bar{\frac{1}{4}}\bar{\frac{1}{4}}\bar{\frac{1}{4}})$ positions,
 and the two bases in each irrep, respectively.
The Bloch bases are defined as a Fourier transformation of the orbitals on $2a$
\begin{equation}
    |\phi_{\alpha\mathbf{k}}\rangle = \frac{1}{\sqrt{N}} \sum_R e^{i\mathbf{k\cdot R}} |a_{\alpha \mathbf{R}}\rangle
\end{equation}
where $|a_{\alpha\mathbf{R}}\rangle$ is the $\alpha$-th orbital in the lattice $\mathbf{R}$, 
 and $\alpha$ is a composite index consisting of irreps, sites, and bases of irreps.
The equation for tight-binding model is given by
\begin{equation}
    \hat{g}\hat{H}(\mathbf{k})\hat{g}^{-1} = \hat{H}(g\mathbf{k})
\end{equation}
Substitute \cref{eq:224-P,eq:224-n,eq:224-C3,eq:224-m} to it, we get a solution as
\begin{widetext}
    \begin{align}
        H(\mathbf{k}) &= \Delta \mu_z\tau_0\sigma_0 + f_x(\mathbf{k})\mu_0\tau_x\sigma_0 + f_y(\mathbf{k})\mu_0\tau_y\sigma_0
                       + f_x^z(\mathbf{k})\mu_y\tau_x\sigma_z + f_y^z(\mathbf{k})\mu_y\tau_y\sigma_z 
                       + f_x^x(\mathbf{k})\mu_y\tau_x\sigma_x + f_y^x(\mathbf{k})\mu_y\tau_y\sigma_x \nonumber \\
                      &+ f_x^y(\mathbf{k})\mu_y\tau_x\sigma_y + f_y^y(\mathbf{k})\mu_y\tau_y\sigma_y 
    \end{align}
    where
    \begin{equation}
        f_x(\mathbf{k}) = t_1 \cos\left(\frac{k_x+k_y+k_z}{2}\right)\cos\frac{k_x}{2}\cos\frac{k_y}{2}\cos\frac{k_z}{2}
                        + t_2 \sin\left(\frac{k_x+k_y+k_z}{2}\right)\sin\frac{k_x}{2}\sin\frac{k_y}{2}\sin\frac{k_z}{2}
    \end{equation}
    \begin{equation}
        f_y(\mathbf{k}) =-t_1 \sin\left(\frac{k_x+k_y+k_z}{2}\right)\cos\frac{k_x}{2}\cos\frac{k_y}{2}\cos\frac{k_z}{2}
                        + t_2 \cos\left(\frac{k_x+k_y+k_z}{2}\right)\sin\frac{k_x}{2}\sin\frac{k_y}{2}\sin\frac{k_z}{2}
    \end{equation}
    \begin{equation}
        f_x^z(\mathbf{k}) = \lambda_1 \cos\left(\frac{k_x+k_y+k_z}{2}\right)\sin\frac{k_x}{2}\sin\frac{k_y}{2}\cos\frac{k_z}{2}
                          + \lambda_2 \sin\left(\frac{k_x+k_y+k_z}{2}\right)\cos\frac{k_x}{2}\cos\frac{k_y}{2}\sin\frac{k_z}{2}
    \end{equation}
    \begin{equation}
        f_y^z(\mathbf{k}) =-\lambda_1 \sin\left(\frac{k_x+k_y+k_z}{2}\right)\sin\frac{k_x}{2}\sin\frac{k_y}{2}\cos\frac{k_z}{2}
                          + \lambda_2 \cos\left(\frac{k_x+k_y+k_z}{2}\right)\cos\frac{k_x}{2}\cos\frac{k_y}{2}\sin\frac{k_z}{2}
    \end{equation}
    and
    \begin{equation}
        f^x_{x/y}(k_xk_yk_z) = f^z_{x/y}(k_yk_zk_x)\qquad 
        f^y_{x/y}(k_xk_yk_z) = f^z_{x/y}(k_zk_xk_y)
    \end{equation}
\end{widetext}
Here $\Delta$, $t_1$, and $t_2$ are the non-spin-orbital-coupling parameters,
 and $\lambda_1$, $\lambda_2$ are the spin-orbital-coupling parameters.
If we set $t_1=-t_2=1$ and $\Delta=1-\delta$ ($\delta>0$) two band inversions are created at $(000)$ and $(\pi\pi\pi)$.
The band inversions are stable under spin-orbital-coupling because the spin-orbital-coupling terms vanish at $(000)$ and $(\pi\pi\pi)$.
We plot the band structure and density of states with $t_1=-t_2=1$, $\Delta=0.8$, $\lambda_1=\lambda_2=0.2$
 in \figref{fig:224WTI}.
On one hand, the density of states in \figref[b]{fig:224WTI} shows that this model is fully gapped at half filling.
On the other hand, from the Fu-Kane criterion we get $z_{2\mathrm{w},1}=z_{2\mathrm{w},2}=z_{2\mathrm{w},3}=1$, $z_{4}=0$.
Therefore, this model corresponds to a weak TI with proper parameters.

\clearpage
\newpage
\section{A guide for the tables} \label{sec:table}

In this appendix, we present a users' guide for our main results, i.e., \cref{tab:eLC,tab:SI2TOP}, where the independent eLCs and the mapping from SI to TCI invariants in all SGs with nontrivial SI groups are tabulated, respectively.
With these data, the diagnosis of topology for a given material reduces to three steps of searching in the ``dictionary'' 
(i) applying standard first principle calculations to obtain the occupied BR at high symmetry momenta,
(ii) checking whether the band structure is fully gapped, which can be done by either calculating density of states or checking the compatibility relation \cite{Bradlyn2017,Vergniory2017,Elcoro2017},
and (iii) using \cref{tab:Indicator,tab:formula} to calculate the indicators and then looking for the corresponding TCI invariants in \cref{tab:SI2TOP}.
A concrete example of using these tables for SnTe is given in the main text.

As byproducts of our work, the independent eLCs and the possible TCI invariant combinations for SGs with trivial SI groups are tabulated in \cref{tab:eLC2,tab:TOP2}, respectively.
Although these states have no indicators and so can not be diagnosed from symmetry data, we think they can yet be regarded as an useful reference for future study in the sense that they are all the possible TCIs that can be realized by layer construction.

In the following we explain the notations in these tables.

\begin{table*}
\begin{tabular}{|>{\centering}p{2.2cm}|>{\centering}p{8cm}|c|}
\hline 
Lattice & SGs & $\mathbf{a}_{1}$, $\mathbf{a}_{2}$, $\mathbf{a}_{3}$\\
\hline 
\hline 
Triclinic primitive & 1, 2 & $\left(100\right)$, $\left(010\right)$, $\left(001\right)$\\
\hline 
Monoclinic primitive & 3, 4, 6, 7, 10, 11, 13, 14 & $\left(100\right)$, $\left(010\right)$, $\left(001\right)$\\
\hline 
Monoclinic base-centred & 5, 8, 9, 12, 15 & $\left(\frac{1}{2}\frac{1}{2}0\right)$, $\left(\bar{\frac{1}{2}}\frac{1}{2}0\right)$,
$\left(001\right)$\\
\hline 
Orthorhombic primitive & 16, 17, 18, 19, 25, 26, 27, 28, 29, 30, 31, 32, 33, 34, 47, 48, 49, 50, 51, 52, 53, 54, 55, 56, 57, 58, 59, 60, 61, 62 & $\left(100\right)$, $\left(010\right)$, $\left(001\right)$\\
\hline 
\multirow{2}{2.2cm}{Orthorhombic base-centered} & 20, 21, 35, 36, 37, 63, 64, 65, 66, 67, 68 & $\left(\frac{1}{2}\frac{1}{2}0\right)$, $\left(\bar{\frac{1}{2}}\frac{1}{2}0\right)$,
$\left(001\right)$\\
\cline{2-3}
 & 38, 39, 40, 41 & $(100)$, $\left(0\frac{1}{2}\frac{1}{2}\right)$, $\left(0\bar{\frac{1}{2}}\frac{1}{2}\right)$ \\
\hline 
Orthorhombic body-centred & 23, 24, 44, 45, 46, 71, 72, 73, 74 & $\left(\bar{\frac{1}{2}}\frac{1}{2}\frac{1}{2}\right)$, $\left(\frac{1}{2}\bar{\frac{1}{2}}\frac{1}{2}\right)$,
$\left(\frac{1}{2}\frac{1}{2}\bar{\frac{1}{2}}\right)$\\
\hline 
Orthorhombic face-centred & 22, 42, 43, 69, 70 & $\left(0\frac{1}{2}\frac{1}{2}\right)$, $\left(\frac{1}{2}0\frac{1}{2}\right)$,
$\left(\frac{1}{2}\frac{1}{2}0\right)$\\
\hline 
Tetragonal primitive & 75, 76, 77, 78, 81, 83, 84, 85, 86, 89, 90, 91, 92, 93, 94, 95, 96, 99, 100, 101, 102, 103, 104, 105, 106, 111, 112, 113, 114, 115, 116, 117, 118, 123, 124, 125, 126, 127, 128, 129, 130, 131, 132,
133, 134, 135, 136, 137, 138 & $\left(100\right)$, $\left(010\right)$, $\left(001\right)$\\
\hline 
Tetragonal body-centred & 79, 80, 82, 87, 88, 97, 98, 107, 108, 109, 110, 119, 120, 121, 122, 139, 140, 141, 142  & $\left(\bar{\frac{1}{2}}\frac{1}{2}\frac{1}{2}\right)$, $\left(\frac{1}{2}\bar{\frac{1}{2}}\frac{1}{2}\right)$,
$\left(\frac{1}{2}\frac{1}{2}\bar{\frac{1}{2}}\right)$\\
\hline 
Trigonal primitive & 146, 148, 155, 160, 161, 166, 167 & $\left(\frac{2}{3}\frac{1}{3}\frac{1}{3}\right)$, $\left(\bar{\frac{1}{3}}\frac{1}{3}\frac{1}{3}\right)$,
$\left(\bar{\frac{1}{3}}\bar{\frac{2}{3}}\frac{1}{3}\right)$\\
\hline 
Hexagonal primitive & 143, 144, 145, 147, 149, 150, 151, 152, 153, 154, 156, 157, 158, 159, 162, 163, 164, 165, 168, 169, 170,
171, 172, 173, 174, 175, 176, 177, 178, 179, 180, 181, 182, 183, 184, 185, 186, 187, 188, 189, 190, 191, 192, 193,
194  & $\left(100\right)$, $\left(010\right)$, $\left(001\right)$\\
\hline 
Cubic primitive & 195, 198, 200, 201, 205, 207, 208, 212, 213, 215, 218, 221, 222, 223, 224 & $\left(100\right)$, $\left(010\right)$, $\left(001\right)$\\
\hline 
Cubic face-centred & 196, 202, 203, 209, 210, 216, 219, 225, 226, 227, 228 & $\left(0\frac{1}{2}\frac{1}{2}\right)$, $\left(\frac{1}{2}0\frac{1}{2}\right)$,
$\left(\frac{1}{2}\frac{1}{2}0\right)$\\
\hline 
Cubic body-centred & 197, 199, 204, 206, 211, 214, 217, 220, 229, 230 & $\left(\bar{\frac{1}{2}}\frac{1}{2}\frac{1}{2}\right)$, $\left(\frac{1}{2}\bar{\frac{1}{2}}\frac{1}{2}\right)$,
$\left(\frac{1}{2}\frac{1}{2}\bar{\frac{1}{2}}\right)$\\
\hline 
\end{tabular}

\raggedright{}\protect\caption{\label{tab:lattice}
The primitive cell settings in SGs with nontrivial TCI SI groups. 
Settings for centered lattices (base-centered, body-centered, face-centered) follow
  \onlinecite{aroyo_bilbao_2009,Hahn2002},
 and the primitive lattice bases are written in the conventional coordinate.
}

\end{table*}

\textit{eLC.}
Each eLC is represented by a layer generating it,
 and the layer notation $(hkl;d)$ consists of 
 Miller indices in \emph{conventional} lattice
 and its position. 
General position is represented by the symbol ``$d_0$''.
Due to the mismatch between conventional and primitive cells,
 the vector
 $\mathbf{g} = h \mathbf{G}_1 + k \mathbf{G}_2 + l\mathbf{G}_3$,
 where $\mathbf{G}_{i=1,2,3}$ are the bases of the conventional reciprocal lattice,
 may not be the minimal reciprocal lattice in its direction,
 and the position $d$ is defined with respect to the minimal vector.
To be specific, the layer represented by $(hkl;d)$ is given by
 \begin{equation}
 \left(hkl;d\right) =
 \left\{ \mathbf{r}|\mathbf{r}\cdot \mathbf{g}^\prime=2\pi(d+q),\quad
  q\in\mathbb{Z},\ 0\le d<1\right\}
 \end{equation}
 where $\mathbf{g}^\prime$ is the minimal reciprocal vector in $\mathbf{g}$'s 
 direction.

\textit{SI group.}
For brevity, the SI group 
 $\mathbb{Z}_p \times \mathbb{Z}_q \times \cdots$
 will be written as $\mathbb{Z}_{p,q,\cdots}$.
% and the component whose odd values indicate strong TI is printed in red.
%For example the SI group in SG \#2 is printed as 
% $\mathbb{Z}_{2,2,2,4}$,
% where the red component corresponds to the $z_{4i}$ factor.

\textit{TCI invariant.}
In the first row of each SG section we list all the possible TCI invariants, wherein the weak invariants are denoted as 
 ``weak'', while all the other TCI invariants are represented by the correspinding symmetry elements.
It should be noticed that the weak invariants are defined with respect to the primitive cell. 
%For example, the 2-fold rotation index $\delta_{2r}$ in SG \#12 is represented by 
% $2^{010}$.

\textit{Convention-independent invariant.}
The convention-independent invariants (see \cref{sec:conv} for detail) for glide, rotation, inversion, screw, and $S_4$ symmetries are printed in double-stroke font and colored with blue.

\textit{Lattice.}
The lattice setting follows the international table \cite{Hahn2002,aroyo_bilbao_2009},
 and the convention for primitive cell is tabulated in \cref{tab:lattice}.

\textit{Symmetry element.}
Mirror plane is denoted as $m^{hkl}$, glide plane is denoted as $g_{t_1 t_2 t_3}^{hkl}$,
 $C_{n=2,4,6}$-rotation axis is denoted as $n^{uvw}$, inversion center is denoted as $i$, 
 $C_{n=2,4,6}$-screw axis with a $q/n$ translation along the axis is denoted as $n_q^{uvw}$,
 and $S_4$ center is denoted as $\bar{4}^{uvw}$, 
Here $uvw$ are the \emph{conventional} indices of the corresponding axis direction, 
 $hkl$ are the \emph{conventional} Miller indices of the corresponding mirror or glide plane,
 and $t_1 t_2 t_3$ are the \emph{conventional} coordinates of the corresponding glide vector.
In order to keep the table compact, negative indice $-u$ is printed as $\bar{u}$.
%For example, in SG \#12, 
% $2^{010}$ represents the 2-fold rotation along $\mathbf{A}_2$,
% and $m^{010}$ represents the mirror perpendicular with $\mathbf{G}_2$.
%Here $\mathbf{A}_{i=1,2,3}$ ($\mathbf{G}_{i=1,2,3}$) are the conventional 
% direct (reciprocal) lattice bases.
Analysis in \cref{sec:conv} shows that for identical symmetry elements
 locating at different positions it is enough to keep only one to define the 
 corresponding TCI invariant. 
In this work, we take the one with minimal $x_1$, and if two have same
 $x_1$ we take the one with minimal $x_2$, and then if two have same
 $x_2$ we take the one with minimal $x_3$, where $0\le x_{i=1,2,3}<1$ are in \emph{primitive} cell.

\textit{Additional symmetry element.}
Due to the discussion in \cref{sec:Layer}, screw or glide elements can exist even in symmorphic SGs,
 which is referred as ``additional symmetry elements'' in \onlinecite{Hahn2002}.
For example, in SG \#12 ($C2/m$), screw axis $2^{010}_1$ can be got by the rotation $2^{010}$ followed by a translation $\mathbf{a}_1$.

\textit{Mirror Chern number.}
As mirror Chern number $C_\mathrm{m}$ can be any integer, here we set a cutoff $p$ for each mirror and list only the TCIs with $-p<C_\mathrm{m}\le p$.
$p$ is properly chosen such that TCIs with higher $C_\mathrm{m}$ can be constructed from the listed TCIs.
This cutoff is printed in the subscript of the corresponding mirror element, such as $m_{(p)}^{010}$.
There are two kinds of mirror planes as shown in \figref[a,b]{fig:Cm}.
For the former we print both $C_{\mathrm{m},0}$ and $C_{\mathrm{m},\pi}$
 while for the latter we print only $C_{\mathrm{m},0}$ since $C_{\mathrm{m},\pi}$ always equals to zero.
In order to make the table more compact, negative mirror Chern number $-n$ is printed as $\bar{n}$.

\clearpage
\newpage

%\subsection{Elementary layer constructions in space groups with nontrivial indicator groups}
\LTcapwidth=0.9\textwidth
\begin{longtable*}{c|c| cccccccccccccccccccc }
\caption{eLCs in all SGs with nontrivial SI groups. \label{tab:eLC}}\\
\hline
\hline
\hline
{\textbf{eLC}} & {\textbf{SI}} & \multicolumn{ 20 }{c}{\textbf{Invariants}} \\
\hline
\endfirsthead
\multicolumn{ 22 }{c} {{\bfseries \tablename\ \thetable{} -- continued}}\\
\hline
{\textbf{eLC}} & {\textbf{SI}} & \multicolumn{ 20 }{c}{\textbf{Invariants}} \\
\hline
\endhead
\hline \multicolumn{ 22 }{r}{{Continued on next page}} \\
\endfoot
\hline \hline
\endlastfoot
\noalign{\vskip0.03cm}
\multicolumn{ 22 }{c}{Space group \#2 : $P\bar{1}$ } \\
\hline
\noalign{\vskip0.03cm}
 $(hkl;d)$ & $\mathbb{Z}_{2,2,2,4}$& weak & $i$\\
\hline
\noalign{\vskip0.03cm}
$ 001 ; 0 $& 0012& 001& $1$ \\
$ 001 ; \frac{1}{2} $& 0010& 001& $0$ \\
$ 010 ; 0 $& 0102& 010& $1$ \\
$ 100 ; 0 $& 1002& 100& $1$ \\
\hline
\noalign{\vskip0.03cm}
\multicolumn{ 22 }{c}{Space group \#10 : $P2/m$ } \\
\hline
\noalign{\vskip0.03cm}
 $(hkl;d)$ & $\mathbb{Z}_{2,2,2,4}$& weak & $m_{(2)}^{010}$& $2^{010}$& $i$\\
\hline
\end{longtable*}
%There are  306 kinds of nontrivial eLCs in total.

\clearpage

%\subsection{Elementary layer constructions in space groups with trivial indicator groups}
\LTcapwidth=0.45\textwidth
\begin{longtable}{c| cccccccccccccccccccc }
\caption{eLCs in all SGs with trivial SI groups. \label{tab:eLC2}}\\
\hline
\hline
\hline
{\textbf{eLC}} & \multicolumn{ 20 }{c}{\textbf{Invariants}} \\
\hline
\endfirsthead
\multicolumn{ 21 }{c} {{\bfseries \tablename\ \thetable{} -- continued}}\\
\hline
{\textbf{eLC}} & \multicolumn{ 20 }{c}{\textbf{Invariants}} \\
\hline
\endhead
\hline \multicolumn{ 21 }{r}{{Continued on next column}} \\
\endfoot
\hline \hline
\endlastfoot
\noalign{\vskip0.03cm}
\multicolumn{ 21 }{c}{Space group \#1 : $P1$ } \\
\hline
\noalign{\vskip0.03cm}
 $(hkl;d)$ & weak \\
\hline
\noalign{\vskip0.03cm}
$ 001 ; d_0 $& 001\\
$ 010 ; d_0 $& 010\\
$ 100 ; d_0 $& 100\\
\hline
\noalign{\vskip0.03cm}
\multicolumn{ 21 }{c}{Space group \#3 : $P2$ } \\
\hline
\noalign{\vskip0.03cm}
 $(hkl;d)$ & weak & $2^{010}$\\
\hline
\noalign{\vskip0.03cm}
$ 001 ; 0 $& 001& $1$ \\
\hline
\end{longtable}
%There are  323 kinds of nontrivial eLCs in total.

\clearpage

%\subsection{Mappings from indicators to invariants in space groups with nontrivial indicator group}
\LTcapwidth=0.9\textwidth
\begin{longtable*}{c| cccccccccccccccccccc }
\caption{Mapping from SI to TCI invariants in all SGs with nontrivial
 SI groups. \label{tab:SI2TOP}}\\
\hline
\hline
\hline
{\textbf{SI}} & \multicolumn{ 20 }{c}{\textbf{Invariants}} \\
\hline
\endfirsthead
\multicolumn{ 21 }{c} {{\bfseries \tablename\ \thetable{} -- continued}}\\
\hline
{\textbf{SI}} & \multicolumn{ 20 }{c}{\textbf{Invariants}} \\
\hline
\endhead
\hline \multicolumn{ 21 }{r}{{Continued on next page}} \\
\endfoot
\hline \hline
\endlastfoot
\noalign{\vskip0.03cm}
\multicolumn{ 21 }{c}{Space group \#2 : $P\bar{1}$ } \\
\hline
\noalign{\vskip0.03cm}
$\mathbb{Z}_{2,2,2,4}$& weak & $i$\\
\hline
\noalign{\vskip0.03cm}
0000& 000& 0\\
0002& 000& \textcolor{blue}{$\mathds{1}$}\\
0010& 001& 0\\
0012& 001& 1\\
0100& 010& 0\\
0102& 010& 1\\
\hline
\end{longtable*}
%There are  3133 kinds of nontrivial topologies
% and  478 kinds of nontrivial SIs in total.
\clearpage

%\subsection{Nonequivalent layer constructions in space groups with trivial indicator group}
\LTcapwidth=0.45\textwidth
\begin{longtable}{ cccccccccccccccccccc }
\caption{TCI invariant combinations in all SGs with trivial
 SI groups. \label{tab:TOP2}}\\
\hline
\hline
\hline
\multicolumn{ 20 }{c}{\textbf{Invariants}} \\
\hline
\endfirsthead
\multicolumn{ 20 }{c} {{\bfseries \tablename\ \thetable{} -- continued}}\\
\hline
\multicolumn{ 20 }{c}{\textbf{Invariants}} \\
\hline
\endhead
\hline \multicolumn{ 20 }{r}{{Continued on next column}} \\
\endfoot
\hline \hline
\endlastfoot
\noalign{\vskip0.03cm}
\multicolumn{ 20 }{c}{Space group \#1 : $P1$ } \\
\hline
\noalign{\vskip0.03cm}
 weak \\
\hline
\noalign{\vskip0.03cm}
000\\
001\\
010\\
011\\
100\\
101\\
\hline
\end{longtable}
%There are  961 kinds of nontrivial topologies
% and  0 kinds of nontrivial SIs in total.

\end{document}